\documentclass[aps,pre,preprint,superscriptaddress]{revtex4-1}

\usepackage{url}
\usepackage{amssymb}
\usepackage{mathtools}
\usepackage{graphicx}
\usepackage{xcolor}

\usepackage{subcaption}
\usepackage{float}

\newcommand{\dint}[1]{\mathrm{d}{#1}}
\newcommand{\ddif}[2]{\frac{\mathrm{d}{#1}}{\mathrm{d}{#2}}}
\newcommand{\dpart}[2]{\frac{\mathrm{\partial}{#1}}{\mathrm{\partial}{#2}}}
\newcommand{\dparth}[3]{\frac{\mathrm{\partial}^{#3}{#1}}{\mathrm{\partial}{#2}^{#3}}}

\newcommand{\ierfc}[2]{\mathrm{i}^{#2}\mathrm{erfc}(#1)}

\newcommand{\tlt}{\tau}
\newcommand{\kBT}{k_\mathrm{B} T}	
		
\newcommand{\td}{t_\mathrm{crit}}
\newcommand{\tk}{\tau_\mathrm{cor}}
\date{\today}

\begin{document}
\title{Transient nucleation driven by solvent evaporation}
	\author{René de Bruijn}
	\email{r.a.j.d.bruijn@tue.nl}
	\affiliation{Department of Applied Physics and Science Education, Eindhoven University of Technology, P.O. Box 513, 5600 MB Eindhoven, The Netherlands}
	\affiliation{Institute for Complex Molecular Systems, Eindhoven University of Technology, P.O. Box 513, 5600 MB Eindhoven, The Netherlands}
	\author{Jasper J. Michels}
	\affiliation{Max Planck Institute for Polymer Research, Mainz, Germany}
	\author{Paul van der Schoot}
	\affiliation{Department of Applied Physics and Science Education, Eindhoven University of Technology, P.O. Box 513, 5600 MB Eindhoven, The Netherlands}
\begin{abstract}
We theoretically investigate homogeneous crystal nucleation in a solution containing a solute and a volatile solvent. The solvent evaporates from the solution, thereby continuously increasing the concentration of the solute. We view it as an idealized model for the far-out-of-equilibrium conditions present during the liquid-state manufacturing of organic electronic devices. Our model is based on classical nucleation theory, taking the solvent to be a source of the transient conditions in which the solute drops out of solution. Other than that, the solvent is not directly involved in the nucleation process itself. We approximately solve the kinetic master equations using a combination of Laplace transforms and singular perturbation theory, providing an analytical expression for the nucleation flux, predicting that (i) the nucleation flux lags slightly behind a commonly used quasi-steady-state approximation, an effect that is governed by two counteracting effects originating from the solvent evaporation: while a faster evaporation rate results in an increasingly larger influence of the lag time on the nucleation flux, this lag time itself we find to decrease with increasing evaporation rate, (ii) the nucleation flux and the quasi-steady-state nucleation flux are never identical, except trivially in the stationary limit and (iii) the initial induction period of the nucleation flux, which we characterize with a generalized induction time, decreases weakly with the evaporation rate. This indicates that the relevant time scale for nucleation also decreases with increasing evaporation rate. Our analytical theory compares favorably with results from numerical evaluation of the governing kinetic equations.
\end{abstract}
\maketitle

\section{Introduction}
Organic electronic devices have received a significant amount of interest over the past decades in a large part due to their solution processability, enabling low-cost and large-scale manufacturing \cite{Richter2017MorphologyViewpoint}. These devices are promising candidates for a wide range of applications including (but not limited to) batteries, displays, transistors and organic photovoltaics (OPV) \cite{Mei2013,Ling2018,Janssen2019}. The (dry) active layer in OPV devices, for instance, typically consists of a blend of a polymeric electron donor and small molecular acceptor with a very complex morphology \cite{Gaspar2018RecentCells,McDowell2018SolventCells}. The blend film is prepared by depositing a solution containing these species onto a substrate, which consequently dries due to solvent evaporation that, e.g., in spin-coating processing is enhanced \cite{McDowell2018SolventCells}. The complex (phase-separated) morphology emerges spontaneously during this drying process. Optimal design of organic electronic devices is hampered by incomplete understanding of the impact of the processing conditions on the emergent morphologies \cite{Zhang2020KeyCoating,Michels2021PredictiveCoating}.

Different morphologies emerge depending on the type of demixing and the processing conditions, resulting in different functionalities \cite{Richter2017MorphologyViewpoint}. The interplay between the processing conditions and the emergent morphology has been studied extensively for the case of liquid-liquid phase separation, not only experimentally \cite{Bobbert2016,Franeker2015spincoating,Franeker2015cosolvents}, but also by means of molecular simulations \cite{Bobbert2016,Lee2020} and phase-field simulations \cite{Schaefer2016StructuringEvaporation,Schaefer2019MorphologyDestabilisation,Negi2018SimulatingInvestigation}. If the demixing involves the formation of a crystalline solid phase in a liquid host, the understanding of the effect of the processing conditions on the phase-separated morphology is much less complete \cite{vanFraneker2015PolymerFormation, Zhang2020KeyCoating}. Since the level of crystallinity and the grain morphology of a semiconducting thin film govern charge carrier mobility and modulate the trap density, this lack of insight arguably slows down material development and process optimization \cite{Zhu2020EfficientProperties,Virkar2010OrganicTransistors,Wu2020Fine-TuningCells}. This holds irrespective of the material type, that is, irrespective of whether it is organic, inorganic or hybrid, and whether the film comprises of a single component or a blend \cite{Zhu2020EfficientProperties,Tennyson2019HeterogeneitySemiconductors}. Hence, there is a need for predictive models that provide not only generic~\cite{Zhang2020KeyCoating,Michels2021PredictiveCoating}, but also quantitative descriptions of the influence of evaporation on crystal nucleation and growth.

Accurate modeling of liquid-solid demixing requires a proper understanding of nucleation of crystallites under the out-of-equilibrium conditions associated with the casting of thin films from an evaporating solvent. Our understanding of nucleation phenomena is largely based on the phenomenological classical nucleation theory (CNT) in one form or another, presuming stationary solution conditions. CNT posits that nucleation is a balance between the gain in free energy for transitioning from a metastable to a stable phase and a free energy cost for the required formation of an interface between these two phases \cite{Kashchiev2000}. The key result of CNT is that the stationary nucleation flux or rate, \textit{i.e.}, the number of stably growing nuclei that emerge per unit volume per unit time, is related to the maximum of the nucleation free energy barrier $\Delta W_*$ as $J \propto \exp(-\Delta W_*/\kBT)$, where the constant of proportionality depends on the precise growth kinetics \cite{Kashchiev2000}. Here, $k_\mathrm{B}$ is Boltzmann's constant and $T$ the absolute temperature. 

Underlying assumptions of CNT include invariant ambient conditions and constant concentration of free solute particles, the latter of which is obviously an approximation. This kind of description agrees qualitatively with experiments but often yields quantitatively wrong predictions, sometimes deviating by many orders of magnitude \cite{Auer2001PredictionColloids,Filion2010CrystalTechniques,Filion2011SimulationPersists}. Attempts to improve upon this state of affairs range from what often are \textit{ad hoc} corrections to any of the assumptions underlying CNT \cite{Reiss1997ResolutionEntropy,Reiss1998RoleNucleation,Drossinos2003ClassicalRevisited,Aasen2020CurvatureTheory,Aasen2023FreeSpinodal} or improving CNT itself \cite{Pan2004DynamicsModel,Peters2006ObtainingMaximization,Lechner2011ReactionEnsemble} using more sophisticated techniques such as (dynamical) density functional theory \cite{Oxtoby1998NucleationTransitions,Oxtoby1988NonclassicalTransition}\cite{Reguera2005TheSystems,Duran-Olivencia2018GeneralNucleation,Lutsko2013ClassicalNucleation,Lutsko2018SystematicallyTheory}, to applying projector-operator techniques to derive formally exact equations of motion \cite{Kuhnhold2019DerivationMarkovian}. Although these approaches often (but not always) yield improved accuracy with respect to experiments or simulations, they come at the expense of a significantly more complicated description and most of these approaches in fact reduce to CNT in some limiting case \cite{Vehkamaki2006ClassicalSystems}. Moreover, recent work suggests that the discrepancy between experiments and simulations or theory might also be due to a misinterpretation of the experimental findings \cite{Wohler2022HardExperiment}. 

Clearly, a stationary CNT description cannot be valid during the far-out-of-equilibrium solution deposition of organic thin films, as the processing precludes the emergence of stationary conditions. To shed light on how such processing conditions might influence the nucleation of crystallites in organic electronic devices, we study the effect of solvent evaporation in the context of CNT, making use of analytical theory and numerical methods. For clarity and conciseness, we focus on a liquid solution that contains a single solute and a single solvent that evaporates out of the system, where we treat the evaporation rate as a free parameter and only the solute is explicitly involved in the (unary) nucleation process. Current understanding of nucleation under such conditions is almost exclusively based on the so-called quasi-steady-state approximation, which presumes that the stationary CNT description remains valid if we replace all quantities by their current-time quantities within what may be called an adiabatic ansatz \cite{Kashchiev2000}. However, this kind of \textit{ad-hoc} approach represents an uncontrolled approximation as the effect of solvent evaporation is not explicitly taken into account but dealt with \textit{a posteriori}. 

Notable exceptions to such an \textit{ad-hoc} approach are, among others, the works of Shneidman \cite{Shneidman1994NucleationResults,Shneidman2007HeatingProblem} and those of Trinkhaus and Yoo \cite{Trinkaus1987NucleationSupersaturation}, which use a variety of techniques to approximately solve the relevant master equation, while explicitly including the non-stationarity of the problem at hand. These studies have focused on a variety of experimentally relevant origins of non-stationarity, such as the depletion of monomers \cite{Shneidman2011Time-dependentNucleation}, finite-rate temperature quenches \cite{Shneidman2007HeatingProblem} and periodically modulated external pressures in the context of the nucleation of cavities in a liquid \cite{Shneidman1994NucleationResults}. These authors have shown that non-stationary conditions may have a significant effect on the nucleation flux because their impact is not captured by the quasi-steady-state approximation even if the non-stationarity is not very pronounced. Still, the quasi-steady-state flux, that is, the flux predicted by CNT presuming the quasi-steady-state approximation to hold, does remain one of the central quantities in the description of Shneidman and others albeit renormalized and not evaluated at the current time as is done in the quasi-steady-state approximation, but at some time in the past~\cite{Shneidman2010}. 

In the present contribution, we apply the continuum description for the nucleation kinetics and study the time-dependent nucleation flux starting from a situation where the solution only contains freely dissolved solute particles. It turns out that the mathematical description for our model, if expressed in appropriately scaled variables, can be mapped onto that of Shneidman~\cite{Shneidman2010}, for which an (approximate) analytical solution is known if the nucleation barrier changes slowly or (nearly) linearly in time~\cite{Shneidman2010}. In fact, by applying singular perturbation theory we find an approximate expression for the nucleation flux that compares favorably with a numerical solution of the governing kinetic equations that we also pursued, recalling that the nucleation flux is defined as the number of supercritical nuclei that appear per unit time per unit volume. We find that the nucleation flux can indeed be expressed in terms of a (renormalized) quasi-steady-state nucleation flux, evaluated at some shifted time. Considering the specific case where the initial solution contains only free solute molecules, we conclude that the nucleation flux is initially vanishingly small and reaches a (renormalized) quasi-steady state only after some time. We quantify the associated lag time by introducing a generalized induction time, which is a generalization of the time lag to arbitrary cluster sizes that acts as a measure for the time required for a nucleus to grow from a single monomer to a size beyond that of the critical nucleus. The lag time decreases non-trivially with increasing evaporation rate, in agreement with our numerical calculations.

The remainder of this paper is structured as follows. In Section \ref{sec:CNT}, we summarize classical nucleation theory for simple solutions, and extend it to the case where the solvent evaporates out of the solution in Section \ref{sec:TNT}. In Section \ref{sec:conditional} we discuss the set of equations that govern the nucleation kinetics and introduce the simplifications we require to solve them. In Section \ref{sec:clustersizedistribution}, we derive an approximate formulation for the nucleation cluster size distribution under solvent evaporation. In Section \ref{sec:nucleationflux} we derive the nucleation flux from the cluster size distribution and discuss its implications. Next, in Section~\ref{sec:inductiontime}, we introduce a generalized induction time and study how the relevant time scale of the nucleation process is influenced by the non-stationary conditions and how it depends on the evaporation rate. We compare our analytical calculations with a numerical integration of the relevant kinetic master equation in Section~\ref{sec:casestudy}, showing on the whole good agreement. Finally, in Section \ref{sec:discussionandconclusion}, we discuss our results and conclude our work. We include a list of symbols in Section \ref{app:tablesymbols}.

\section{Classical Nucleation Theory}\label{sec:CNT}
\begin{figure}[bt]
    \centering
    \includegraphics[width = 0.6\textwidth]{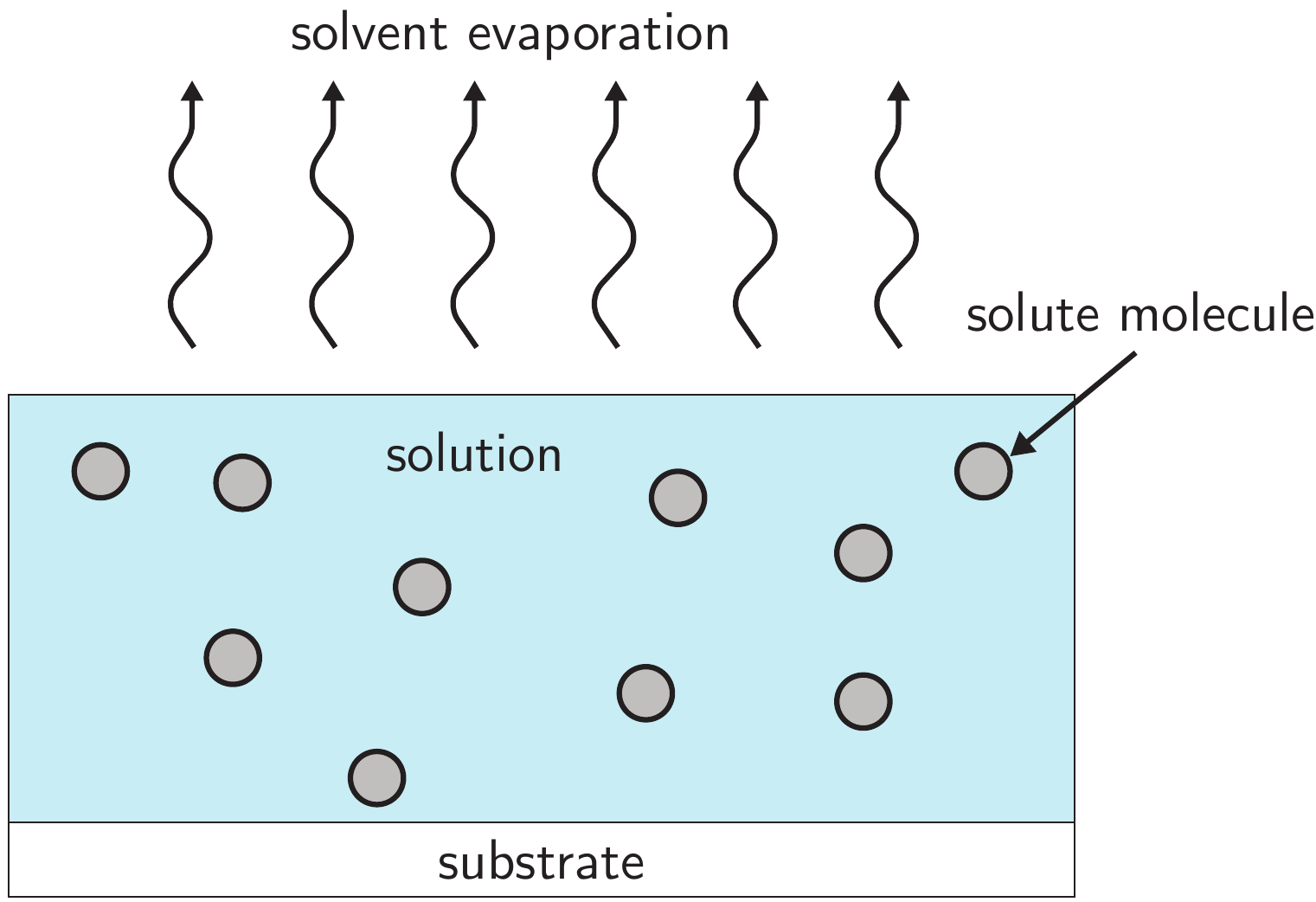}
    \caption{Schematic representation of our model. The fluid solution has a thin-film geometry and solvent evaporates at the top liquid-gas interface indicated by the wavey arrows. Solute molecules are schematically represented by spherical particles, which can aggregate into larger clusters consisting of multiple solute molecules.}
    \label{fig:schematic}
\end{figure}

We consider isothermal homogeneous nucleation in a thin film, consisting of a fluid containing a single solute in a volatile solvent that evaporates out of the solution. See Fig.~\ref{fig:schematic} for a schematic image of the model. Experimentally, maintaining isothermal conditions requires some form of temperature control, since the temperature of the thin film decreases due to solvent evaporation and the (latent) heat of vaporization that is required. This can be achieved, \textit{e.g.}, by using a substrate with a large heat capacity and a solute that conducts heat sufficiently fast or by actively maintaining the temperature of the substrate onto which the solution is cast~\cite{Bornside1989SpinModel}. We presume that the solvent acts as a so-called carrier fluid: a background fluid in which only the solute component is involved in the (unary) nucleation process~\cite{Kashchiev2000}. In our model description, solvent evaporation drives nucleation as it continuously increases the local solute concentration and with that the supersaturation. We also presume that the evaporation is sufficiently slow so that we can ignore any spatial inhomogeneities in the fluid film, originating from the accumulation of solute at the evaporating surface~\cite{Schaefer2017DynamicSolutions}. If this homogeneity approximation is not warranted, we could, in principle, take the spatial dependence of the supersaturation into account~\cite{Reguera2003HomogeneousGradient,Reguera2003HomogeneousFlow}. We stress, however, that in this case homogeneous nucleation is likely preempted by heterogeneous nucleation at the solution-air interface~\cite{Kashchiev2000}. We return to this issue at the end of this paper.

Following the standard description of CNT, we identify nuclei of the emerging solid phase using the number of solute particles $n$ in a cluster as the sole reaction coordinate. In our analysis, we treat $n$ as a continuous variable and insist that the clusters evolve via the attachment and detachment of single solute particles (monomers) only. As usual, we presume these stochastic processes to be Markovian, even though recent work suggests that in reality this might not necessarily be true~\cite{Kuhnhold2019DerivationMarkovian}. We choose to ignore this for two reasons. First, a non-Markovian description requires explicit knowledge of the so-called memory kernels~\cite{Kuhnhold2019DerivationMarkovian}, which as far as we are aware are presently unknown for our model system, even in the absence of evaporation. Second, if solvent evaporation has an effect on nucleation in solution, we expect it to already present itself in CNT. Hence, for us a CNT approach provides a convenient way to evaluate the role of externally imposed non-stationarity of the cluster size distribution. As we shall see, even within the framework of CNT our calculations turn out to be not trivial.

We expect the time evolution of the dimensionless cluster number densities $\rho(n,t)$ for $n>1$ to be described by the continuity equation in $n$-space,~\cite{Kashchiev2000,Kalikmanov2013}
\begin{equation}\label{eq:MasterEqContinuum}
    \dpart{\rho(n,t)}{t} = - \dpart{J(n,t)}{n} + K(n,t),
\end{equation}
where $J(n,t)$ is the cluster flux density in $n$-space, \textit{i.e.}, the number of nuclei of size $n$ that emerge per unit of time, and $K(n,t)$ a source term expressed per unit of time that enters due to solvent evaporation. We render the cluster densities $\rho(n,t)$ dimensionless via implicit multiplication with the molecular volume of a solute molecule $v_0$. (A list of symbols can be found in Section \ref{app:tablesymbols}.) The time evolution of the monomer density $\rho(n=1,t)$ is not described by Eq.~\eqref{eq:MasterEqContinuum}, but is typically included as a boundary condition. In the absence of evaporation, the most commonly used condition is to maintain a fixed concentration of monomers~\cite{Kashchiev2000}. This means that either the depletion of monomers is negligible during nucleation or that any cluster greater than some maximum size is removed from the solution and the monomers that they consist of are reinserted into the system, a common but somewhat artificial construction~\cite{Kashchiev2000}. If depletion of monomers is relevant, mass-conserving boundary conditions can be applied~\cite{Kashchiev2000}. We discuss the appropriate boundary conditions under evaporative conditions in more detail at a later stage.

We can obtain the cluster flux density $J(n,t)$ from the microscopic cluster dynamics. Often, the precise mechanism according to which monomers attach to a growing cluster is either known, or at least some reasonable ansatz can be made. In contrast, the detachment mechanism is typically not known or difficult to explicitly model~\cite{Kashchiev2000,Kalikmanov2013}. To circumvent this problem, we can make use of a construction known as the {\it constrained equilibrium approximation}, which is based on enforcing detailed balance. See, \textit{e.g.}, the works of Ruckenstein, Wu, Kashchiev and others~\cite{Ruckenstein1990ALiquids,Wu1996NucleationTheory,Slezov1998CommentsTheory,Schenter1999DynamicalNucleation,Kashchiev2000,Kalikmanov2013}. For this, we consider a hypothetical equilibrium where the cluster flux density $J(n,t)$ must be zero for all $n$ and hence where the cluster size distribution must be static $\rho(n,t)\equiv F_\mathrm{eq}(n)$ at least in the absence of evaporation. In such a hypothetical equilibrium, the dimensionless cluster size distribution $F_\mathrm{eq}(n)$ must be related to the nucleation barrier via a Boltzmann weight, enabling us to express the detachment rate in terms of the attachment rate and the ``equilibrium'' cluster size distribution~\cite{Kashchiev2000}. In the presence of evaporation the ``equilibrium'' cluster size distribution does depend on time and is quasi static, so $F_\mathrm{eq}(n)\rightarrow F_\mathrm{eq}(n,t)$. 

Following the standard prescription within CNT, we presume that this construction, which is obviously artificial since nucleation is an inherently out-of-equilibrium process, remains valid. Upon taking the continuum limit of the (discrete) microscopic nucleation rate, where we only include the Markovian contributions and presume clusters to grow via the attachment and detachment of monomers only, we find the cluster flux density $J(n,t)$ to obey \cite{Kashchiev2000,Kalikmanov2013}
\begin{equation}\label{eq:Fluxinnspace}
    J(n,t) = - k^+(n,t) F_\mathrm{eq}(n,t) \dpart{}{n}\left[\frac{\rho(n,t)}{F_\mathrm{eq}(n,t)}\right],
\end{equation}
where $k^+(n,t)$ is the (microscopic) attachment frequency for a monomer to an $n$-mer, which depends on the (time-dependent) monomer concentration and may depend on the cluster size via, \textit{e.g.}, the surface area in the case of reaction or interfacial-limited growth or with the linear dimensions (radius) of a cluster for diffusion-limited growth~\cite{Kashchiev2000}. Since our model does not depend on the functional form of the attachment frequency, we do not yet restrict it to any specific attachment model. The instantaneous ``equilibrium''  cluster density $F_\mathrm{eq}(n,t)$ follows Boltzmann statistics, and is therefore given as 
\begin{equation}\label{eq:ConstraintEquilibrium}
     F_\mathrm{eq}(n,t) = \rho(1,t)\mathrm{e}^{-\beta \Delta W(n,t)}.
\end{equation}
Here, $\beta$ denotes as usual the reciprocal thermal energy $1/\kBT$ and $\Delta W(n,t)$ is the thermodynamic work required to form an $n$-mer, that is, a cluster consisting of $n$ solute particles. 
The thermodynamic work depends on time on account of the impact of solvent evaporation that continually changes the number density even if their number remains constant. The set of equations given by Eqs.~\eqref{eq:MasterEqContinuum} and~\eqref{eq:Fluxinnspace} in the absence of the (evaporation) source term $K(n,t)$ is commonly referred to as the Zeldovich equation~\footnote{The validity and accuracy of this equation, specifically related to treating $n$ as a continuous variable, are discussed by \cite{Wu1992continuum}}~\cite{Zeldovich1943}. 

Within the framework of classical nucleation theory, the dimensionless work $\beta\Delta W(n,t)$ to form an $n$-mer from $n$ monomers has two main contributions and reads
\begin{equation}\label{eq:NucleationFreeEnergy}
    \beta \Delta W(n,t) = - n \beta \Delta \mu(t) + \beta \gamma_\infty A(n).
\end{equation}
The first term encodes that in an over-saturated solution the lowest free energy state corresponds to the condensed, solid state, not the metastable dissolved state. The free energy difference \textit{per monomer} between these states is given by the difference of their chemical potentials, $\Delta\mu$ \cite{Kashchiev2000}. We presume that a well-defined and sharp interface exists between the two phases, also known as the capillary approximation. The existence of an interface has a free energy cost associated with it, which is given by the second term in Eq.~\eqref{eq:NucleationFreeEnergy}, where $\gamma_\infty$ is the interfacial tension of a flat interface, and $A(n) = A_0 n^{2/3}$ the surface area of the nucleus of aggregation number $n$ with $A_0$ a unspecified geometric constant with units of area that depends on the cluster shape. Note that we tacitly assume the interfacial tension not to depend strongly on the crystal face exposed to the solvent, so we need not take the crystal structure into account.

Extrapolated to $n=1$, Eq.~\eqref{eq:NucleationFreeEnergy} tells us that the free energy to form a monomer $\beta \Delta W(1,t)$ must be non-zero. Since a monomer need not be formed, we correct Eq.~\eqref{eq:NucleationFreeEnergy} by subtracting from it the free energy cost associated with the formation of a monomer to come to a thermodynamically consistent description. Although this monomer correction has been subject to criticism, see, e.g., \cite{Kashchiev2000, Wu1996NucleationTheory}, we argue that we require it in the case of solvent evaporation to ensure that the number of solute monomers is conserved, as we show below. Obviously, for sufficiently small $n$, the free energy Eq.~\eqref{eq:NucleationFreeEnergy} cannot be valid because the concept of an interface breaks down. Nevertheless, we apply it for all cluster sizes, as for small $n$ the details of the nucleation free energy have a negligible influence on the quantities we are interested in~\cite{Kashchiev2000}. 

The chemical potential difference between a monomer in the dissolved and in the solid state we take to be that in a dilute solution, \textit{i.e.},
\begin{equation}\label{eq:ChemPotDifF}
    \beta\Delta \mu(t) = \ln\left[\frac{\rho(1,t)}{\rho_\infty}\right] = \ln S(t),
\end{equation}
which is generally reasonable for the production processes we study~\cite{Kashchiev2000}. Here, $\rho_\infty$ is the solubility limit and $S(t) \equiv \rho(1,t) / \rho_\infty$ the supersaturation. Only if $S(t)>1$ the condensed phase is more stable than the dissolved phase. For condensed phases, we do not need to consider the effect of the Laplace pressure in Eq.~\eqref{eq:ChemPotDifF} on account of their (near) incompressiblity~\cite{Kashchiev2000}. 
We now have all the ingredients describing the free energy $\Delta W$ required to form a cluster of $n$ solute particles. The most relevant properties of this free energy barrier are (i) the maximum height $\Delta W_* = \Delta W_*(t)$, (ii) the critical nucleus size $n_*=n_*(t)$ associated with this maximum height, which we obtain by finding the value of $n$ for which $\Delta W$ is stationary, and (iii) the so-called Zeldovich factor $Z=Z(t)$, which we define below and is a measure for the width of the nucleation barrier. 

Within our model, these quantities obey the following expressions:
\begin{equation}\label{eq:maxNucleationBarrier}
    \beta\Delta W_* = 3 \varepsilon^{-2} \left[ 1- \left(3n_*^{-2/3}+ 2n_*^{-1}\right)\right],
\end{equation}
where we define $\varepsilon \equiv 3 n_*^{-1/3} \sigma^{-1/2}$ for reasons of conventience with $\sigma \equiv \beta \gamma_\infty A_0$ a dimensionless interfacial tension, recalling that $A_0$ is a geometric constant, 
\begin{equation}\label{eq:criticalclustersize}
    n_* = \left(\frac{2 \sigma}{3 \ln S(t)}\right)^{3},
\end{equation}
the critical nucleus size and 
\begin{equation}\label{eq:zeldovichfactor}
    Z \equiv \sqrt{- \frac{1}{2 \pi} \dparth{\beta \Delta W}{n}{2}}\Bigg|_{n=n_*} = \sqrt{\frac{\sigma}{\pi}}\frac{1}{3}n_*^{-2/3} = \frac{1}{\sqrt{\pi}}\frac{1}{\varepsilon n_*},
\end{equation}
the above-mentioned Zeldovich factor. From the definition of the Zeldovich factor in Eq.~\eqref{eq:zeldovichfactor} we read off that it must be inversely proportional to the width of the `critical region', \textit{i.e.}, the width of the cluster size range around the critical nucleus, where thermal fluctuations dominate the growth of clusters because the barrier is locally nearly flat~\cite{Kashchiev2000}. For future reference we notice that the parameter $\varepsilon$ can be seen as a measure for the (square root of) the reciprocal barrier height. 

Our final model ingredient is the evaporation source term $K(n,t)$. In our idealized model, the number of molecules in a cluster only changes due to the presence of a gradient in the (nucleation) cluster flux density $J(n,t)$, and not directly due to solvent evaporation. Consequently, we can derive an expression for the evaporation source term most easily by considering the case where cluster growth is completely absent. For this purpose, we express the dimensionless cluster density as $\rho(n,t) = v_0 N(n,t)/V(t)$ in terms of the number of clusters $N(n,t)$ of size $n$ at time $t$, $V(t)$ the time-dependent volume of our (thin-film) solution and the molecular volume of a solute molecule $v_0$ introduced above. If we set the cluster flux density equal to zero, $J(n,t) = 0$, the number of monomers in a cluster does not change, implying that $N(n,t) = N(n)$. The cluster density still changes due to changes in the volume of the solution. This allows us to express $K(n,t)$ in terms of $V(t)$, where at time $t=0$ the evaporation starts and the film has the volume $V(0)$, because  then $\partial \rho(n,t) / \partial t = K(n,t)$ and $\rho(n,t) = v_0 N(n)/V(t)$. 
The source term we obtain can be written as 
\begin{equation}\label{eq:EvaporationSourceTerm}
    K(n,t) = -\rho(n,t) \frac{V(0)}{V(t)}\dpart{}{t}\left(\frac{V(t)}{V(0)}\right) = \rho(n,t) \dpart{}{t} \ln \chi(t),
\end{equation}
where, for convenience, we define the reciprocal dimensionless volume of the film $\chi(t)$ as
\begin{equation}\label{eq:inversevolume}
    \chi(t) = \frac{V(0)}{V(t)}.
\end{equation}
To make the functional dependence of $\chi(t)$ on time $t$ explicit, we would in principle need to set up a model for the solvent evaporation. We leave it unspecified for now and introduce our solvent evaporation model in a later stage. We shall presume that Eq.~\eqref{eq:EvaporationSourceTerm} holds even if nucleation takes place and the number of clusters of a certain size is time dependent.

To make headway, we take the initial state to already represent a supersaturated solution ($\Delta \mu(0) > 0$). This contrasts with the under-saturated initial state ($\Delta \mu(0) < 0$) that arguably would be the starting point in actual experiments, where supersaturation is achieved after time due to evaporation. However, insisting on a finite supersaturation at time zero allows us to relatively straightforwardly evaluate the time needed for nucleation to respond to evaporation-induced changes in the solute density. We note that we need not set the initial supersaturation to some infinitesimal value beyond the binodal, so $S(0^+)=1^+$, recalling that for $S(0)=1$ the nucleation barrier diverges; see Eq.~\eqref{eq:maxNucleationBarrier}. For nucleation barriers in excess of a few hundred times the thermal energy $\kBT$, relevant just beyond the binodal, nucleation is an exceedingly improbable and slow process, which, when combined with the relatively small volumes relevant for thin films, on the order of milli- to microlitres~\cite{Danglad-Flores2018DepositionAnalysis,Gumpert2023PredictingPhotovoltaics}, results in negligible nucleation~\cite{Kashchiev2000}. As a result, there is some critical supersaturation below which the expected number of nucleation events in a thin film is (much) lower than expected on the time scale of the experiments~\cite{Kozisek2011FormationVolumes,Tiwari2016StochasticSelf-assembly}. Hence, we can justifiably set the initial time $t = 0$ at or slightly below this critical supersaturation. The actual critical supersaturation can be estimated as the supersaturation when the equality $J(t) \times V(t) \times t_\mathrm{exp} = 1$ is satisfied. Here, $J$ is the nucleation flux, implicitly dependent on the supersaturation $S(t)$, $V$ the volume of the solution and $t_\mathrm{exp}$ the experimental time scale. It can therefore only be estimated if the kinetic and thermodynamic properties of the nucleating solute are known. For the sake of simplicity, we shall presume that the initial supersaturation is a free parameter. 

If the degree of supersaturation is stationary $S(t) = S$, \textit{i.e.}, in the absence of monomer depletion and solvent evaporation, when $\rho(1,t) = \rho > \rho_\infty$, $K(n,t) = 0$ and $\partial \rho(n,t)/\partial t =0$, the steady-state solution of Eqs.~\eqref{eq:MasterEqContinuum} and \eqref{eq:Fluxinnspace} for the case of a sufficiently large nucleation barrier $\beta \Delta W_* \gg 1$ reads \cite{Kashchiev2000}
\begin{equation}\label{eq:critflux}
    J_\mathrm{SS} \equiv \rho Z k^+_* e^{-\beta \Delta W_*} = Z k^+_* F_\mathrm{eq}(n_*).
\end{equation}
Here, $k^+_*$ denotes the attachment frequency of a monomer to the critical nucleus, $Z$ the Zeldovich factor, given by Eq.~\eqref{eq:zeldovichfactor}, and $F_\mathrm{eq}(n)$ the hypothetical equilibrium distribution given by Eq.~\eqref{eq:ConstraintEquilibrium}. Under transient conditions, Eq.~\eqref{eq:critflux} is no longer the actual solution to Eqs.~\eqref{eq:MasterEqContinuum} and \eqref{eq:Fluxinnspace}. If, however, solvent evaporation is sufficiently slow on the time scale of nucleation, then the so-called Quasi-Steady-State (QSS) approximation yields a reasonably accurate nucleation flux. This corresponds to letting all quantities in Eq.~\eqref{eq:critflux} depend on time. Since this does not actually account for how precisely solvent evaporation influences the nucleation rate, we focus next on solving Eqs.~\eqref{eq:MasterEqContinuum} and \eqref{eq:Fluxinnspace} explicitly while allowing for non-stationary effects. This means that we shall go beyond the QSS approximation. Nevertheless, we shall see that the nucleation flux obtained from the QSS approximation remains a central quantity in our results, away from the conditions where it holds.

\section{Transient nucleation theory}\label{sec:TNT}
In order to investigate the effect of solvent evaporation on (transient) nucleation, we find it useful to rewrite our governing equations into a form better suited for analytical treatment. This requires a number of steps. We first introduce the normalized density of clusters or distribution of clusters of size $n$ at time $t$ as 
\begin{equation}\label{eq:defnu}
\nu(n,t) \equiv \frac{\rho(n,t)}{F_\mathrm{eq}(n,t)},
\end{equation}
where $F_\mathrm{eq}(n,t)$ is given by Eq.~\eqref{eq:ConstraintEquilibrium}. As such, $\nu(n,t)$ expresses deviations from the instantaneous ``equilibrium'' cluster size distribution. We obtain the time-evolution equation for the normalized cluster density by combining Eqs.~\eqref{eq:MasterEqContinuum}, \eqref{eq:Fluxinnspace}, \eqref{eq:ConstraintEquilibrium}, \eqref{eq:NucleationFreeEnergy}, \eqref{eq:EvaporationSourceTerm} and \eqref{eq:defnu}, reading
\begin{equation}\label{eq:MasterEqContReduced1}
    \dpart{\nu(n,t)}{t} + \nu(n,t)\dpart{}{t}\left[\ln\left(\chi^{-1}(t) F_\mathrm{eq}(n,t)\right)\right] =  F_\mathrm{eq}^{-1}(n,t) \dpart{}{n}\left[ k^+(n,t) F_\mathrm{eq}(n,t) \dpart{\nu(n,t)}{n}\right]
\end{equation}
with $\chi(t)$ defined by Eq.~\eqref{eq:inversevolume}. 
Invoking the no-depletion boundary condition that conserves the \textit{number} of solute molecules (``monomers'') in free solution, results in a boundary condition for the monomer density $\rho(1,t) = \rho(1,0) \chi(t)$ that depends on time on account of evaporation, marking the deviation from standard CNT. This in turn allows us to simplify the second term on the left hand side of Eq.~\eqref{eq:MasterEqContReduced1} by defining the quantity 
\begin{equation}\label{eq:zeta}
    \zeta(n,t) \equiv \dpart{}{t}\left[\ln\left(\chi^{-1}(t) F_\mathrm{eq}(n,t)\right)\right] = -\dpart{}{t}\beta \Delta  W(n,t) =  \left(n-1\right)\dpart{}{t}\beta\Delta \mu(t),
\end{equation}
where we make use of the fact that the nucleation free energy Eq.~\eqref{eq:NucleationFreeEnergy} depends on time only through the chemical potential difference of the \textit{monomers} in the two phases, $\Delta \mu(t)$. The subtraction of unity in the factor $n-1$ accounts for the monomer correction to the nucleation barrier discussed in the previous section. 

For the sake of clarity, we rewrite Eq.~\eqref{eq:MasterEqContReduced1} in the form of a diffusion-advection equation or a Fokker-Plank equation if we interpret $\nu(n,t)$ as a scaled probability distribution function, 
\begin{equation}\label{eq:MasterEqContReduced2}
    \dpart{\nu(n,t)}{t} + \nu(n,t)\zeta(n,t) =   \dpart{}{n}\left[k^+(n,t) \dpart{\nu(n,t)}{n} \right] + g(n,t) \dpart{}{n}\nu(n,t),
\end{equation}
where
\begin{equation}\label{eq:detGrowthRate}
g(n,t) \equiv - k^+(n,t)\dpart{\beta \Delta  W(n,t)}{n}.
\end{equation}
Eq.~\eqref{eq:MasterEqContReduced2} shows that cluster growth has a diffusive contribution, which according to the standard interpretation of Fokker-Plank equations represents how, at the microscopic level, thermal fluctuations affect cluster growth, and an advective contribution that is driven by the generalized force $-\partial\beta \Delta  W(n,t)/\partial n$~\cite{Dhont2003AnColloids}. We interpret this advective contribution as the deterministic part of the cluster growth, as is usual in CNT~\cite{Zeldovich1943,Kalikmanov2013}. Taking the shape of the nucleation free energy landscape Eq.~\eqref{eq:NucleationFreeEnergy} into account, we note that $g(n,t)$ is negative for $n < n_*$, implying shrinking clusters, and positive for $n > n_*$, implying growing clusters, consistent with our interpretation of the deterministic part of our kinetic equation.

Next, we use Eq.~\eqref{eq:MasterEqContReduced2} and introduce a characteristic time scale for nucleation in our model. This time scale represents an estimate for the average residence time of a nucleus within the critical region, here referred to as the lifetime of a critical nucleus. To find a suitable expression, let us first consider the case where $\zeta(n,t)=0$ and evaporation does not take place. We remind the reader that the Zeldovich factor $Z$ (Eq.~\eqref{eq:zeldovichfactor}) can be interpreted as the reciprocal of the width of the critical region in $n$-space, where the nucleation free energy barrier is essentially flat, resulting in a very small deterministic growth rate $g(n)$. As a result, the size of a nucleus is in effect dominated by thermal fluctuations, encoded in the diffusive contribution to Eq.~\eqref{eq:MasterEqContReduced2}. 
Hence, for the growth of clusters in the critical region we can ignore the deterministic contribution. Considering that nucleation only occurs if the nucleation barrier is sufficiently large, Eq.~\eqref{eq:zeldovichfactor} implies that the width of the critical region must be very much smaller than $n_*$. Therefore, the attachment frequencies $k^+(n)$, which act as diffusion coefficients in $n$-space, can be considered independent of the cluster size $n$ and set equal to $k^+(n) = k^+(n_*)$. The resulting equation, appropriate only in the critical region of width $Z^{-1}$ around $n_*$, is a standard one-dimensional diffusion equation in $n$-space with constant diffusion coefficient. We conclude that the resulting average residence time in the critical region must be equal to
\begin{equation}\label{eq:lifetime}
    \td \equiv 2 \pi^{-1} k_+(n_*)^{-1} Z^{-2}.
\end{equation}
Here, the factor two accounts for the fact that half the clusters in the critical region grow and half shrink. The factor $\pi^{-1}$ is a common but non-standard constant of proportionality that emerges in CNT~\cite{Kashchiev2000,Wu1996NucleationTheory}. 

Under conditions where evaporation does take place and $\zeta(n,t)> 0$, the mean residence time becomes a function of time. Hence, we apply a QSS-type of approximation and define the characteristic time on an instantaneous basis as 
\begin{equation}\label{eq:QSSlifetime}
\td(t) = 2 \pi^{-1} k_+(n_*,t)^{-1} Z(t)^{-2}.
\end{equation} 
Below we show that this indeed turns out to be the relevant characteristic time scale, as it emerges naturally from Eq.~\eqref{eq:MasterEqContReduced2}. Since both this time scale and the characteristic ``length'' scale, given by the size of the critical nucleus $n_*(t)$, drift with time, we argue that the $(n,t)$-coordinate space used in Eq.~\eqref{eq:MasterEqContReduced2} is not suited for direct study. For this reason, we switch to a more natural reaction coordinate space, inspired by the work of Shneidman~\cite{Shneidman2010}. First, we introduce the reduced size coordinate $x=n/n_*(t)$, with $n_*(t)$ the critical cluster size that depends on the instantaneous supersaturation $S(t)$ according to Eq.~\eqref{eq:criticalclustersize}. The derivatives in Eq.~\eqref{eq:MasterEqContReduced2} transform to $\partial/\partial n \to n_*^{-1}\partial/\partial x $ and to $\partial/\partial t\big|_n \to \partial/\partial t\big|_x + (\partial x/\partial t)\big|_n \partial/\partial x$, where the subscript indicates which variable is to be held constant while taking the derivative with respect to time. The latter implies that we introduce a co-moving coordinate system where $x$ and $t$ are considered as independent variables. Next, we introduce the reduced attachment frequencies $\bar{k}^+(x,t)= k^+(x,t)/k^+(1,t)$ by dividing all attachment frequencies $k^+(x,t)$ by the critical attachment frequency $k_+^*(t)\equiv k^+(1,t)$ evaluated at the critical cluster size $n = n_*$ (or $x = 1$) and make use of Eq.~\eqref{eq:zeldovichfactor} and Eq.~\eqref{eq:lifetime}, to find, 
\begin{align}\label{eq:MasterEqSolf}
    \td(t)\dpart{\nu(x,t)}{t} & + \td(t)\nu(x,t)\zeta(x,t) = \\  
    \nonumber  
    &\quad\dpart{}{x}\left[\frac{1}{2} \bar{k}^+(x,t) \varepsilon^2\dpart{\nu(x,t)}{x}\right] + g(x,t) \dpart{}{x}\nu(x,t),
\end{align}
where we redefine the deterministic growth rate as
\begin{equation}\label{eq:detgrowthx}
    g(x,t) \equiv - \frac{1}{2} \bar{k}^+(x,t) \varepsilon^2\left(\dpart{\beta \Delta W(x,t)}{x}\right) + x \td(t)\left(\dpart{\ln n_*(t)}{t}\right).
\end{equation}
The nucleation free energy barrier in the $x,t$ coordinate system reads
\begin{equation}\label{eq:NucBarrierxt}
    \beta \Delta W(x,t) = 3 \varepsilon^{-2} \left(\left[3 x^{2/3} - 2 x\right] - \left[3 n_*^{-2/3}(t) - 2 n_*^{-1}(t)\right]\right),
\end{equation}
where the terms in the second square brackets on the right-hand side are due to the monomer correction introduced in the previous section, which turn out to not affect the growth rate Eq.~\eqref{eq:detgrowthx}. Recall that the parameter $\varepsilon$ is a measure for the reciprocal of the nucleation barrier height, see Eq.~\eqref{eq:maxNucleationBarrier}, which corresponds to $x=1$ in Eq.~\eqref{eq:NucBarrierxt}. 

Next, we introduce a reduced time ${\tlt}$, making use of the time-dependent lifetime of a critical nucleus $\td(t)$ in Eq.~\eqref{eq:QSSlifetime} that has emerged naturally from the transformation to the new reaction coordinate space in Eq.~\eqref{eq:MasterEqSolf}. Based on this equation, we argue that the convenient time variable would be one that yields $\dint \tlt = \dint t/\td(t)$ because it accounts for the difference between the rate at which time passes for the observer and the time experienced by a cluster. The reduced, or integrated time $\tlt$ that explicitly takes the drift of the unit of time $\td(t)$ into account~\cite{Crank1975TheDiffusion}, reads
\begin{equation}\label{eq:scaledtime}
\tlt = \int_0^{t}\frac{\mathrm{d} t'}{\td(t')}.
\end{equation}
For stationary nucleation processes, $\td(t) =\td$, so $\tlt$ encodes the time that passes in units of the lifetime of a critical nucleus $\td$. For non-stationary nucleation processes this interpretation remains valid, where Eq.~\eqref{eq:scaledtime} simply takes into account that the unit of time drifts with time, emphasizing the highly non-linear character of the problem at hand. 

The dominant contribution of the non-stationary solution conditions in Eq.~\eqref{eq:MasterEqSolf} due to solvent evaporation is $\zeta(x,\tlt)$, recalling that it is defined as the (negative) time derivative of the nucleation barrier. Using the definition of this quantity in Eq.~\eqref{eq:zeta}, we now may write it in terms of the $(x,\tlt)$ coordinate system as
\begin{equation}\label{eq:defI}
    \zeta(x,\tlt) = -\dpart{}{\tlt} \beta \Delta W(x,\tlt) = \left(n_*(\tlt) \dpart{}{\tlt} \ln \chi(\tlt) \right) \left[x-n_*^{-1}(\tlt)\right] \equiv \omega(\tlt) \psi(x,\tlt),
\end{equation}
with $\omega (\tlt)= n_*(\tlt) \partial \ln \chi(\tlt) / \partial \tlt $ a `non-stationarity coefficient' or `non-stationarity index'~\cite{Shneidman2010} that measures how strongly the non-stationary conditions affect the nucleation \textit{of a critical nucleus} of the solid phase of the solute. 
The non-stationarity index (or coefficient) depends on time but, as we shall argue below, only weakly so if the evaporation is sufficiently slow. The function $\psi(x,\tlt) = (x-n_*^{-1}(\tlt))$ is the cluster size-dependent part of $\zeta(x,\tlt)$, and follows straightforwardly from Eq.~\eqref{eq:zeta}. 
The physical interpretation of the non-stationarity coefficient is straightforward, if we consider the ratio of the time derivative of the monomer concentration and the monomer concentration itself, $\rho(1,\tlt)^{-1}\partial\rho(1,\tlt)/\partial \tlt = \dpart{}{\tlt} \ln \chi(\tlt)$, and compare this with Eq.~\eqref{eq:EvaporationSourceTerm} expressing the rate at which the density of clusters changes due to evaporation. See also Eq.~\eqref{eq:MasterEqContinuum}. This allows us to introduce the time scale of solvent evaporation as 
\begin{equation}\label{eq:timescale}
    t_\mathrm{evap}(\tlt) = \left(\partial \ln \chi(\tlt) / \partial \tlt\right)^{-1}
\end{equation}
Hence, using Eqs.~\eqref{eq:scaledtime},~\eqref{eq:defI}~and~\eqref{eq:timescale} we can now express the non-stationarity coefficient in units of the actual time $t$ as 
\begin{equation}\label{eq:nonstatcoefficient}
\omega(t) \equiv n_*(t)\td(t)/t_\mathrm{evap}(t).
\end{equation} 
Clearly, the quantity $\omega$ measures the relative magnitude of the time scales for evaporation and nucleation. Multiplication with the critical nucleus size $n_*(t)$ accounts for the fact that the evaporation timescale enters via the time derivative of the chemical potential difference \textit{per monomer}. For the remainder of this paper, unless explicitly stated otherwise, we consider all quantities to be in the $\{x,\tlt\}$ co-ordinate system.

\section{Mathematical model}\label{sec:conditional}

The set of equations given by the nucleation master equation in reduced form \eqref{eq:MasterEqSolf}, the growth rate $g(x)$ \eqref{eq:detgrowthx}, and the rate of change of the nucleation barrier due to solvent evaporation \eqref{eq:defI}, describe our model in the $\{x,\tlt\}$ coordinate space. This set of equations we obtain from the $\{n,t\}$ coordinate space equations without any approximations, except for the introduction of the (very customary) no-depletion boundary condition. As it turns out, our model in \textit{reduced form} maps onto the model by Shneidman~\cite{Shneidman2010}, who studied nucleation in a single-component model without solvent evaporation, but where the supersaturation changes due to, \textit{e.g.}, an externally applied time-dependent pressure or temperature. 

That the models map onto each other is actually not trivial, as our starting point, the continuity equation Eq.~\eqref{eq:MasterEqContinuum}, contains an evaporation source term $K$, whereas the work by Shneidman only considers cases without source term. The fact that the models are equivalent once expressed in reduced form is because we divide the cluster densities $\rho$ by the hypothetical equilibrium cluster densities $F_\mathrm{eq}$. Both densities are affected similarly by the evaporation source term, for which reason it drops out of the equations. Hence, we argue that our work generalizes the single-component work of Shneidman for conserved particle numbers to mixtures where the amount of solvent is \textit{not} conserved, whereas the amount of nucleating solute is. Because the remainder of this work uses nucleation master equation \eqref{eq:MasterEqSolf} in reduced form, we can apply the methods and results of Shneidman to our case~\cite{Shneidman2010}. 

Exact analytical solutions for equations of the form of Eq.~\eqref{eq:MasterEqSolf}, \textit{i.e.}, Fokker-Planck equations with size- and time-dependent coefficients, are, in general, not known~\cite{Risken1996Fokker-PlanckEquation}. However, following Ref.~\cite{Shneidman2010}, very accurate approximate solutions can still be obtained if we introduce some simplifications that are valid in experimentally relevant limits. 
We are able to simplify Eq.~\eqref{eq:MasterEqSolf} considerably by presuming that (i) the nucleation barrier is much larger than the thermal energy, so $\Delta W_* \gg \kBT$, which by itself is a prerequisite for the validity of CNT, and (ii) solvent evaporation is sufficiently slow for the following conditions to hold
\begin{equation}\label{eq:constraints}
    \frac{1}{n_*}\left|\dpart{n_*}{\tlt}\right| \equiv  3 \varepsilon^2 |\omega| \ll 1, \qquad \frac{1}{\varepsilon}\left|\dpart{\varepsilon}{\tlt}\right| \equiv \varepsilon^2 |\omega| \ll 1, \qquad \left|\frac{1}{\omega}\dpart{\omega}{\tlt}\right| \ll 1.
\end{equation}
Under these constraints, we can consider the critical cluster size, $n_*$, the reciprocal of the square root of the nucleation barrier, $\varepsilon$ and the non-stationarity index, $\omega$, to be so weakly time-dependent \textit{in the co-moving co-ordinate system} that they are essentially constant. These constraints also mean that the time derivative of the free energy barrier is, for all intents and purposes, independent of time $\zeta(x,\tlt) \equiv \zeta(x)$, and the same holds for the reduced attachment frequencies $\bar{k}^+(x,\tau)\equiv \bar{k}^+(x)$~\footnote{Due to the second-order reaction kinetics, we expect the attachment frequencies to be given by
$k^+(x,\tlt) = \rho(1,\tlt) \kappa(x,\tlt)$ with $\kappa(x,\tlt)$ a reaction coefficient associated with monomer attachment to a cluster of size $x$. As a result, the reduced attachment frequencies are $\bar{k}^+(x,\tlt)=\kappa(x,\tlt)/\kappa(1,\tlt)$. If $n_*$ and $\varepsilon$ can be treated as essentially independent of time, then, for all relevant models for the aggregation kinetics known to us, this holds for $\kappa(x,\tlt)/\kappa(1,\tlt)$ as well~\cite{Turnbull1949RateSystems,Kashchiev2000,Kalikmanov2013}}. 
Considering that a high nucleation barrier implies a small numerical value for its reciprocal square root $\varepsilon$, as can be concluded from Eq.~\eqref{eq:maxNucleationBarrier}, we find from Eq.~\eqref{eq:constraints} that such an approach is warranted for the variables $n_*$ and $\varepsilon$ if $\omega < \mathcal{O}(\varepsilon^{-2})$, and we justify our treatment of $\zeta$ as a constant \textit{a posteriori}. 

With these simplifications, we may for scaled cluster sizes $x \ll (\varepsilon^{2} \omega)^{-1}$ neglect the contribution in Eq.~\eqref{eq:detgrowthx} from the second term, reading $x (\partial\ln n_* / \partial\tlt) \sim  - x \varepsilon^2 \omega$ in reduced reaction coordinates. Since the clusters grow essentially unidirectionally for $x\gg 1$, and go downhill in the free energy landscape, our results limited to $x < (\varepsilon^{2} \omega)^{-1}$ are not affected by the omission of this term. We stress that since $\varepsilon^{2} \omega$ is assumed to be very small, this approach is justified even if the scaled cluster size $x$ is not particularly small. For larger clusters, where this contribution is not small, its omission would result in an overprediction of the rate at which clusters grow. Explicitly including this term for clusters for which we cannot neglect this contribution to Eq.~\eqref{eq:MasterEqSolf}, turns out to make analytical treatment exceedingly more difficult. We consider these large clusters to be outside the scope of this work, which primarily focuses on the nucleation of clusters under solvent evaporation and not on their subsequent growth to arbitrary sizes. In this late stage regime, our nucleation model, which treats clusters as non-interacting entities, becomes less accurate anyway, as the likelihood for clusters growing into each other increases with cluster size. To study this situation, one could treat nucleation and the subsequent growth of clusters as separate processes and also explicitly correct for excluded volume of clusters, \textit{e.g.}, by combining this work with the Johnson-Mehl-Avrami-Kolmogorov model~\cite{Cahn1995TheDomain}. At present, this model cannot account for the changing volume of the thin film solution, for which it would require appropriate adjustment.

Under these assumptions, taking $\varepsilon$, $n_*$ and $\omega$ to be constant and $\zeta(x)$ and $\bar{k}^+(x)$ to be independent of time, Eq.~\eqref{eq:MasterEqSolf} reduces to
\begin{align}\label{eq:MasterEqSolve}
    \dpart{\nu(x,\tlt)}{\tlt} & + \nu(x,\tlt)\zeta(x) = \\  \nonumber  &\quad\frac{1}{2}\varepsilon^2\dpart{}{x}\left[\bar{k}^+(x)\dpart{\nu(x,\tlt)}{x}\right] + g(x)\dpart{}{x}\nu(x,\tlt),
\end{align}
where we find the deterministic growth rate of the clusters $g(x)$ to be time-independent in the current coordinate system, and given by
\begin{equation}\label{eq:detgrowthx2}
    g(x) = -  \frac{1}{2}\varepsilon^2 \bar{k}^+(x) \left(\dpart{\beta \Delta W}{x} \right) = - 3(x^{1/3}-1)\bar{k}^+(x).
\end{equation}
Here, we inserted Eq.~\eqref{eq:NucBarrierxt} and recall that $g(x)<0$ for $x<1$ and $g(x)>0$ for $x>1$. See also our earlier discussion in the previous section. 
We call attention to the cancellation of the factor $\varepsilon^2$ in Eq.~\eqref{eq:detgrowthx2}, as it is multiplied with the factor $\varepsilon^{-2}$ in Eq.~\eqref{eq:NucBarrierxt}. This means that $g(x)$ represents the dominant mechanism for cluster growth over the whole size domain, except for sizes near that of the critical cluster, where it becomes very small.

Eq.~\eqref{eq:MasterEqSolve} is the central equation of this work. Although derived for homogeneous nucleation within the CNT framework, we point out that its extension to heterogeneous nucleation or allowing for more complex nucleation barriers is relatively straightforward. In Appendix ~\ref{app:heterogeneousNucleation} we outline how evaporation modifies heterogeneous nucleation within CNT. We solve Eq.~\eqref{eq:MasterEqSolve} invoking our initial and boundary conditions that in the co-moving coordinate system attain the following form
\begin{equation}\label{eq:boundaryConditions}
    \nu(x,0) = \frac{\rho(x,0)}{F_\mathrm{eq}(x,0)} = \delta(x-1/n_*), \qquad \nu(n_*^{-1},\tlt) = 1 \qquad \lim\limits_{x\rightarrow \infty}\nu(x,\tlt) = 0.
\end{equation}
We reiterate that the first condition ascertains that only monomers are present in the solution at $\tlt = 0$, the second one that the number of free solute molecules remains fixed and the third that the reduced cluster size distribution $\nu(x,\tlt)$ approaches zero for large cluster sizes $x\rightarrow \infty$. Recall that in our model experiment we instantaneously quench into a metastable state with an arbitrary degree of supersaturation at $\tlt = 0$, which also marks the point where solvent evaporation commences. The second condition follows from Eqs.~\eqref{eq:ConstraintEquilibrium} and \eqref{eq:defnu} when expressed in $x,\tlt$ co-ordinates, and only holds if the ``monomer correction'' is applied to the CNT free energy making it thermodynamically consistent. The third condition enforces the number of solute molecules per volume, and consequently the cluster density, to remain finite, noting also that the (hypothetical) ``equilibrium'' system described by Eq.~\eqref{eq:ConstraintEquilibrium} predicts an (unphysical) diverging cluster density in the limit of very large clusters.

In the following two sections, we discuss the steps required to derive an analytical expression for the nucleation flux $J(x,\tlt)$. In the next section, we first approximately solve Eq.~\eqref{eq:MasterEqSolve} using a combination of Laplace transforms and singular perturbation theory, and obtain an analytical expression for the nucleation cluster size distribution $\hat{\nu}(x,s)$ in the Laplace domain. This quantity we use in the subsequent section~\ref{sec:nucleationflux} to derive the analytical expression for the nucleation flux $J(x,\tlt)$ in the time domain. We show how this last quantity exhibits two distinct regimes in time, with an early time regime identified with the response of the nucleation flux to the initial quench at $\tlt = 0$. The duration of this early time regime can be characterized by a lag or induction time. The late-time regime emerges after this lag time. We discuss how the nucleation flux is affected by solvent evaporation at both early and late times, and show that in the late-time regime the nucleation flux can be described by a QSS-like flux. To shed light on the effect of solvent evaporation on the early-time behavior, we calculate a generalized induction time in the subsequent section, which we find to depend non-trivially on the solvent evaporation rate via the non-stationarity coefficient $\omega$. Finally, before we conclude this work, we compare our analytical results with numerical solutions of the \textit{discrete} nucleation master equation, explicitly introducing a simple model for solvent evaporation and the attachment frequencies.

\section{Cluster size distribution}\label{sec:clustersizedistribution}
In order to solve Eq.~\eqref{eq:MasterEqSolve}, and to find out how the nucleation flux is affected by solvent evaporation, we take the Laplace transform with respect to the time variable $\tlt$ and solve for the cluster size distribution $\nu(x,\tlt)$. We solve the resulting ordinary differential equation perturbatively using $\varepsilon \ll 1$ as an asymptotic expansion parameter, \textit{i.e.}, presuming that the nucleation barrier is asymptotically high. We use $\varepsilon$ rather than $\varepsilon^2$ as the expansion parameter, as terms of order $\varepsilon$ turn out to be relevant in some parts of the asymptotic analysis. The fact that the highest order derivative in Eq.~\eqref{eq:MasterEqSolve} occurs in combination with this very small parameter necessitates the use of singular perturbation techniques~\cite{Nayfeh2011IntroductionTechniques}. 
In the asymptotic limit where this parameter goes to zero, we obtain a lower-order, overconstrained PDE that can no longer satisfy all the boundary conditions given in Eq.~\eqref{eq:boundaryConditions} unless it does so by coincidence. We use a method known as matched asymptotic expansions to deal with this problem~\cite{Nayfeh2011IntroductionTechniques}. Because our calculations follow the prescription of Shneidman, we only discuss the main steps here, and refer to that work~\cite{Shneidman2010} and to Appendix~\ref{app:technicalDetails} for details. 

Taking the Laplace transform of Eq.~\eqref{eq:MasterEqSolve} yields
\begin{align}\label{eq:MasterEqSolveLaPlace}
    s \hat{\nu}(x,s) - \nu(x,0) + \hat{\nu}(x,s)\zeta(x) = \frac{1}{2} \varepsilon^2 \ddif{}{x}\left[\bar{k}^+(x)\ddif{\hat{\nu}(x,s)}{x}\right] + g(x)\ddif{\hat{\nu}(x,s)}{x},
\end{align}
where the hats indicate that we are dealing with Laplace-transformed quantities. 
The initial conditions imposed by Eq.~\eqref{eq:boundaryConditions} dictate that $\nu(x,0)$ must be zero in Eq.~\eqref{eq:MasterEqSolveLaPlace} for all reduced cluster sizes $x > 1/n_*$.
We remind the reader that the (deterministic) cluster growth rate $g(x)$, defined in Eq.~\eqref{eq:detgrowthx2}, is to leading order of $\mathcal{O}(\varepsilon^0)$, but changes sign (and vanishes) at $x = 1$. This suggests that the diffusive contribution in Eq.~\eqref{eq:MasterEqSolveLaPlace}, which is proportional to the small parameter $\varepsilon^2$, can only be important near $x=1$, that is, near the maximum of the free energy barrier where $g(x)$ is very small. As discussed in Section~\ref{sec:TNT}, the width of this region around $x=1$ is inversely proportional to the Zeldovich factor and hence proportional to $\varepsilon$ as is made explicit in Eq.~\eqref{eq:zeldovichfactor}. This means that this width becomes asymptotically small as $\varepsilon$ approaches zero. 

\begin{figure}[hbt]
        \includegraphics[width=0.99\textwidth]{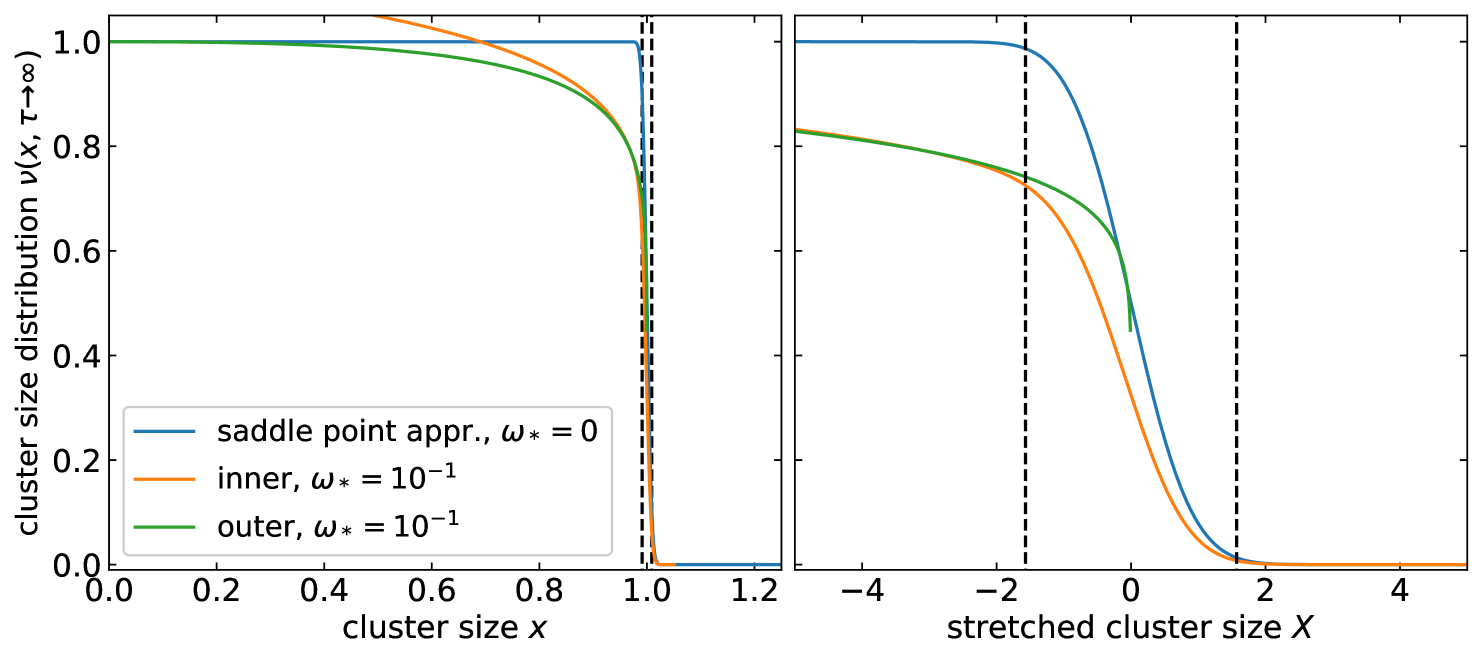}
    \caption{Left: The cluster size distribution $\nu(x,\tlt\to \infty)$ under stationary conditions ($\omega_* = 0$, blue) \cite{Kashchiev2000}, and under non-stationary conditions for $\omega_* = 10^{-1}$ and $\varepsilon = 10^{-2}$, with the outer solution for $x < 1$ (green) given in Eq.~\eqref{eq:smallersolution} and the inner solution around $x = 1$ (orange), given by the Laplace inverse of Eq.~\eqref{eq:inner}. The outer solution for $x > 1$ is not shown, as it is zero everywhere. The vertical dashed lines indicate edges of the critical region. Right: data and colors same as left, shown as a function of the stretched variable $X = (x-1)/\varepsilon$ and zoomed in on the critical region. Here we assume that the attachment frequencies vary very slowly with $x$ in the subcritical ($x < 1$) region for simplicity, so set $\bar{k}^+(x) = 1$.}
    \label{fig:nucleationdistribution}
\end{figure}

From standard perturbation theory we deduce that the ``critical'' region of width $\mathcal{O}(\varepsilon^{1})$ around $x=1$ corresponds to a so-called transition layer~\cite{Nayfeh2011IntroductionTechniques}. Hence, we expect the solution of Eq.~\eqref{eq:MasterEqSolve} to change rapidly in this very narrow region, which must be important at any order in the perturbation expansion. This we make evident in the left panel of Fig.~\ref{fig:nucleationdistribution}, which shows that in the absence of solvent evaporation the cluster size distribution (blue curve), obtained from Eq.~\eqref{eq:MasterEqContinuum} using a saddle-point approximation, indeed varies very weakly with cluster size, except near this transition layer~\cite{Kashchiev2000}. This allows us to divide our $x$-domain into three regions and find separate solutions valid for each of these. In the transition layer, where the diffusive contribution in Eq.~\eqref{eq:MasterEqSolveLaPlace} cannot be neglected, we obtain a so-called inner solution valid for $|x - 1| \lesssim \mathcal{O}(\varepsilon)$. In addition, there are two outer solutions for $x < 1$ and $x > 1$ away from the transition layer, where the terms multiplied with the small perturbation parameter $\varepsilon$ can be safely neglected. To resolve how the cluster size distribution changes in the asymptotically narrow transition region, we introduce a stretched variable $X = (x-1)/\varepsilon$~\cite{ShiSeinfeldOkuyama1990}, allowing us to study how the inner solution varies with the reaction coordinate $x$~\cite{Nayfeh2011IntroductionTechniques}, as we show in the right panel of Fig.~\ref{fig:nucleationdistribution}.

Below, we obtain the leading-order solution in all three regions by expanding $\hat{\nu}(x,s)$ in terms of the small variable $\varepsilon$, the reciprocal square root of the nucleation barrier, and only retaining the zeroth order ($\mathcal{O}\left(\varepsilon^0\right)$) term in powers of this variable. Due to the singular nature of the parameter expansion, the zeroth order term still explicitly depends on the expansion parameter $\varepsilon$. We do not include higher order terms in our expansion, as we believe it would produce only a very minor improvement to the zeroth order solution yet significantly complicate the mathematical treatment. This we base on the known higher-order solutions for the case of stationary nucleation~\cite{Hoyt1993HigherNucleation}. The piece-wise solutions valid in the three separate regions in $x$ we finally obtain by an appropriate matching procedure, demanding that the outer and inner solution should match at some intermediate point. The existence of the latter is evident from Fig.~\ref{fig:nucleationdistribution}, corresponding to the point where the inner and outer solution are (nearly) indistinguishable~\cite{Nayfeh2011IntroductionTechniques}. In principle, we can combine the inner and outer solutions to construct a composite solution that is uniformly valid for all values of the reduced cluster size $x$. As it turns out, this is neither required nor useful in order to find the nucleation flux in the supercritical regime ($x > 1$), hence we refrain from doing so. For more details on the singular perturbation method we refer, \textit{e.g.}, to \cite{Nayfeh2011IntroductionTechniques}.

Focusing first on the outer solutions, where, as mentioned, we can neglect the term multiplied by $\varepsilon^2$, we can straightforwardly solve Eq.~\eqref{eq:MasterEqSolveLaPlace} subject to the boundary conditions given in Eq.~\eqref{eq:boundaryConditions}, resulting in 
\begin{equation}\label{eq:smaller}
    \hat{\nu}_\mathrm{outer}(x < 1,s) = \frac{1}{s}\exp\left[s \tlt_\mathrm{d}(x) + \tlt_\omega(x)\right],
\end{equation}
and
\begin{equation}\label{eq:larger}
    \hat{\nu}_\mathrm{outer}(x > 1,s) = 0.
\end{equation}
Here, we find that the dimensionless time for a monomer to grow to a cluster of size $x < 1$ is related to the negative growth rate, since $g(x<1) < 0$, as 
\begin{equation}\label{eq:decaytime}
\tlt_\mathrm{d}(x) = -\int_{1/n_*}^{x} \frac{\dint x'}{g(x')},
\end{equation}
and
\begin{equation}\label{eq:tau_omega}
    \tlt_\mathrm{\omega}(x) = \int_{1/n_*}^{x} \frac{\zeta(x')}{g(x')}\dint x',
\end{equation}
a time encoding the effects of non-stationarity on the growth of subcritical clusters. Eq.~\eqref{eq:larger} is a consequence of the far-field boundary condition $\hat{\nu}(x \to \infty,s) = 0$ in Eq.~\eqref{eq:boundaryConditions}. Note that the solution Eq.~\eqref{eq:larger} does \textrm{not} imply that the actual cluster size density $\rho(x,\tlt)$ is equal to zero but that it is vanishingly smaller than the hypothetical equilibrium density for all times. This is also the case for stationary nucleation, as can be seen from the blue curve in Fig.~\ref{fig:nucleationdistribution}. Note also that the indicated solutions for $x\lessgtr 1$ distinguish between the two different outer solutions away from the critical zone around $x=1$. We refer to Appendix~\ref{app:technicalDetails} for details. In the time domain Eq.~\eqref{eq:smaller} results in
\begin{equation}\label{eq:smallersolution}
    \nu(x<1,\tlt) = e^{\tlt_\omega(x)}H(\tlt -\tlt_\mathrm{d}(x)),
\end{equation}
where $H(\dots)$ is the usual Heaviside step function. This we plot in Fig.~\ref{fig:nucleationdistribution} for $x < 1$ (green) in the quasi-stationary limit that emerges in the late time $\tlt \to \infty$, as a function of the (dimensionless) cluster size $x$ (left) and the stretched variable $X$ (right).

Let us now deal with the inner solution, valid for $|x-1| \lesssim \mathcal{O}(\varepsilon)$. To resolve the function in this very narrow region measuring a width of order $\varepsilon$, we switch to the stretched variable $X = (x-1)/\varepsilon$. Inserting this in Eq.~\eqref{eq:MasterEqSolveLaPlace}, expanding all functions in powers of $\varepsilon$ and only retaining the zeroth order contributions ($\mathcal{O}\left(\varepsilon^{0}\right)$) yields 
\begin{equation}\label{eq:innerequation}
\dparth{\hat{\nu}_\mathrm{inner}(X,s)}{X}{2} + 2 X \dpart{\hat{\nu}_\mathrm{inner}(X,s)}{X} - 2 \left[s + \omega_*\right] \hat{\nu}_\mathrm{inner}(X,s) = 0.
\end{equation}
Here, we made use of the fact that to lowest order in $\varepsilon$, $\zeta(X) \approx \zeta(x=1) \equiv \omega_*$, which using Eq.~\eqref{eq:defI} reads
\begin{equation}\label{eq:omegastar}
    \omega_* = \omega(1-n_*^{-1})
\end{equation}
and is a measure for the rate of change of the nucleation barrier at the critical size, see Eq.~\eqref{eq:defI} and the list of symbols~\ref{app:tablesymbols}. This quantity turns out to play a central role in our work, and measures how strongly solvent evaporation affects nucleation. We simplify the reduced attachment frequencies $\bar{k}^+(X) \approx 1$, where we tacitly presume that the attachment frequencies depend sufficiently weakly on the cluster size $x$, at least on the scale that is relevant for the inner solution. As far as we are aware, this seems to be true for all current microscopic models describing attachment of monomers to a growing cluster~\cite{Kashchiev2000,Turnbull1949RateSystems,Kalikmanov2013}. 

Eq.~\eqref{eq:innerequation} turns out to be a special case of the ordinary differential equation for the confluent hypergeometric function. The relevant solution we require for our model is given by~\cite{Shneidman1987,Shneidman1988,ShiSeinfeldOkuyama1990}
\begin{equation}\label{eq:inner}
    \hat{\nu}_\mathrm{inner}(X,s) = A(m)\ierfc{X}{m} + B(m)\ierfc{-X}{m},
\end{equation}
where $m = s + \omega_*\in \mathbb{Z}$ is a shifted Laplace ``frequency'' or variable, $A(m)$ and $B(m)$ yet to be determined functions and $\ierfc{X}{m}$ the repeated integral of the complementary error function~\cite{2010NISTFunctions}
\begin{equation}\label{eq:ierfc}
    \mathop{\ierfc{z}{n}}=\int_{z}^{\infty}\mathop{\ierfc{t}{n-1}}\,\dint{t}=\frac{2}{\sqrt{\pi}}\int_{z}^{\infty}\frac{(t-z)^{n}}{n!}e^{-t^{2}}\,\dint{t}.
\end{equation}
Here, `$\mathrm{i}^n$' indicates that the complementary error function is integrated $n$ times. Other solutions to Eq.~\eqref{eq:innerequation}, \textit{i.e.}, for $m \notin \mathbb{Z}$, turn out to be irrelevant when applying the inverse Laplace transform to return to the time domain, so we do not need to consider them. See, \textit{e.g.}, \cite{Shneidman1987,Shneidman1988,ShiSeinfeldOkuyama1990}.

We obtain the functions $A(m)$ and $B(m)$ by demanding that the inner and outer solutions are identical in the so-called distinguished limit $\varepsilon\to0$ at a suitably chosen `intermediate' value of the $x$-coordinate. At this intermediate value of the coordinate both the inner and outer solutions are valid. The existence of such value is a consequence of Kaplun's expansion theorem~\cite{Nayfeh2011IntroductionTechniques}. A common strategy is to choose the matching coordinate (slightly) outside the inner region, corresponding to the region where the inner and outer solutions overlap in Fig.~\ref{fig:nucleationdistribution}~\cite{ShiSeinfeldOkuyama1990,Shneidman1987,Shneidman1988}\cite{Nayfeh2011IntroductionTechniques}. This results in the matching conditions \cite{Nayfeh2011IntroductionTechniques}
\begin{equation}\label{eq:matchlower}
    \lim\limits_{x\to1^{-}} \hat{\nu}_\mathrm{outer}(x<1,s) = \lim\limits_{X\to-\infty} \hat{\nu}_\mathrm{inner}(X,s),
\end{equation}
and
\begin{equation}\label{eq:matchupper}
    \lim\limits_{x\to1^{+}} \hat{\nu}_\mathrm{outer}(x>1,s) = \lim\limits_{X\to\infty} \hat{\nu}_\mathrm{inner}(X,s),
\end{equation}
where the limits expressed in terms of the reduced size coordinate $x$ and the stretched coordinate $X$ emerge as we apply the distinguished limit $\varepsilon\to0$.

From these matching conditions we obtain
\begin{equation}\label{eq:constA}
    A(m) = \frac{1}{2}\frac{1}{m-\omega_*}\varepsilon^{m} \Gamma(m+1)\exp\left[m \tlt_\mathrm{stat} - \omega_* \left( \tlt_\mathrm{stat} -  \tlt_\mathrm{non-stat}\right)\right],
\end{equation}
with 
\begin{equation}\label{eq:defCconstant}
    \tlt_\mathrm{stat} = \int_{1/n_*}^{1}\dint x' \left(\frac{1}{g(x')} - \frac{1}{x'-1}\right) - \ln(1-n_*^{-1}),
\end{equation}
and 
\begin{equation}\label{eq:defDconstant}
    \tlt_\mathrm{non-stat} = \int_{1/n_*}^{1}\dint x' \left(\frac{\psi(x')}{\psi(1)}\frac{1}{g(x')} - \frac{1}{x'-1}\right) - \ln(1-n_*^{-1}),
\end{equation}
where $\psi(x) = 1 -n_*^{-1}$, referring to Eq.~\eqref{eq:defI}, and
\begin{equation}\label{eq:constB}
    B(m) = 0.
\end{equation}
Details of the calculations can be found in the Appendix~\ref{app:technicalDetails}. The solution for $B(m)$ follows straightforwardly because the outer solution for $x>1$ equals $\hat{\nu}_\mathrm{outer}(x,s)=0$. The constants $\tlt_\mathrm{stat}$ and $\tlt_\mathrm{non-stat}$ are (dimensionless) times, for which we provide a physical interpretation at a later stage in this paper. In some specific cases, these constants can be simplified and expressed in terms of elementary functions but here they depend on the (as yet) unspecified scaled attachment frequency $\bar{k}^+(x)$ via the deterministic growth rate $g(x)$. We note that the constant $\tlt_\mathrm{stat}$ also emerges if we were to study the stationary case with $\omega_* = 0$, whereas the constant $\tlt_\mathrm{non-stat}$ only emerges in the non-stationary case. In Fig.~\ref{fig:nucleationdistribution}, we plot the solution for $|x -1| \lesssim \mathcal{O}(\varepsilon)$ (orange) after transformation to the time domain in the quasi-stationary limit that emerges at $\tlt \to \infty$, as a function of the (dimensionless) cluster size $x$ (left) and the stretched variable $X$ (right).

\begin{figure}[tbh]
    \begin{subfigure}[b]{\textwidth}
    \includegraphics[width=\textwidth]{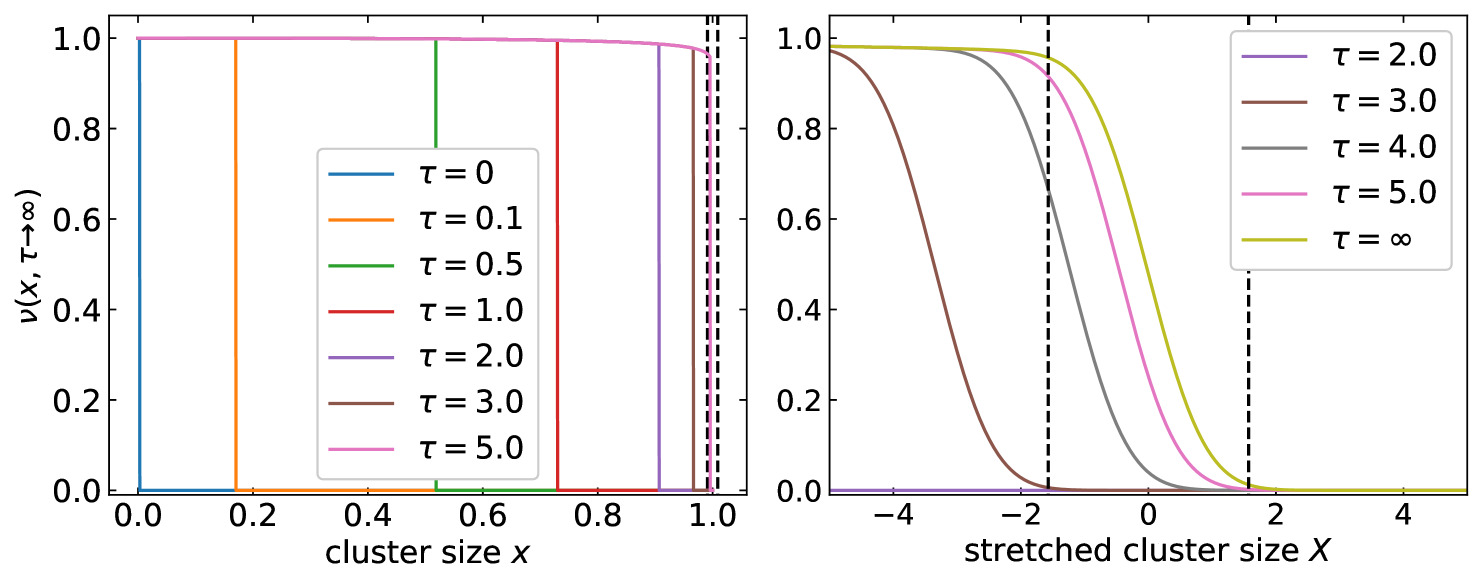}
    \end{subfigure}
    ~
    \begin{subfigure}[b]{\textwidth}
    \includegraphics[width=\textwidth]{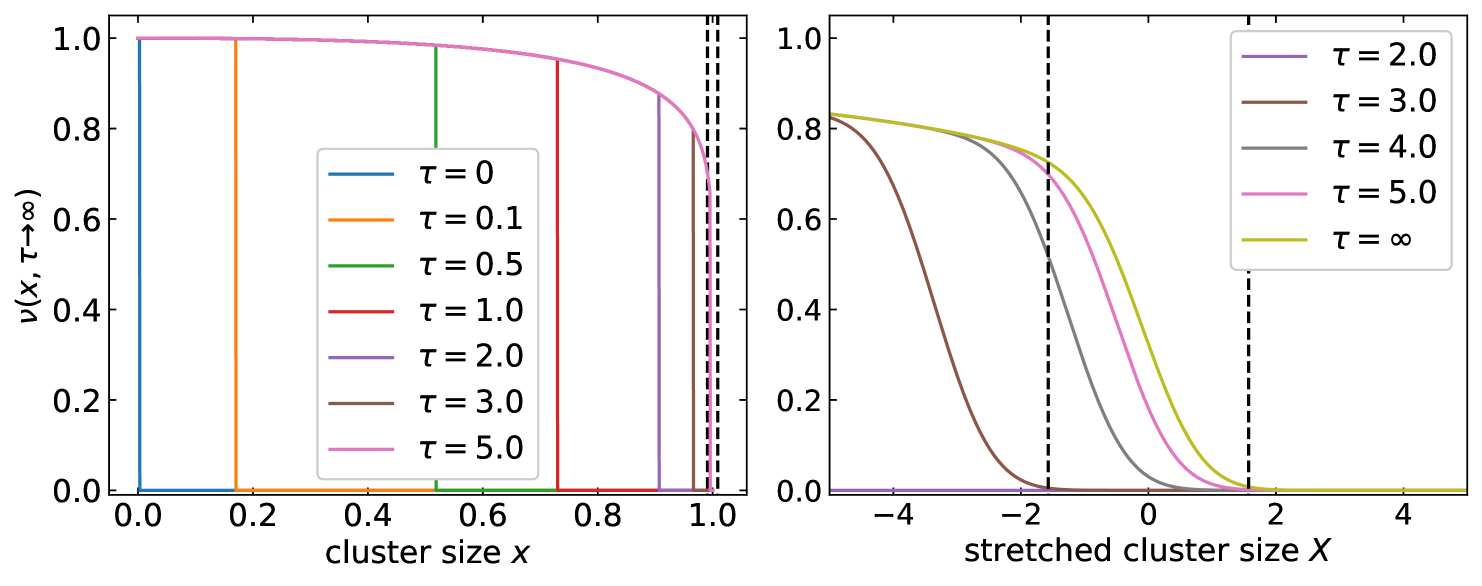}
    \end{subfigure}    
    \caption{The cluster size distribution as a function of the cluster size for various size for the asymptotic inverse barrier height $\varepsilon = 10^{-2}$, the reduced cluster frequencies $\bar{k}^+(x) = 1$ and non-stationarity indexes $\omega_* = 0.01$ (top) and $\omega_* = 0.1$ (bottom). Top left and bottom left: The subcritical cluster size distribution at various moments in time after the initial quench. As the wavefront reaches the critical cluster size, the subcritical size distribution $\nu(x,\tau)$ no longer changes. After this time, the wavefront enters the inner region, shown in the right figures.}
    \label{fig:sizedistributiontime}
\end{figure}

As already advertised, we refrain from constructing a uniformly valid, composite solution and focus on the properties of the piece-wise solutions. In Fig.~\ref{fig:sizedistributiontime} we show how the cluster size distribution $\nu(x,\tlt)$ depends on time and on the rate of solvent evaporation via the non-stationarity index $\omega_*$. Focusing first on the subcritical region for $x < 1$ using Eq.~\eqref{eq:smallersolution}, Fig.~\ref{fig:sizedistributiontime} shows that the clusters size densities are populated in such a way to represent a sharp wavefront, the propagation of which is related to the time required for a monomer to grow to a cluster of size $x$ $\tlt_\mathrm{d}(x)$. We associate the subcritical growth time $\tlt_\mathrm{d}(x)$ with a partial lag time accounting only for cluster growth up to a (subcritical) size $x < 1$. For vanishingly weak evaporation, so $\omega_* \approx 0$, $\nu(x,\tlt)$ approaches unity in the subcritical region for all values for $x<1$. For the case of faster solvent evaporation, $\omega_* > 0$ this no longer holds, as the cluster densities tend to lag behind the ``equilibrium'' cluster size distribution. This is encoded in time $\tlt_\omega(x)$, which implicitly depends on $\omega_*$. Physically, this means that the cluster densities cannot immediately accommodate the changes in time due to the changing nucleation barrier, causing them to lag behind the equilibrium cluster size distribution. This results in a value for $\nu(x,\tlt)$ smaller than unity. Note that for constant $\omega_*$, the subcritical cluster size distribution $\nu(x,\tlt)$ is independent of time after all subcritical clusters are populated, \textit{i.e.}, after the subcritical lag time has passed, $\tlt > \tlt_\mathrm{d}(x=1-\varepsilon)$, which must be evaluated at the edge of the critical region, after which the diffusive growth mechanism is most important \cite{ShiSeinfeldOkuyama1990}.

Near and in the critical region, the cluster size distribution is described by the inner solution Eq.~\eqref{eq:inner}, which does not permit an analytical Laplace inversion for $\omega_* \neq 0$. The cluster size densities remain close to zero until the subcritical clusters are populated, after which the (near-)critical cluster densities become populated as well. Here, the cluster growth front is no longer infinitely sharp and propagates with a different velocity, as the growth of clusters is now dominated by the diffusive contribution in Eq.~\eqref{eq:MasterEqSolf}. For very long times and for constant $\omega_*$, the critical cluster size distribution $\nu_\mathrm{inner}(x,\tlt)$ becomes independent of time as well. 

For subcritical and critical clusters the quantity $\nu(x,\tlt) = \rho(x,\tlt)/F_\mathrm{eq}(x,\tlt)$ is the relevant physical descriptor, since the cluster densities $\rho(x,\tlt)$ and the hypothetical cluster distribution $F_\mathrm{eq}(x,\tlt)$ turn out to be of similar magnitude~\cite{Shneidman1987,Shneidman1988,ShiSeinfeldOkuyama1990,Kashchiev2000}. This is not true in the supercritical region. Here, $F_\mathrm{eq}$ diverges with decreasing value of $\varepsilon$ as $\dpart{F_\mathrm{eq}}{x} \sim \mathcal{O}(\varepsilon^{-2})$.  At the same time, the cluster density $\rho(x,\tlt)$ is a decreasing function of $x$, which follows from the physical understanding that clusters that cross the nucleation barrier continue to grow unidirectionally. Hence, the quantity $\nu(x,\tlt)$ is a poor descriptor of the physical process in the supercritical region, because in the limit of an asymptotically high barrier, where $\varepsilon \to 0$, we must have $\nu(x > 1,\tlt) = 0$. The only information this expression contains is that $\rho(x,\tlt)$ is asymptotically infinitely smaller than $F_\mathrm{eq}$, while most of the relevant physics is now actually encoded in the cluster density $\rho(x,\tlt)$ itself and not in the hypothetical cluster distribution $F_\mathrm{eq}$. So a composite solution valid over the whole range of cluster sizes $x$ constructed from the piece-wisely valid outer and inner solutions can also not yield correct results in the supercritical region, capturing only the information in the hypothetical cluster distribution $F_\mathrm{eq}$.

As a result, the relevant physics that is captured by $\rho(x> 1,\tlt)$ remains obfuscated. This is especially problematic for quantities that depend on the cluster densities $\rho(x,\tlt)$ via the normalized cluster densities $\nu(x,\tlt)$, such as the nucleation flux $J(x,\tlt)$. Consequently, in the supercritical regime we cannot derive the nucleation flux from $\nu(x,\tlt)$. In Refs.~\cite{Shneidman1987,Shneidman1988,Shneidman2010}\cite{ShiSeinfeldOkuyama1990} a method was introduced to circumvent this problem, using what we know of how clusters, in our model, grow. In the next section, where we focus on the supercritical region where clusters grow steadily to macroscopic size, we apply this method to our model, which in effect extends the validity of the inner solution to the supercritical region. As it turns out, this is most easily done in the Laplace domain, and we only return to the time domain at the end of our analysis.

\section{Nucleation Flux}\label{sec:nucleationflux}
The definition of the nucleation flux (per unit of time), Eq.~\eqref{eq:Fluxinnspace}, can be written in the co-moving variables $x$ and $\tlt$ as
\begin{equation}\label{eq:Fluxinxspace}
    J(x,\tlt) = - 2 Z^{-1} \varepsilon \pi^{-1/2} \bar{k}^+(x) F_\mathrm{eq}(x,\tlt) \dpart{}{x} \nu(x,\tlt),
\end{equation}
which we obtain by multiplying Eq.~\eqref{eq:Fluxinnspace} by the lifetime of a cluster Eq.~\eqref{eq:lifetime} to render the expression in terms of the scaled time $\tlt$, and making it dimensionless. This expression relates the nucleation flux for a cluster with size $x$ to the reduced cluster size distribution $\nu(x,\tlt)$. As discussed in the previous section, we cannot directly use this equation in the supercritical region where $\nu(x>1,\tlt)$ is zero, as $\nu(x>1,\tlt)$ is a very poor descriptor of the physics at hand in the asymptotic analysis that we apply. This is a delicate issue that has been identified and discussed in detail by Okuyama and co-workers~\cite{ShiSeinfeldOkuyama1990} as well as by Shneidman~\cite{Shneidman1987,Shneidman1988}, who avoid the problem and introduce a method to obtain the supercritical nucleation flux from the inner solution indirectly by making use of what we understand from how, within our model, clusters grow.

According to Eq.~\eqref{eq:MasterEqContReduced2}, clusters grow unidirectionally in the supercritical region involving an advection-type process. As a result, if we obtain the nucleation flux at time $\tlt$ for some cluster size $x_0>1$ in the supercritical region and if we know the growth rate, then mass conservation dictates that the nucleation flux at some size $x_1 > x_0$ must be given by $J(x_1,\tlt) = J(x_0,\tlt - \tlt'(x_0,x_1))$, where $\tlt'(x_0,x_1)$ is the time it takes for a cluster to grow from size $x_0$ to size $x_1$. This time is related to the growth rate Eq.~\eqref{eq:detgrowthx2} according to
\begin{equation}\label{eq:innertoouter}
\tau'(x_0,x_1) = \int_{x_0}^{x_1} \frac{\dint x'}{g(x')}.
\end{equation}

In order to use the inner solution for this approach, we must choose our initial cluster size $x_0 > 1$ inside the supercritical region, yet sufficiently close to the critical region that it remains (reasonably) accurate, meaning $\varepsilon < x_0-1 \ll 1$~\cite{Shneidman1987}. Arguably, using such an intermediate size $x_0$ is similar in spirit to the matching procedure we carried out above, but now applied to the nucleation flux rather than the cluster density distribution $\nu(x,\tlt)$. Physically, this intermediate size corresponds to a cluster size where the diffusive contribution to cluster growth is sufficiently small that we can propagate cluster growth deterministically. The work of Okuyama and co-workers~\cite{ShiSeinfeldOkuyama1990} and that of Shneidman~\cite{Shneidman1987,Shneidman1988,Shneidman2010} seem not to agree on the appropriate choice for $x_0$. In the former work, the authors argue that the choice for the intermediate point $x_0$ must in one way or another influence the final outcome of the calculation. The latter works, on the other hand, claim that the exact value for $x_0$ is irrelevant and set the value of $x_0$ equal to unity. For the case of an asymptotically large barrier, the difference between these two approaches becomes vanishingly small, so we opt to set $x_0\to 1^+$. 

In what follows, we first derive an integral equation that expresses the non-stationary nucleation flux in terms of the stationary flux valid in the inner region. Next, we extend the solution to the supercritical region, where it turns out that the procedure described above only needs to be applied to the stationary flux, for which we can rely on earlier work~\cite{ShiSeinfeldOkuyama1990,Shneidman1987,Shneidman1988,Shneidman2010}. Finally, we solve the obtained integral equation and extract an explicit expression for the nucleation flux. 
It appears to be convenient to remain in the Laplace domain in order to find the nucleation flux. The definition of the nucleation flux Eq.~\eqref{eq:Fluxinxspace} can be rewritten in terms of the stretched cluster size $X = \varepsilon^{-1}(x-1)$ 
giving
\begin{equation}\label{eq:Fluxinnspace2}
    \widehat{J}(X,s) = \mathcal{L}\left\{\td J(X,t)\right\} = - 2 Z^{-1} \pi^{-1/2} \mathcal{L}\left\{\bar{k}^+(X) F_\mathrm{eq}(X,\tlt) \dpart{}{X} \nu(X,\tlt)\right\},
\end{equation}
where $\mathcal{L}\{\dots \}$ indicates the Laplace transform in the time domain. (See also the list of symbols in Section~\ref{app:tablesymbols}.)

To demonstrate that the non-stationary nucleation flux can be expressed in terms of the stationary nucleation flux, directly linking stationary and non-stationary nucleation phenomena, we first seek to simplify the expression for the equilibrium size distribution $F_\mathrm{eq}(x,\tlt)$ so that as a consequence we can simplify Eq.~\eqref{eq:Fluxinnspace2}. 
From Eq.~\eqref{eq:ConstraintEquilibrium} we conclude that in the $\{X,\tlt\}$-coordinate system the hypothetical equilibrium size distribution $F_\mathrm{eq}(X,\tlt) \propto \exp -\beta \Delta W(X,\tlt)$ depends on the stretched cluster size $X$ and the time $\tlt$ via the nucleation free energy only. The inner solution is valid only (very) close to the critical size corresponding to $X=0$, near the maximum of the nucleation barrier. As a result, we can approximate the nucleation free energy $\beta \Delta W(X,\tlt)$ given in Eq.~\eqref{eq:NucBarrierxt} by a parabolic function, which in the current variables results in the simple form $\beta \Delta W(X,\tlt) \approx \beta \Delta W (X=0,\tlt) - X^2$. This we presume to remain accurate in our extrapolation procedure and is similar in spirit to the saddle-point approximation fundamental to many of the well-established results in classical nucleation theory~\cite{Kashchiev2000,Kalikmanov2013}. Next, we focus on the time-dependent nucleation barrier $\beta \Delta W (X=0,\tlt)$ associated with the critical cluster size, and remind the reader that the time derivative of the nucleation barrier $\zeta$, defined in Eq.~\eqref{eq:defI}, can in our case be treated as a constant, see Sec.~\ref{sec:conditional}. Hence, we can replace the nucleation barrier $\beta \Delta W(X=0,\tlt)$ by the expansion
\begin{equation}
    \beta \Delta W(X=0,\tlt) = \beta \Delta W(0,0) + \dpart{\beta \Delta W(0,\tlt)}{\tlt}\Bigg|_{\tlt=0}\tlt = \beta \Delta W(0,0) - \omega_* \tlt,
\end{equation}
where $\omega_* \equiv \zeta(1)$. This turns out to be an exact expression as higher order terms vanish given our presumption of a constant first derivative. 

Inserting the resulting approximation $F_\mathrm{eq}(X,\tlt)=F_\mathrm{eq}(0,0)\exp \left(X^2 + \omega_* \tlt \right)$ in Eq.~\eqref{eq:Fluxinnspace2}, we find for the nucleation flux
\begin{equation}\label{eq:Fluxinxspace3}
    \widehat{J}(X,s) =  - 2 Z^{-1} \pi^{-1/2} F_\mathrm{eq}(0,0)e^{X^2}\bar{k}^+(X)\mathcal{L}\left\{\exp\left\{ \omega_* \tlt\right\}  \dpart{}{X} \nu(X,\tlt)\right\}.
\end{equation}
From Eq.~\eqref{eq:critflux} and Eq.~\eqref{eq:lifetime} we conclude that our dimensionless steady-state nucleation flux can be written as $\td J_\mathrm{SS} = 2  Z^{-1} \pi^{-1} F_\mathrm{eq}(0,0)$, allowing us to simplify Eq.~\eqref{eq:Fluxinxspace3} to
\begin{equation}\label{eq:Fluxinxspace4}
    \widehat{J}(X,s) =  - \sqrt{\pi}  J_\mathrm{SS} e^{X^2}\bar{k}^+(X)\mathcal{L}\left\{\exp\left\{ \omega_* \tlt\right\}  \dpart{}{X} \nu(X,\tlt)\right\}.
\end{equation}
Eq.~\eqref{eq:Fluxinxspace4} suggests taking a shifted Laplace transform, where the Laplace variable $s$ is replaced by $s - \omega_*$, resulting in 
\begin{equation}\label{eq:laplacedomflux}
    \widehat{J}(X,s) = -\sqrt{\pi} J_\mathrm{SS} \mathrm{e}^{X^2}\bar{k}^+(X)\dpart{\hat{\nu}(X,s-\omega_*)}{X}.
\end{equation}
The shifted Laplace frequency differs from the earlier-introduced shifted Laplace frequency $m= s + \omega_*$, which emerged in the expression for the inner solution for the cluster size distribution in the critical region $|x - 1| \ll 1$, Eq.~\eqref{eq:inner}. The different shifts in the Laplace frequency for the cluster size distribution and the nucleation flux show that they are affected differently by the solvent evaporation. This also becomes evident in our results: As shown in Fig.~\ref{fig:sizedistributiontime} the normalized cluster distribution becomes quasi-static in the late time regime, whereas, as we shall find, the nucleation flux always depends on time explicitly, even for long times. The shift in the Laplace parameter required for the nucleation flux cancels that in the cluster size distribution, implying that we must replace all $m$ frequencies by $s$ frequencies in the expressions for the inner solution Eqs.~\eqref{eq:inner}~and~\eqref{eq:constA}. 

Inserting the inner solution $\hat{\nu}_\mathrm{inner}(X,s)$, we obtain
\begin{align}\label{eq:laplacedomflux2}
    \widehat{J}(X,s) &= -\sqrt{\pi}  J_\mathrm{SS}\bar{k}^+(X)\mathrm{e}^{X^2} \times \\
    &\quad\dpart{}{X}\left[\left(\frac{\exp\{-\omega_*\left(\tlt_\mathrm{stat} - \tlt_\mathrm{non-stat}\right)\}}{2 (s-\omega_*)}\right)\varepsilon^{s} \Gamma(s+1)\exp\{s \tlt_\mathrm{stat}\} \ierfc{X}{s}\right], \nonumber
\end{align}
with $\ierfc{X}{s}$ again the repeated integral of the complementary error function, Eq.~\eqref{eq:ierfc}.
In the absence of evaporation, we have $\omega_* = 0$ and our expression reduces to the known result for the nucleation flux under stationary conditions $\widehat{J}_\mathrm{stat}(X,s)$~\cite{Shneidman1987,Shneidman1988,ShiSeinfeldOkuyama1990}
with
\begin{equation}\label{eq:laplacedomstat}
    \widehat{J}_\mathrm{stat}(X,s) = -e^{X^2} \bar{k}^+(X) \sqrt{\pi} J_\mathrm{SS}\frac{1}{2 s}\varepsilon^{s} \Gamma(s+1)\exp\{s \tlt_\mathrm{stat}\} \dpart{}{X}\ierfc{X}{s}.
\end{equation}
It transpires that Eq.~\eqref{eq:laplacedomflux2} factorizes into the product of a term describing the non-stationarity of the problem and the stationary nucleation flux, 
\begin{equation}\label{eq:laplacedomflux3}
    \widehat{J}(X,s) = \left[\frac{s}{(s - \omega_*)}\exp\{-\omega_*\left(\tlt_\mathrm{stat} - \tlt_\mathrm{non-stat}\right)\}\right] \widehat{J}_\mathrm{stat}(X,s).
\end{equation}

As we are interested in the nucleation flux in the time domain, we perform the inverse Laplace transform using the convolution theorem. We cast the convolution integral in terms of the quasi-steady state or QSS nucleation flux and the time-derivative of the stationary nucleation flux by applying integration by parts. We refer to Appendix~\ref{app:technicalDetails} for details of the calculations. 
Returning from the stretched cluster size $X$ to the reduced cluster size $x$, the resulting convolution-type equation expresses the non-stationary nucleation flux
\begin{equation}\label{eq:RealDomain}
    J(x,\tlt) = \int_0^{\tlt} \dint \tau' J_\mathrm{QSS}(\tau' + \tk) \dpart{}{\tlt} \frac{J_\mathrm{stat}(x,\tlt-\tau')}{J_\mathrm{SS}}
\end{equation}
in terms of the \textit{normalized} nucleation flux under \textit{stationary} conditions $J_\mathrm{stat}(x, \tlt)/J_\mathrm{SS}$, where $J_\mathrm{SS}$ is steady-state nucleation rate in the absence of solvent evaporation defined in Eq.~\eqref{eq:critflux}. Further, $\tk \equiv \tlt_\mathrm{non-stat} - \tlt_\mathrm{stat}$ is an emergent timescale where $\tlt_\mathrm{non-stat}$ and $\tlt_\mathrm{stat}$ are defined in Eqs.~\eqref{eq:defCconstant} and \eqref{eq:defDconstant}.  $J_\mathrm{stat}(x,\tlt)$ is the Laplace inverse of Eq.~\eqref{eq:laplacedomstat}. The quasi-steady-state nucleation flux $J_\mathrm{QSS}(\tlt)$ reads
\begin{equation}\label{eq:QSS}
    J_\mathrm{QSS}(\tlt) = 2  Z^{-1} \pi^{-1} F_\mathrm{eq}(x=1,\tlt) = J_\mathrm{SS}\frac{F_\mathrm{eq}(x=1,\tlt)}{F_\mathrm{eq}(x=1,0)} = J_\mathrm{SS}e^{\omega_* \tlt},
\end{equation}
using again the definition of the steady-state flux Eq.~\eqref{eq:critflux} and setting $J_\mathrm{QSS}(\tlt =0) \equiv J_\mathrm{SS}$. 

Equation \eqref{eq:RealDomain} expresses the non-stationary nucleation flux as the integrated product of the quasi-steady nucleation flux and the \textit{rate of change} of the stationary nucleation flux. It follows from Eq.~\eqref{eq:RealDomain} that the stationary and non-stationary nucleation processes are intrinsically connected, as could also have been expected given the fact that the quasi steady-state approximation provides a reasonable extension of CNT in the case of slowly changing ambient conditions~\cite{Kashchiev2000}. Physically, we can understand Eq.~\eqref{eq:RealDomain} as follows: The nucleation flux cannot instantaneously respond to changes in the ambient conditions and `relaxes' from some initially vanishing flux towards a final state in a finite amount of time, which gives rise to a time lag. This is true in the absence of evaporation. For the non-stationary case, so in the presence of evaporation, the final state drifts with time, and we must correct for this drifting ``final'' state. Eq.~\eqref{eq:RealDomain} expresses this by treating all the nucleation events before the present time $\tlt$ as separate events, starting at continually changing initial conditions. The non-stationary nucleation flux now corresponds to the sum (integral) of all these nucleation events and the subsequent growth of clusters to a predefined size $x$. 

The stationary nucleation flux in Eq.~\eqref{eq:RealDomain} can be obtained from the inverse Laplace transform of Eq.~\eqref{eq:laplacedomstat}, and reads~\cite{Shneidman1987,Shneidman1988,ShiSeinfeldOkuyama1990}
\begin{equation}\label{eq:stationaryFlux}
    J_\mathrm{stat}(x,\tlt) = J_\mathrm{SS}\exp\left[-\exp\left(\tlt - (\tlt'_\mathrm{i}(x_0) + \tau'(x_0,x))\right)\right],
\end{equation}
where $\tlt_\mathrm{i}'(x_0)$ represents the time for a \textit{single} nucleus to grow from the monomer state to the size $x_0 \geq 1 $. The timeshift $\tau'(x_0,x)$ is given by Eq.~\eqref{eq:innertoouter}, and must be introduced to propagate the inner solution into the supercritical region. It represents the time it takes to grow clusters from size $x_0$ to size $x$, presuming that only the deterministic growth rate is important for the size $x_0$ and that mass conservation holds. As we discussed at the start of this section, we set $x_0 = 1^+$. 
We refer to the quantity $\tlt_\mathrm{i}(x) = \tlt_\mathrm{i}'(1^+) + \tau'(1^+,x)$ as the incubation time, not to be confused with the induction time: the former describes the growth time of single clusters, whereas the latter is a statistic related to the nucleation flux, describing the delay period for the nucleation flux to respond to changing in the ambient conditions~\cite{Wu1992timelag}. The two quantities are related, which we make evident at a later stage of this article. The incubation time up to size $x$ we find to be identical to the expression derived by Shneidman~\cite{Shneidman1987,Shneidman1988} as 
\begin{equation}\label{eq:definductiontime}
    \tlt_\mathrm{i}(x) = \mathcal{P}\left(\int_{1/n_*}^{x}\frac{\dint x'}{g(x')}\right) + \ln\left(6 \beta \Delta W_*\right) - 2 \tlt_\mathrm{stat}.
\end{equation}
Here, $\mathcal{P}(\cdots)$ denotes the principal value of $\cdots$ and $g(x')$ is the deterministic growth rate, defined by Eq.~\eqref{eq:detGrowthRate}. The principle value notation was introduced by Shneidman~\cite{Shneidman1987} to correctly account for the divergence of the integrand at the critical cluster size for $x=1$. This is a consequence of combining the (non-diverging) subcritical and supercritical growth times into a single integral. The constant $\tlt_\mathrm{stat}$ we already defined in Eq.~\eqref{eq:defCconstant}. Irrespective of the model used for the attachment frequencies, Eq.~\eqref{eq:definductiontime} increases with increasing cluster size $x$.

The incubation time expresses the combination of contributions due to both the deterministic part of the growth, captured by the rate $g(x')$, and thermal fluctuations, the remaining terms. We refer to Refs.~\cite{Shneidman1987,Shneidman1992} for a more in-depth discussion on the incubation time, but stress that it can be rewritten and explicitly decomposed in three contributions: (i) the time for a cluster to grow from a monomer to the lower boundary of the critical region (related to the subcritical growth time $\tau_\mathrm{d}$ in Eq.~\eqref{eq:decaytime}), (ii) the time it takes to cross the critical region (Eq.~\eqref{eq:lifetime}), and (iii) the time to grow from a critical to a supercritical cluster of size $x$. Incidentally, this also implies that $\tlt_\mathrm{i}(x)\geq 1$. It is important to point out that the expression that we obtained for the stationary nucleation flux, Eq.~\eqref{eq:stationaryFlux}, is only accurate for $x > 1$ on account of the use of the growth time Eq.~\eqref{eq:innertoouter}~\cite{Shneidman1987, Shneidman1988}. Hence, our result for the non-stationary nucleation flux is, by construction, also only valid for $x>1$. This is no limitation for the present work, as we are only interested in the nucleation of stably growing, supercritical clusters.

Before we continue with integrating Eq.~\eqref{eq:RealDomain}, we need to point out an inconsistency in Eq.~\eqref{eq:stationaryFlux}: it does not correctly account for the very early times, resulting in a non-zero nucleation flux at the initial time $\tlt = 0$. As far as we are aware, this inconsistency is present in all results on the kinetics of classical nucleation theory obtained using similar singular perturbation techniques, under both stationary and non-stationary ambient conditions\cite{ShiSeinfeldOkuyama1990,Hoyt1993HigherNucleation,Shneidman1987,Shneidman1992,Shneidman2010}. It has been attributed to be a direct consequence of the use of a singular perturbation approach~\cite{Shneidman1987}. For stationary nucleation, this inconsistency is generally not a practical issue, as the value of Eq.~\eqref{eq:stationaryFlux} at the initial time $\tlt = 0$ is many orders of magnitude smaller than unity, resulting in a negligibly small correction. 

Finally, we can explicitly integrate Eq.~\eqref{eq:RealDomain} by combining it with Eqs.~\eqref{eq:QSS}~and~\eqref{eq:stationaryFlux}. This produces our result for the nucleation flux for homogeneous solute nucleation from solution that we write in a form that facilitates interpretation,
\begin{equation}\label{eq:nucleationFluxResult}
    J(x,\tlt) = J_\mathrm{tQSS}(x,\tlt)Q\left(\omega_* + 1, \exp\left\{-(\tlt - \tlt_\mathrm{i}(x))\right\}\right) - J_0(x),
\end{equation}
with
\begin{equation}\label{eq:correctedNucleationFluxResult}
    J_\mathrm{tQSS}(x,\tlt) = \Gamma(\omega_* +1) J_\mathrm{QSS}(\tlt - \tlt_\mathrm{i}(x) + \tk)
\end{equation}
a renormalized, time-shifted quasi-steady-state (tQSS) nucleation, $\Gamma(\dots)$ denotes the gamma function, and $Q(n,z) = \Gamma(n,z)/\Gamma(n)$ the regularized gamma function with $\Gamma(n,z)$ the upper incomplete gamma function of the variables $n$ and $z$~\cite{2010NISTFunctions}. As it turns out, the prefactor in Eq.~\eqref{eq:correctedNucleationFluxResult} can be absorbed in the time shift, resulting in the time shift $\tlt - \tlt_\mathrm{i}(x) + \tk + \omega_*^{-1} \ln \omega_*\Gamma(\omega_*)$, making explicit that the evaporation rate actually influences the delay or lag time. We return to this in the next section and show that this evaporation-induced time shift also emerges naturally from a (generalized) induction time. The quantity $J_0(x)$ in Eq.~\eqref{eq:nucleationFluxResult} is equal to the first term in Eq.~\eqref{eq:nucleationFluxResult} evaluated at $\tlt = 0$ and follows from the lower integration boundary in Eq.~\eqref{eq:RealDomain}. These expressions for the nucleation flux also apply for the case of heterogeneous nucleation, introducing minor modifications only. See Appendix.~\ref{app:heterogeneousNucleation} for details.

Eq.~\eqref{eq:correctedNucleationFluxResult} is the first main result of this paper. It applies under the presumptions of (i) a high nucleation barrier, so its reciprocal value $\varepsilon^2$ must be very small, and (ii) that the non-stationarity index $\omega_*$ is constant in the co-moving reference frame and satisfies $\omega_* \varepsilon^2 \ll 1$. For $\omega_* = 0$, $J_\mathrm{tQSS}(x,\tlt) = J_\mathrm{QSS}(0) = J_\mathrm{SS}$. The shift in time present in Eq.~\eqref{eq:correctedNucleationFluxResult} arguably represents some form of memory effect, which we argue makes sense because the nucleation flux describes the rate of emergence of clusters of size $x$; these nuclei do not form instantaneously at this size but grow from a single monomer to a supercritical nucleus under conditions that change continuously due to the on-going solvent evaporation. 

For the remainder of this article, we shall for simplicity set $J_0(x) = 0$. This is allowed for two reasons. First, for high nucleation barriers and $\tlt=0$, the regularized gamma function $Q$ entering Eq.~\eqref{eq:correctedNucleationFluxResult} is very many orders of magnitude smaller than unity and can, for all intents and purposes, be considered identical to zero. Second, the reason why $J_0(x)$ in principle should be non-zero at all, originates from the inconsistency in the stationary flux Eq.~\eqref{eq:stationaryFlux}, which is not strictly zero at the initial time $\tlt = 0$. 

This argument does only hold for values of $\omega_*$ that are not too high, otherwise the value of the regularized gamma function $Q$ can no longer be neglected at $\tlt =0$. This we find to be related to a vanishing and ultimately negative (``virtual") induction time, which we argue in the next section to only occur if $\omega_* \gtrsim 5.3$, although the exact value depends on the incubation time $\tlt_\mathrm{i}$. As a result, for $\omega_* > 5.3$, $J_0(x)$ can no longer be justifiably set equal to zero. It turns out that this issue cannot be remedied simply by accounting for a non-zero $J_0(x)$, as the resulting expression still lacks a lag time. This we return to in detail in the next section. Hence, we believe that the origin of the problem actually lies in Eq.~\eqref{eq:stationaryFlux}, which does not accurately account for the (very) early times, as discussed above. We postpone an in-depth discussion for which values our results remain accurate to the next sections and first discuss the implications of Eq.~\eqref{eq:nucleationFluxResult}.

\begin{figure}[tb]
    \begin{subfigure}[b]{0.49\textwidth}
        \includegraphics[width=\textwidth]{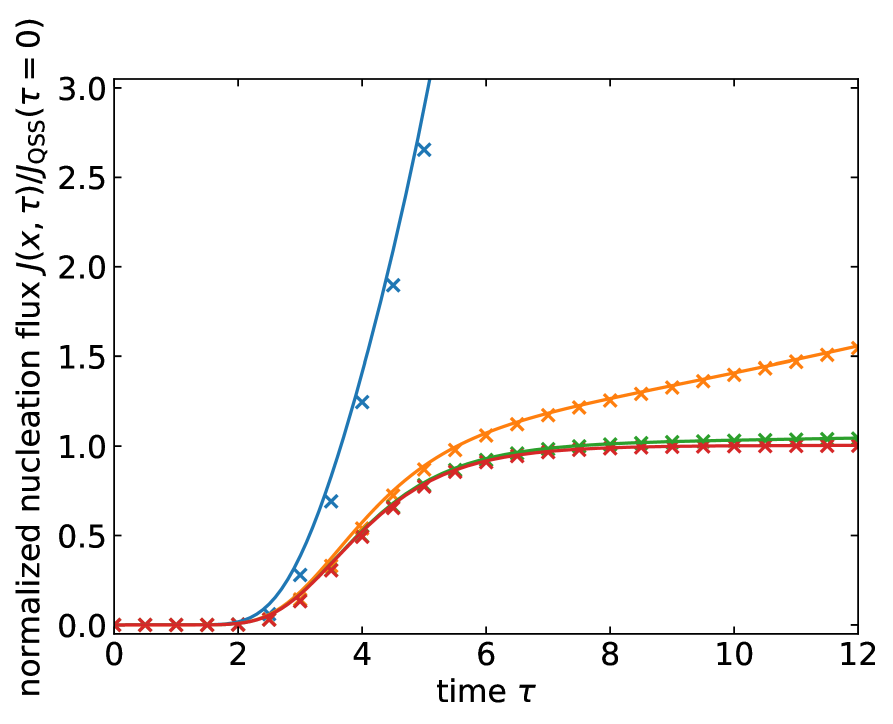}
    \end{subfigure}
    \hfill
    \begin{subfigure}[b]{0.49\textwidth}
        \includegraphics[width=\textwidth]{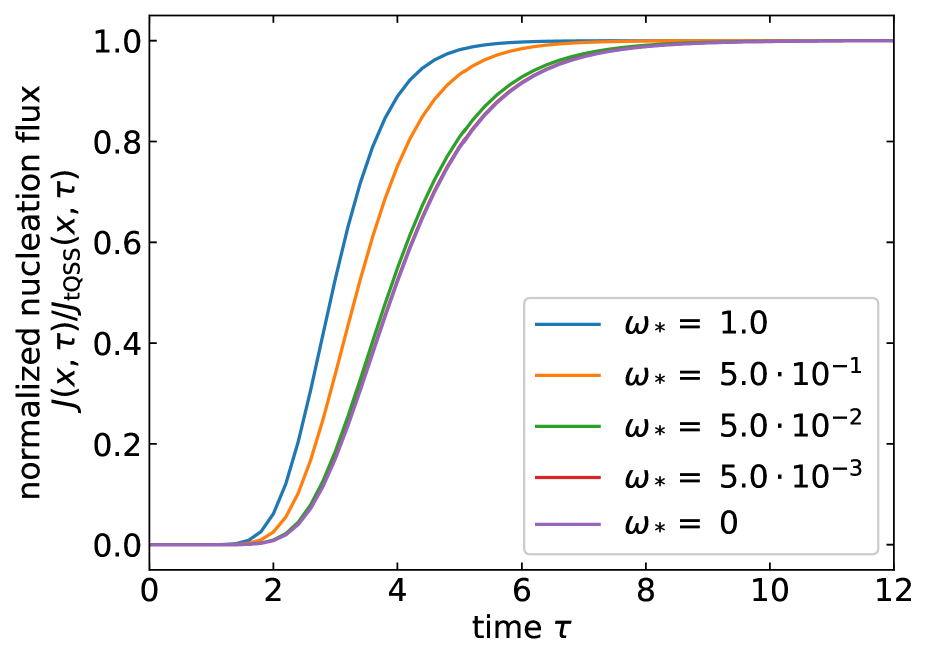}
    \end{subfigure}
    \caption{Left: The nucleation rate Eq.~\eqref{eq:nucleationFluxResult} evaluated at reduced cluster size $x = 1.5$ and normalized by quasi-steady-state nucleation flux $J_\mathrm{QSS}(\tlt=0)$ (see Eq.~\eqref{eq:QSS}) as a function of the scaled time $\tlt$. The critical cluster size is set to $n_* = 152$, and the initial nucleation barrier to $\Delta W_*(\tlt=0) = 70$. We include data for the non-stationarity index values of $\omega_*  = 0.005$ (red), $0.05$ (green) $0.5$ (orange) and $1$ (blue), discussed in more detail in Section~\ref{sec:casestudy}. Crosses are numerical solution of the discrete master equations for identical parameter values with a time interval $\Delta \tau = 1/4$ between the data points, discussed in more detail in Section~\ref{sec:casestudy}. Right: The nucleation rate Eq.~\eqref{eq:nucleationFluxResult} normalized by the renormalized, time-shifted quasi-steady-state nucleation rate Eq.~\eqref{eq:correctedNucleationFluxResult}, both evaluated at the same reduced cluster size $x = 1.5$. The color coding is the same as in the left panel.}
    \label{fig:nucleationFlux}
\end{figure}

Eq.~\eqref{eq:nucleationFluxResult} makes explicit both the initial and late time behavior, which we illustrate in Fig.~\ref{fig:nucleationFlux}. Instead of evaluating the dimensionless surface tension and the initial supersaturation we, somewhat arbitrarily, set the values of the critical cluster size at $n_*=152$ and the nucleation barrier $\beta \Delta W_*(\tlt=0) = 70$ at time zero. The surface tension corresponding with these values is consistent with a relevant model system for organic electronics and corresponds with that of fullerene dissolved in carbon disulfide for an initial degree of supersaturation of $S(0) = 2.7$~\cite{Aksenov2005KineticsSolutions}. These values are chosen to ensure that the constraints in Eq.~\eqref{eq:constraints} hold, \textit{i.e.}, ensure that the critical cluster size and the inverse nucleation barrier drift only very weakly with time. For different initial parameter values, we find qualitatively identical results as long as the constraints in Eq.~\ref{eq:constraints} hold and we are cognizant that the non-stationarity index depends implicitly on the initial conditions as $\omega_*\sim \beta \Delta W_*^{a}$, where the value for the exponent depends on the aggregation kinetics, \textit{e.g.}, for reaction-limited aggregation $a=5/2$, and $a=3$ for diffusion-limited aggregation. Hence, we only show results for a single initial condition. 

The left and right panels show the time dependence of the nucleation flux $J(x,\tlt)$ for an increasing rate of evaporation represented by increasing values of the non-stationarity index $\omega_* = 0, 0.005, 0.05, 0.5$ and $1.0$. The model used for evaporation we discuss in Sec.~\ref{sec:casestudy} in more detail. We normalize by the constant quasi steady-state flux $J_\mathrm{QSS}(\tlt = 0)$ (left) and (right) the time-dependent renormalized and time-shifted version of it, $J_\mathrm{tQSS}(\tlt)$, given in Eq.~\eqref{eq:correctedNucleationFluxResult}. As both $J_\mathrm{tQSS}(\tlt)$ and $J(x,\tlt)$ are invariants of $x$ if we express them in terms of a function of the shifted time $\tlt - \tlt_\mathrm{i}(x)$, changing $x$ only affects the time delay it suffices to show our results for a single, arbitrarily chosen scaled cluster size. We set $x = 1.5$. The symbols in the left panel are (color-matched) numerical solutions of the nucleation equations that we discuss in more detail below, showing excellent agreement for the values of $\omega_*$ shown. The agreement deteriorates somewhat with high values of $\omega_*$, which of course is not surprising, given Eq.~\eqref{eq:constraints} demands that $\varepsilon^2 \omega_* \approx ( \beta \Delta W_*)^{-1} \omega_* \ll 1$. We discuss this discrepancy in detail in Sec.~\ref{sec:casestudy}.

The left panel of Fig.~\ref{fig:nucleationFlux}, where the nucleation flux is scaled to the (constant) QSS value at time zero, shows that in the absence of evaporation and $\omega_* = 0$ (purple, indistinguishable from the red curve) a constant asymptote is reached after some induction time. The asymptote is given by the steady-state value $J_\mathrm{QSS}(0) = J_\mathrm{SS}$. In the presence of evaporation, for $\omega_* > 0$, the flux in this late time regime increases exponentially with time, as we in fact expect from Eq~\eqref{eq:correctedNucleationFluxResult}. For very small values of $\omega_*$ there is a weak drift on the time scale shown in the figure, resulting in what looks like a pseudo plateau. For sufficiently rapid solvent evaporation, in particular if $\omega_* \gg 1$, the nucleation flux does not seem to exhibit this kind of pseudo plateau but increases exponentially immediately after the incubation time. Considering that $\omega_*$ is a ratio of time scales, this could be expected. For $\omega_* < 1$ nucleation is more rapid and dictates the kinetics. For $\omega_* > 1$ solvent evaporation is more rapid, so the nucleation kinetics is dictated by solvent evaporation.

The right panel shows that normalization by the time-dependent and rapidly increasing tQSS flux the nucleation flux reduces to a sigmoidal shape. The distinction between the early and late time regimes, which becomes less evident in the left panel for high $\omega_*$ due to the apparent absence of the pseudo-plateau, is still evident from the right graph. From this, we conclude that the quasi steady-state nucleation flux remains the central quantity for the late stages of the nucleation process, except that it is renormalized by a factor $\Gamma(\omega_* + 1)$ and evaluated at some earlier time (see Eq~\eqref{eq:correctedNucleationFluxResult}). The magnitude of this renormalization factor grows rapidly only for $\omega_* > 1$, but remains within $12 \%$ from unity if the evaporation time scale is much larger than the nucleation time scale ($0 < \omega_* < 1$).

Let us now focus specifically on the early-time behavior. The right panel in Fig.~\ref{fig:nucleationFlux} shows that initially the normalized nucleation flux remains vanishingly small for some time, after which it steeply rises to reach the tQSS value. That this must be so is actually evident by considering the limiting behavior of the regularized gamma function in Eq.~\eqref{eq:nucleationFluxResult} $\lim_{z\to 0}Q(n,z) = 1$ for $\tlt \to \infty$. This time delay may be considered a lag or induction time that evidently depends on the magnitude of $\omega_*$: with increasing evaporation rate, the curves become steeper and the time lag shortens. Clearly, solvent evaporation influences the early-time behavior but only if it is sufficiently fast, considering the fact that the curves obtained for $\omega_* \leq 0.05$ overlap. 

For vanishing solvent evaporation $\omega_* \to 0$, Eq.~\eqref{eq:nucleationFluxResult} reduces to a transient time-shifted quasi-steady-state nucleation flux, realizing that $Q(\omega_* + 1,z) \sim \exp(-z) \left(1+\mathcal{O}(\omega_*)\right)$~\cite{2010NISTFunctions}, yielding
\begin{equation}\label{eq:transQSS}
    J(x,\tlt) = J_\mathrm{QSS}(\tlt - \tlt_\mathrm{i}(x) + \tk)\exp\left[-\exp\left\{-\tlt + \tlt_\mathrm{i}(x)\right\}\right],
\end{equation}
expressed as a product of the late-time quasi-steady-state nucleation flux $J_\mathrm{QSS}(\tlt - \tlt_\mathrm{i}(x) + \tk)$ evaluated at some shifted time, and a so-called double exponential that describes the initial transient regime, \textit{i.e.}, the sigmoidal curves shown in the right panel of Fig.~\ref{fig:nucleationFlux}. For stationary nucleation $J_\mathrm{QSS}(\tlt - \tlt_\mathrm{i}(x) + \tk) = J_\mathrm{SS}$, so Eq.~\eqref{eq:transQSS} correctly reproduces Eq.~\eqref{eq:stationaryFlux}.

The late-time time-shifted QSS flux is reached after some initial lag period~\cite{Kashchiev2000,Kalikmanov2013}. Eq.~\eqref{eq:transQSS} suggests that even in the regime where the QSS approximation is justifiable, $\omega_* \to 0$, the QSS nucleation flux should actually be evaluated at some shifted time $\tlt - \tlt_\mathrm{i}(x) + \tk$ incorporating the earlier mentioned memory effect. Such a memory effect is not included in the standard QSS theory, even though it is a natural consequence of the existence of a lag time in the kinetics of nucleation: it simply accounts for the fact that the nucleation flux cannot instantaneously respond to a change in the ambient conditions~\cite{Kashchiev2000}. We find that the `standard' QSS flux $J_\mathrm{QSS}(\tlt)$ and the time-shifted QSS flux obtained from our theory $J_\mathrm{QSS}(\tlt -\tlt_\mathrm{i}(x) + \tk)$ are only identical in the stationary limit for $\omega_*=0$. The relative difference between the two fluxes scales exponentially with $\omega_*(\tlt_\mathrm{i}(x) - \tk)$. Similar arguments apply if we would consider some other source of non-stationarity instead of solvent evaporation~\cite{Shneidman2010}. In other words, the QSS flux is never correct, except for the trivial case of stationary nucleation. 

For sufficiently slow solvent evaporation, however, the difference between the QSS and time-shifted QSS flux does become small. Sufficiently slow here means that the height of the nucleation barrier varies over much longer times than the nucleation time scale. For cluster sizes moderately close to the critical size $x = 1$, we would expect that the incubation time $\tlt_\mathrm{i}(x)$ is of the order of the lifetime of a critical cluster, and, as we shall show in the next section, so is the timescale $\tk$, which is the contribution to the memory effect that has its origin in the evaporation. Hence, in the regime where the QSS flux is expected to be accurate, so if the non-stationarity index $\omega_*$ is very small, the difference between the quasi-steady nucleation flux $J_\mathrm{QSS}$ evaluated at $\tlt$ or at $\tlt - \tlt_\mathrm{i}(x) + \tk$ should be very small. Practically, this means that the impact of the memory effect implicit in Eq.~\eqref{eq:transQSS} should be expected to be small in the regime where the QSS approximation is warranted, explaining why the QSS flux without the time-shift is still being used in the field~\cite{Kashchiev2000}.  In the next section, we make the shift in the early-time behavior observed in Fig.~\ref{fig:nucleationFlux} (right) explicit by introducing the generalized induction time as a measure for the duration of the early-time regime, and find it to depend on the non-stationarity coefficient $\omega_*$ in a highly non-trivial way.

\section{Induction time}\label{sec:inductiontime}
A common approach to quantify the duration of the early-time regime under conditions of stationary nucleation is to calculate the lag or induction time that is associated with the initial transient regime before some sort of steady state is reached. Unfortunately, the induction time is a somewhat ambiguous quantity and various definitions have been put forward in the literature~\cite{Wu1992timelag,Kashchiev2000,Wedekind2007NewProcesses}. The three most common definitions are as follows.
\begin{itemize}
    \item[i)] The induction time, denoted $\theta$, is proportional to the life time of a critical cluster $\td$ Eq.~\eqref{eq:lifetime}, with some (non-standard) constant of proportionality of the order unity. The term `lag time' is generally associated with this definition~\cite{Wu1992timelag,Kashchiev2000};
    \item[ii)] The induction time is inversely proportional to the product of the volume of the system under investigation and the steady-state nucleation rate, representative of the time for the first cluster to emerge. This definition is frequently used in mean-first-passage-time analyses in simulations~\cite{Wedekind2007NewProcesses};
    \item[iii)] The induction time, $\theta$, is implicitly defined by the integral equality~\cite{Wu1992timelag,Kashchiev2000}    \begin{equation}\label{eq:inductiontimestationary}
    \lim\limits_{t\to\infty}\int_\theta^t \dint t'  J_\mathrm{SS} = \lim\limits_{t\to\infty}\int_0^t \dint t' J(x,t').
    \end{equation}
\end{itemize}
According to definition iii), the induction time marks the point in time after which integration of a steady-state nucleation rate yields the same number of nuclei per unit of volume as for the actual nucleation rate, starting from the time at which the quench takes place, set at $t \equiv 0$. We opt to use this definition because it generalizes the lag time to arbitrary reduced cluster sizes $x \geq 1$. From an experimental point of view, it is also more straightforward to determine, at least for stationary nucleation. The reason is that it is not limited by the observation of clusters of the critical size. Such clusters are not necessarily directly observable, \textit{e.g.}, because they tend to be too small or present in a concentration that is too low for direct observation~\cite{Spinella1998CrystalSilicon}. 

Of course, the induction time defined via Eq.~\eqref{eq:inductiontimestationary} is only valid under stationary conditions, but has a straightforward generalization to our far out-of-equilibrium and non-stationary setting, 
\begin{equation}\label{eq:inductiontimenonstationary}  
\lim\limits_{\tlt\to\infty}\int_{\theta}^{\tlt} \dint \tlt'  J_\mathrm{tQSS}(x,\tlt') = \lim\limits_{\tlt\to\infty}\int_0^{\tlt} \dint \tlt' J(x,\tlt'),
\end{equation}
where on both sides of the equal sign the limit $\tlt \to \infty$ must be taken simultaneously. This expression we construct by demanding that the integrated nucleation flux (right-hand side), equals the integrated late-time asymptotic to the nucleation flux (left-hand side) if it were initiated at the induction time $\theta$. Here, we presume that $J_0(x) = 0$ in the nucleation flux Eq.~\eqref{eq:nucleationFluxResult}. Notice that the integrals on both sides of the equal sign diverge, and in fact do so identically by construction. The reason is that the nucleation flux $J(x,\tlt')$ approaches $J_\mathrm{tQSS}(x,\tlt')$ asymptotically for $\tlt \to \infty$, see also the discussion near Eq.~\eqref{eq:nucleationFluxResult}. The downside of this definition is that it cannot be obtained without explicitly solving the kinetic nucleation equations. Whilst this is actually possible under stationary conditions~\cite{Wu1992timelag,Shneidman1992}, this is not so for our non-stationary model: our approach requires an explicit solution for the nucleation flux of the master equation. 

We combine our analytical expressions for the nucleation flux Eq.~\eqref{eq:nucleationFluxResult} presuming $J_0(x) = 0$, and the tQSS flux Eq.~\eqref{eq:correctedNucleationFluxResult} with Eq.~\eqref{eq:inductiontimenonstationary} and solve for the induction time $\theta$. To do this, we evaluate the integrals in Eq.~\eqref{eq:inductiontimenonstationary} separately. We evaluate the integral on the right-hand side of Eq.~\eqref{eq:inductiontimenonstationary} by first inserting the expression for the nucleation flux Eq.~\eqref{eq:nucleationFluxResult}. In evaluating the resulting expression we make use of the work by Shneidman~\cite{Shneidman2010}, who calculated a similar integral to find the number density of clusters larger than some cluster size $x$, arriving at
\begin{equation}\label{eq:nucflux}
    \int_0^{\tlt} \dint \tlt' J(x,\tlt') = J_\mathrm{QSS}(0)\exp\left[\omega_*\tk\right]\mathrm{E}_{1-\omega_*}\left(\exp\left[-(\tlt-\tlt_\mathrm{i}(x))\right]\right),
\end{equation}
where $J_\mathrm{QSS}(0) = J_\mathrm{SS}$ given by Eq.~\eqref{eq:QSS} and $\mathrm{E}_{p}$ is the exponential integral of order $p$~\cite{2010NISTFunctions}.
To evaluate the integral on the left-hand side of Eq.~\eqref{eq:inductiontimenonstationary}, we make use of the QSS nucleation flux Eq.~\eqref{eq:QSS} and the tQSS nucleation flux Eq.~\eqref{eq:correctedNucleationFluxResult}, which allows us to write the latter as 
\begin{equation}
    \begin{split}
    J_\mathrm{tQSS}(x,\tlt') & = \Gamma(\omega_* +1) J_\mathrm{QSS}(\tlt' - \tlt_\mathrm{i}(x) + \tk) \\ &  = \Gamma(\omega_* +1)J_\mathrm{QSS}(0) \exp\left[\omega_*\tk\right] \exp\left[\omega_* \left(\tlt' - \tlt_i(x)\right)\right].
    \end{split}
\end{equation}
Integration with respect to $\tlt'$ yields
\begin{equation}\label{eq:integratedtQSSflux}
     \int_\theta^{\tlt} \dint \tlt'  J_\mathrm{tQSS}(x,\tlt')  = \Gamma(\omega_*)J_\mathrm{QSS}(0) \exp\left[\omega_*\tk\right]\left\{\exp\left(\omega_*(\tlt - \tlt_\mathrm{i}(x)\right) - \exp\left(\omega_*(\theta - \tlt_\mathrm{i}(x))\right)\right\},
\end{equation}
where we have made use of the well-known recurrence relation for the Gamma function $z^{-1}\Gamma(z+1) = \Gamma(z)$. 
Combined with Eqs.~\eqref{eq:inductiontimenonstationary}~and~\eqref{eq:nucflux}, and after reordering of the terms, we find
\begin{equation}\label{eq:intermediateinductiontime}
\Gamma(\omega_*)\left\{\exp\left(\omega_*(\theta - \tlt_\mathrm{i}(x)\right)\right\} = \lim\limits_{\tlt\to\infty} \left\{ \Gamma(\omega_*)\exp\left(\omega_*(\tlt - \tlt_\mathrm{i}(x))\right) - \mathrm{E}_{1-\omega_*}\left(e^{-(\tlt-\tlt_\mathrm{i}(x)}\right)\right\},
\end{equation}
with the left hand side independent of the time $\tlt$. 

Both bracketed terms on the right-hand side in this expression diverge but do so with the same rate, and turn out to cancel each other exactly. This follows from the fact that the argument of the exponential integral $\mathrm{E}_{1-\omega_*}$ becomes zero in the late-time limit $\tlt \to \infty$. We conveniently introduce the series expansion for the exponential integral as \cite{2010NISTFunctions}
\begin{equation}\label{eq:expansionEp}
    \mathrm{E}_{p}(z) = z^{p-1} \Gamma(1-p) - \sum_{k=0}^{\infty}\frac{(-z)^k}{k!(1 - p + k)},
\end{equation}
of which the first term is identical to the first term on the right hand side in Eq.~\eqref{eq:intermediateinductiontime}. This reduces Eq.~\eqref{eq:intermediateinductiontime} to 
\begin{equation}\label{eq:inductiontime}
\Gamma(\omega_*)\left\{\exp\left(\omega_*(\theta - \tlt_\mathrm{i}(x)\right)\right\} = \lim\limits_{\tlt\to\infty} \left\{\sum_{k=0}^{\infty}\frac{(-\exp\{-(\tlt-\tlt_\mathrm{i}(x)\})^k}{k!(\omega_* + k)}\right\} = \frac{1}{\omega_*},
\end{equation}
where the last equality follows from the observation that under conditions of solvent evaporation $\omega_* \geq 0$, only the $k=0$ term survives in the limit $\tau \rightarrow \infty$. After inverting Eq.~\eqref{eq:inductiontime}, we find an explicit expression for the induction time as a function of the non-stationarity index $\omega_*$,
\begin{equation}\label{eq:Result0InductionTime}
    \theta(x,\omega_*) = \tlt_\mathrm{i}(x)-\frac{1}{\omega_*}\ln\Gamma(\omega_*) - \frac{1}{\omega_*}\ln\omega_* = \tlt_\mathrm{i}(x)-\frac{1}{\omega_*}\ln\left(\omega_*\Gamma(\omega_*)\right).
\end{equation}
This evaporation-borne contribution to the induction time also emerges in the time shift of the tQSS flux, Eq.~\eqref{eq:correctedNucleationFluxResult}, if we absorb its prefactor $\Gamma(1 + \omega_*)$ into the time shift. Hence, it appears to be a fundamental, non-stationary contribution to the induction or lag time in non-stationary nucleation. 

For the sake of clarity, we express Eq.~\eqref{eq:Result0InductionTime} in terms of the induction time under stationary conditions $\theta_0(x) \equiv \theta(x,\omega_*=0) = \tlt_\mathrm{i}(x) + \gamma_\mathrm{E} \geq 1$~\cite{Shneidman1992} with $\gamma_\mathrm{E} = 0.577\dots$ the Euler–Mascheroni constant~\cite{2010NISTFunctions}, as 
\begin{equation}\label{eq:ResultInductionTime}
    \theta(x,\omega_*) = \theta_0(x) - \left[\gamma_\mathrm{E} +\frac{1}{\omega_*}\ln\left(\omega_*\Gamma(\omega_*)\right)\right].
\end{equation}
This expression indeed reduces to the stationary induction time for the limiting case $\omega_* \to 0$, which can be verified using the expansion~\cite{2010NISTFunctions}
\begin{equation}
    \ln \Gamma(z) = - \gamma_\mathrm{E} z - \ln z + \sum_{k=1}^{\infty} \left[\frac{z}{k} -  \ln\left(1 + \frac{z}{k}\right)\right].
\end{equation}

Eq.~\eqref{eq:ResultInductionTime} is novel and is the second main result of this paper, the first being the nucleation flux Eq.~\eqref{eq:nucleationFluxResult}~and~Eq.~\eqref{eq:correctedNucleationFluxResult}. 
Because $\omega_*$ does not depend on the reduced cluster size $x$, we conclude from Eq.~\eqref{eq:ResultInductionTime} that the shift in the induction time due to solvent evaporation is also independent of that quantity. Consequently, the relative impact of solvent evaporation on the induction time must decrease with increasing $x$ because Eq.~\eqref{eq:ResultInductionTime} is an increasing function of $x$. To shed light on how Eq.~\eqref{eq:ResultInductionTime} depends on the evaporation rate encoded in the quantity $\omega_*$, we focus on the limit $\omega_*\ll 1$. For that purpose we expand Eq.~\eqref{eq:ResultInductionTime} near the stationary case $\omega_* = 0$, yielding to lowest order in powers of $\omega_*$,
\begin{equation}\label{eq:linearizedInductionTime}
    \theta(x,\omega_*) = \theta_0(x) - \frac{\pi^2}{12}\omega_* +\mathcal{O}\left(\omega_*^2\right),
\end{equation}
showing that the function is well-behaved in the limit $\omega_*\rightarrow 0$. 

\begin{figure}[bt]
    \centering
    \includegraphics[width = 0.6\textwidth]{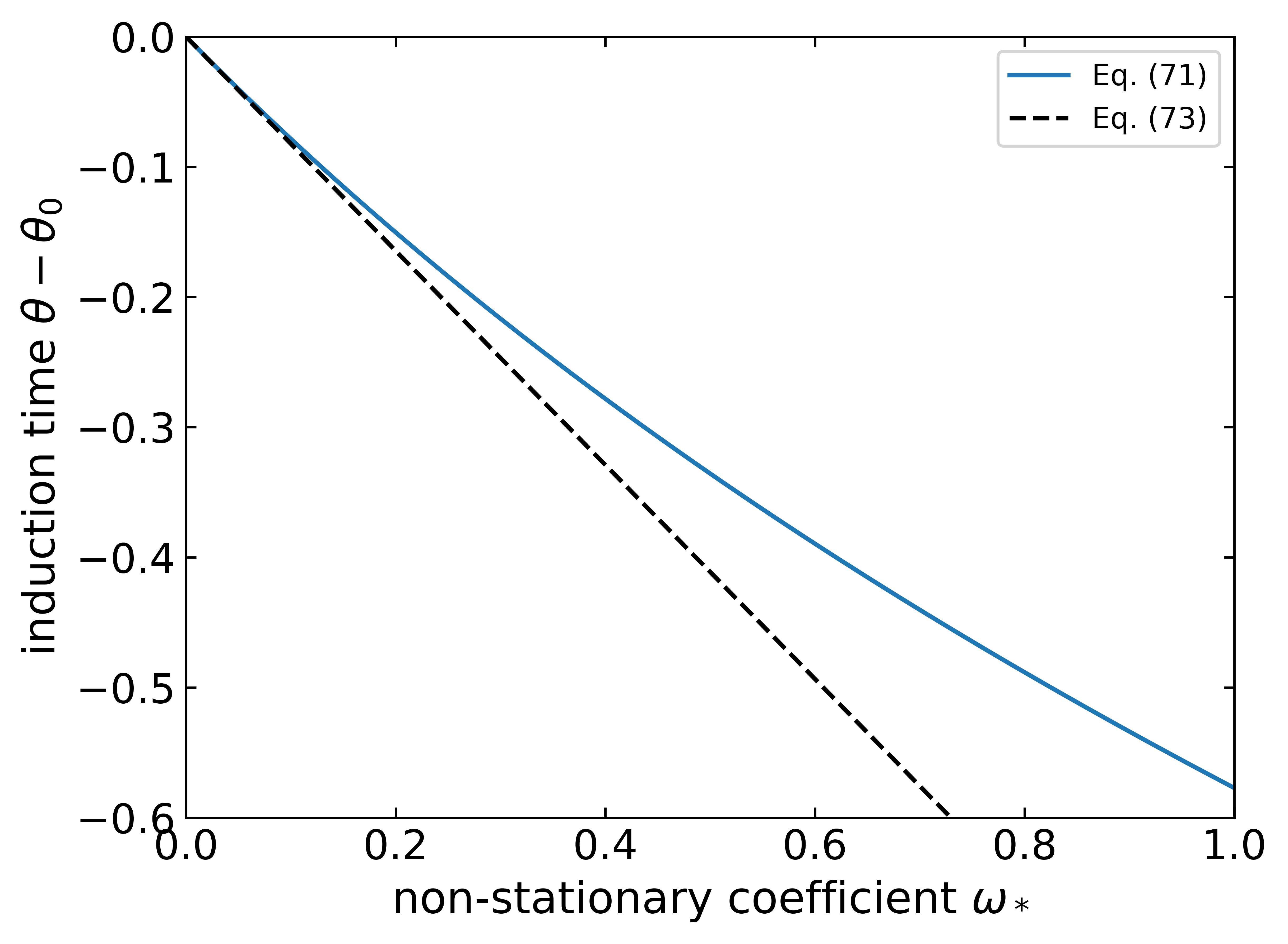}
    \caption{The difference between the induction time at non-zero solvent evaporation $\theta$ and at zero solvent evaporation $\theta_0$ expressed in reduced time units, as a function of the non-stationarity coefficient $\omega_*$. The blue, solid line corresponds with the full result Eq.~\eqref{eq:ResultInductionTime}, the black dashed line is the linearized result Eq.~\eqref{eq:linearizedInductionTime}.}
    \label{fig:ResultInductionTime}
\end{figure}

It seems that the induction time $\theta$ decreases from its stationary value $\theta_0$ with increasing value of $\omega_*$. This agrees with the expectation that solvent evaporation must increase the supersaturation with time, which in turn must induce an increasingly rapid nucleation and hence lead to a reduction of the associated time scale. In Fig.~\ref{fig:ResultInductionTime} we plot the effect of solvent evaporation on the induction time $\theta - \theta_0$ using the non-stationarity index $\omega_*$ as the control parameter. Here, the blue, solid line is the full result of Eq.~\eqref{eq:ResultInductionTime}, and the dashed lines the linearized result Eq.~\eqref{eq:linearizedInductionTime}. Outside of the linearized regime, Eq.~\eqref{eq:linearizedInductionTime} overestimates the actual reduction of the induction time.

From Eq.~\eqref{eq:linearizedInductionTime} and Fig.~\ref{fig:ResultInductionTime} we conclude that the decrease of the induction time is of the order $\omega_*$ for $\omega_* \ll 1$. Hence, as $\theta_0 \geq 1$, a discernible shift is only observable if $\omega_*$ is sufficiently large compared to the induction time under stationary conditions, $\theta_0(x)$. This is only the case if the time scale for solvent evaporation is of similar magnitude as the nucleation time scale (multiplied with the number of molecules in the critical cluster, see Eq.~\eqref{eq:defI}). If that happens to be the case, then for some value for $\omega_*$ the induction time $\theta(x,\omega_*)$ may actually vanish and become negative beyond that value. The root cause of this lies in our expression for the nucleation flux Eq.~\eqref{eq:nucleationFluxResult}, which should be zero for very early times, but is not so, as discussed earlier in Sec.~\ref{sec:nucleationflux}. Apparently, the magnitude of this inconsistency becomes magnified with increasing value of $\omega_*$. We stress that this cannot be amended by invoking a non-zero value for $J_0(x)$ in Eq.~\eqref{eq:nucleationFluxResult}, as this turns out to not affect the value of $\omega_*$ for which the induction time becomes negative. In our implementation, Eq.~\eqref{eq:nucleationFluxResult} incorrectly predicts a non-vanishing and non-negligible nucleation flux for supercritical clusters at $\tlt=0^+$. This physically means that clusters have grown instantaneously from a monomer to a supercritical size, which is physically impossible. 

It is not clear to us how to remedy this issue. Based on our result for the reduced size distribution $\nu(x,\tlt$) in the subcritical region, Eq.~\eqref{eq:smallersolution}, we argue that the smallest physically possible induction time must be equal to or larger than the time required for the subcritical cluster densities to populate, which is approximately equal to the subcritical growth time $\tlt_\mathrm{d}(x=1-\varepsilon)$, see Eq.~\eqref{eq:decaytime}, with $\varepsilon$ the inverse square root of the nucleation barrier. This growth time remains non-negative, finite and independent of solvent evaporation rate implicit in $\omega_*$ \cite{ShiSeinfeldOkuyama1990}. Hence, Eq.~\eqref{eq:ResultInductionTime} should actually be treated with caution before it predicts negative induction times: for values of $\omega_*$ that yield an induction time smaller or approximately equal to the (positively valued) subcritical growth time, Eq.~\eqref{eq:ResultInductionTime} should already be treated with caution. For values of $\omega_*$ that yield an induction time much larger than the subcritical growth time, Eq.~\eqref{eq:ResultInductionTime} should be accurate. 

Taking Eq.~\eqref{eq:ResultInductionTime} at face value, we argue that it provides an upper bound on $\omega_*$ for the validity of our results for the nucleation flux Eq.~\eqref{eq:nucleationFluxResult}. Numerical inversion of Eq.~\eqref{eq:ResultInductionTime} shows that the value of $\omega_*$ for which the induction time becomes negative grows very rapidly with increasing value of $\theta_0(x)$. Considering the lower bound on the stationary induction time, $\theta_0(x) = 1$, we find that beyond $\omega_* \approx 5.3$ the induction time becomes negative. If on the other hand $\theta_0(x) = 10$ we find this to happen for $\omega_* \approx 6 \cdot 10^{4}$. The fact that these limiting values are (much) larger than unity means that we expect the theory to break down under conditions where nucleation is predominated by the effect of solvent evaporation, consistent with Eq.~\eqref{eq:nonstatcoefficient} and the constraints expressed in Eq.~\eqref{eq:constraints}. 

Solvent evaporation need not be fast in an absolute sense for this to occur, only fast relative to nucleation. This is certainly the case very close to the binodal, where both the number of molecules in a critical cluster and the lifetime of a critical cluster diverge (see Eq.~\eqref{eq:lifetime}), causing even slow solvent evaporation to be rapid relative to nucleation. This situation presents itself during a finite quench from the stable to the metastable solution, relevant to experiments. Very close to the binodal, however, nucleation is so slow that it cannot be observed on experimental time scales, so we need not consider this situation. The relevant situation here is how fast solvent evaporation is compared to nucleation, starting from the point that the degree of supersaturation is sufficiently high for nucleation to occur on experimental time scales. This strongly depends on the specific nucleation kinetics, and is hard to estimate \textit{a priori}.

Away from the binodal, these conditions only emerge if solvent evaporation is very rapid. The experimentally relevant conditions of a finite rate quench from the stable to an unstable solution precludes the emergence of this situation, as the initial nucleation occurs always, in some sense, close to the binodal. Setting this aside, such rapid solvent evaporation introduces the additional problem that stratification in the thin film is likely to occur and our model, which presumes a homogeneous solution, does not apply. Local accumulation of solute is strongest at the solution-air interface and heterogeneous nucleation at this interface arguably becomes increasingly important~\cite{Schaefer2017DynamicSolutions}. If heterogeneous nucleation becomes the dominant nucleation mechanism, we can actually use an extension of this work, detailed in Appendix~\ref{app:heterogeneousNucleation}. 

\begin{figure}[t]
    \begin{subfigure}[b]{0.49\textwidth}
        \includegraphics[width=\textwidth]{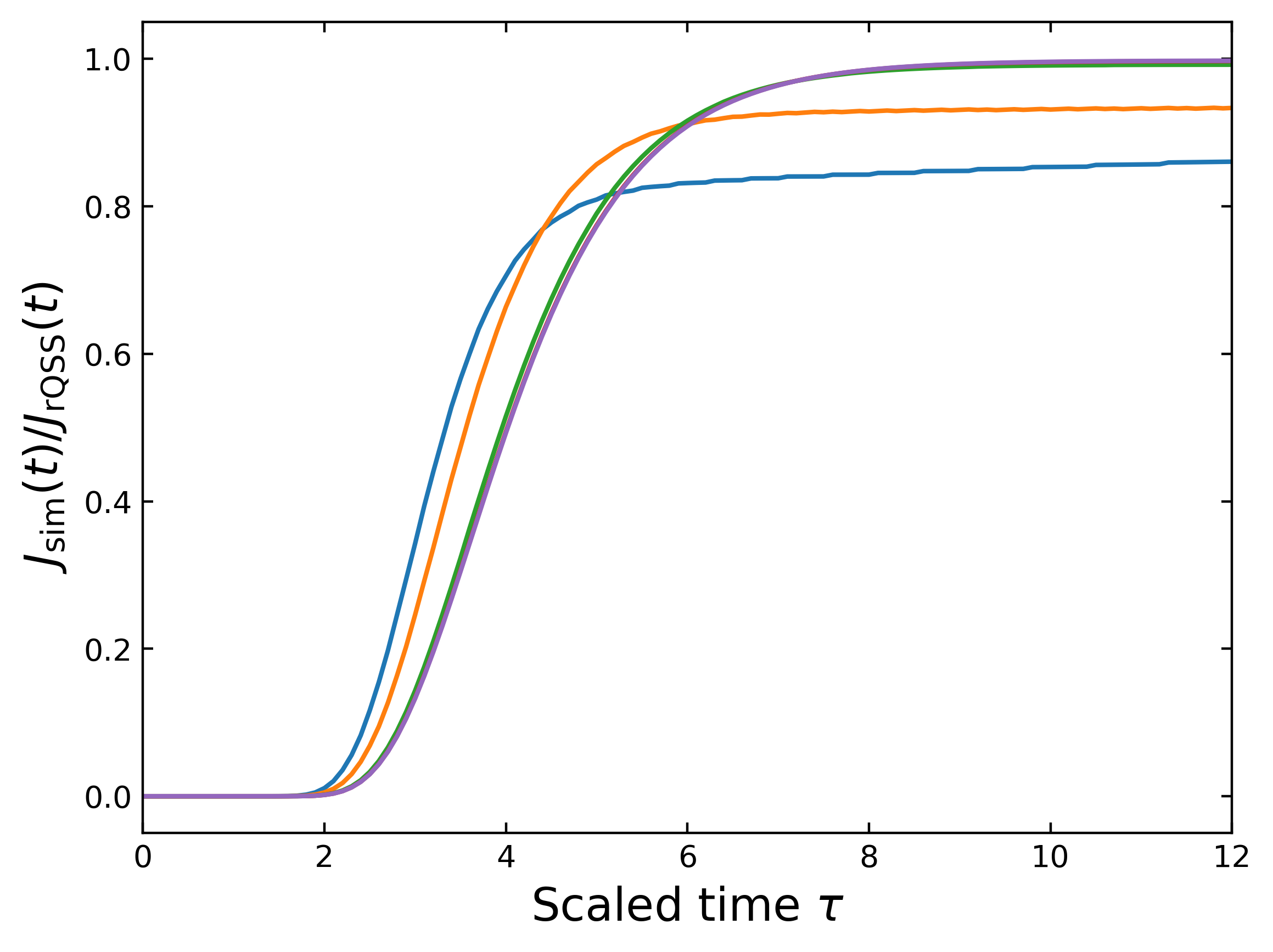}
    \end{subfigure}
    \caption{The nucleation rate $J_\mathrm{sim}(x,\tlt)$ as obtained from numerical calculations at scaled cluster size $x=1.5$, normalized by the tQSS nucleation rate (Eq.~\eqref{eq:correctedNucleationFluxResult}). The initial critical cluster size is set to $n_* = 152$, and the initial nucleation barrier is set to $\Delta W_*(\tlt=0) = 70$. The color correspond to different values for the non-stationarity index $\omega_*(0) = 0$ (purple), $\omega_*(0) = 5 \times 10^{-2}$ (green), $\omega_*(0) = 5\times 10^{-1}$ (orange) and $\omega_*(0) = 1$ (blue).
}
    \label{fig:nucleationFluxStartup}
\end{figure}

A comparison of the theoretical results presented in this section with the numerical results in Fig.~\ref{fig:nucleationFlux} could verify how accurate  Eq.~\eqref{eq:ResultInductionTime} is for high $\omega_*$. In Fig.~\ref{fig:nucleationFluxStartup} we plot the nucleation flux obtained from numerical calculations scaled to the theoretical tQSS flux, Eq.~\eqref{eq:correctedNucleationFluxResult}, allowing for comparison to the results shown in the right panel of Fig.~\ref{fig:nucleationFlux}. Note that the curves approach unity for low $\omega_*$, so the tQSS flux is accurate, but this is not true for the larger values of $\omega_*$ shown, where in the late time (after the induction time) the tQSS overestimates the numerical nucleation flux by about  10 - 20 \%. Qualitatively, the agreement is rather good, however, showing that the induction time must indeed decrease with increasing non-stationarity index $\omega_*$. A quantitative comparison requires the explicit integration of the implicit definition of the induction time in Eq.~\eqref{eq:inductiontimenonstationary} using the numerically obtained nucleation flux. This is only possible if we can extract from the full nucleation flux (including the wind-up period) the late-time nucleation flux (excluding the wind-up period). The latter we find difficult to extract from the numerical calculations. It is now not clear how to extract the late-time nucleation flux from the numerical nucleation flux required to solve Eq.~\eqref{eq:inductiontimenonstationary}. Qualitatively however, the results shown in Fig.~\ref{fig:nucleationFluxStartup} are in good agreement with theoretical result Eq.~\eqref{eq:ResultInductionTime}, where both show that increasing the non-stationarity $\omega_*$ decreases the induction time.

In the next section, we investigate in detail under what conditions our results apply. In particular, we discuss how the rate of solvent evaporation influences the accuracy of our prediction, how the initial supersaturation affects our results and return to the comparison of our results with the predictions of the QSS flux and with numerical calculations.

\section{Limits of the theory}\label{sec:casestudy}
Before summarizing our findings, we discuss in more detail the impact of our assumptions and approximations on our predictions. The predictions presented in the two previous sections are based on an extension of classical nucleation theory, while presuming (i) a nucleation barrier that is considerably larger than the thermal energy, as well as (ii) a solvent evaporation rate that is sufficiently slow to give an approximately constant non-stationarity index $\omega_*$. Concomitantly, this allows us to treat the reciprocal of the nucleation barrier, $\varepsilon^2$ and the critical cluster size, $n_*$, to be constant in our co-moving co-ordinate system. However, strictly speaking, solvent evaporation causes  these quantities to be time dependent. In addition, and this would be assumption (iii), in our calculations nucleation and evaporation start at some initial finite supersaturation, whereas in experiment the initial conditions typically correspond to (under)saturated conditions. 

The question then arises under what practical conditions these three assumptions are justified.
To start addressing this question, assumption (i) is implicit in classical nucleation theory and implies that for the time scales considered the solution should never be close to a spinodal, as the vicinity of a spinodal implies a vanishing interfacial tension and hence a vanishing nucleation barrier~\cite{Rasmussen1982EnergeticsSpinodal,Wedekind2009CrossoverVapor,Binder1984NucleationCriterion}. Given that for a phase transition in which molecules in a dissolved state drop out of solution and form a solid, crystalline state, a spinodal is not thought to exist and we do not expect a low surface tension under any conditions~\cite{Bartell2007DoDecomposition,Sosso2016CrystalSimulations}. We also note that colloidal dispersions tend to have a very low surface tension if the particles are sufficiently large~\cite{deHoog1999MeasurementSuspension,vanderSchoot1999RemarksSystems}, implying that for these the regime where CNT applies may be limited. For such particles a metastable gas-liquid spinodal might be hidden under the binodals, as is known for aqueous solutions of certain proteins~\cite{Asherie1996PhaseSolutions} and is suspected for dispersions of fullerenes under appropriate conditions~\cite{Hagen1993DoesPhase}. This scenario we explicitly exclude from our analysis.

To investigate under what conditions assumption (ii), expressed by the set of conditions Eq.~\eqref{eq:constraints}, remains reasonable, we need to introduce microscopic models for solvent evaporation and monomer attachment. Only then are we able to derive an expression for the non-stationarity index $\omega_*$ in terms of the volumetric evaporation rate and the initial nucleation barrier height, which is related to the initial supersaturation. Using these models, we evaluate under what conditions our analytical expression for the nucleation flux Eq.~\eqref{eq:nucleationFluxResult} should be accurate. It turns out that the constraints set by Eq.~\eqref{eq:constraints} can be expressed in terms of a single constraint, predicting that our analytical predictions are accurate if at the initial time $\tlt=0$ the ratio of the nucleation barrier $\beta \Delta W_*$ and the non-stationarity index $\omega_*$ is much larger than unity. Since we find that the non-stationarity index depends on the initial barrier height $\omega_*\sim \beta \Delta W_*^{5/2}$, this ratio becomes much smaller than unity very close to the binodal. Here, the nucleation flux also vanishes irrespective of the evaporation rate. Hence, to ensure the reliability of our predictions we require that the ratio is larger than unity at the moment that the first nuclei emerge, in agreement with the results on the generalized induction time discussed in the preceding section. We return to this issue in more detail below.

Let us introduce the for us relevant models for molecular aggregation and solvent evaporation. As for the molecular aggregation, two limiting cases can be discerned. On the one hand we have diffusion-limited growth if attachment of a monomer to a cluster is much faster than the diffusion towards it. On the other hand, cluster growth is reaction-limited if attachment is much slower than diffusion. For the growth of clusters of solution-processed organic molecules, either of these regimes or a combination may be relevant, depending on the processing conditions and the properties of solute and solvent~\cite{Tumbek2012AttachmentMica,Yu2019Diffusion-LimitedCells,Bangsund2019FormationFilms}. 

As it turns out, the case of diffusion-limited growth is inconvenient in the presented nucleation model, because it does not permit us to analytically calculate the quantities of interest. Hence, we opt to invoke a reaction-limited growth model~\cite{Kashchiev2000}. Within this model, we can actually calculate all quantities of interest analytically. The reaction-limited growth model posits that the attachment frequencies are proportional to the surface area of a cluster and can be written as~\cite{Turnbull1949RateSystems,Kashchiev2000},
\begin{equation}\label{eq:ballisticmodel}
    k_+(x,t) = k_0 n_*^{2/3} \rho_1(t) x^{2/3},
\end{equation}
where the dimensionless quantity $x^{2/3} = (n/n_*)^{2/3}$ is proportional to the surface area of a cluster, $\rho_1(t)$ is the dimensionless monomer density and $k_0$ a constant with units of reciprocal seconds and is proportional to the attempt frequency. The critical cluster size $n_*$ enters Eq.~\eqref{eq:ballisticmodel} in order to express it in terms of the reduced cluster size $x$, not the actual number of monomers per cluster $n$. The attempt frequency may depend on the cluster size if the energetic cost of forming an activation complex itself depends on the cluster size~\cite{Kashchiev2000}. This we neglect for simplicity. 

Next, we introduce a model for the solvent evaporation. Here, we presume that the evaporation rate is proportional to the concentration of solvent molecules, which seems to be accurate in practice~\cite{Bornside1989SpinModel}. If the initial solute concentration is low and we limit ourselves to the initial stages of the drying process, this then predicts the \textit{volume} of our solution to decrease linearly with time. The resulting monomer density is then given by
\begin{equation}\label{eq:evaporationmodel}
    \rho_1(t)/\rho_1(0) = V(0)/V(t) \equiv \chi(t) = \frac{1}{1-\alpha t}.
\end{equation}
Here, $\alpha$ is the reciprocal of the drying time, a measure for the rate of evaporation. We treat $\alpha$ as a freely adjustable parameter and maintain our assumption that monomer depletion due to aggregation is negligible. This simple description of the evaporation kinetics applies, \textit{e.g.}, in the context of spin-coating of solutions in the initial stages of the drying process~\cite{Franeker2015spincoating,Bornside1989SpinModel}, where the drying time can be expressed in terms of the height $h$ of the film $\alpha^{-1} = h \left(\dint h/\dint t\right)^{-1}$. Experimentally, drying times for micrometer-thick films can be mere seconds~\cite{vanFraneker2015PolymerFormation}.

Using these models for monomer attachment and solvent evaporation, we are now able to investigate for what evaporation rates our analytical result Eq.~\eqref{eq:nucleationFluxResult} remains valid, and under what conditions the constraints Eq.~\eqref{eq:constraints} hold. For this, we first obtain an explicit expression for the scaled time $\tau(t)$ as a function of the actual, physical time $t$ by invoking Eqs.~\eqref{eq:scaledtime}~and~\eqref{eq:lifetime}, and subsequently calculate the non-stationarity index $\omega$, which we no longer assume to be constant in time. Using Eq.~\eqref{eq:ballisticmodel}, we find the life time of a critical cluster Eq.~\eqref{eq:lifetime} to obey the relation
\begin{align}
    \frac{\td(t)}{\td(0)} &= \chi^{-1}(t)\left(1 + \Delta\mu_0^{-1}\ln \chi(t)\right)^{-2}.
\end{align}
scaled to the value at the initial time $t = 0$. Here, $\Delta \mu_0 = \ln \left[\rho(t=0)/\rho_\infty\right] = \ln S(t=0)$ is the chemical potential difference of a monomer in the dissolved and the condensed phase at $t = 0$. 

Next, we insert this expression in the definition of the scaled time Eq.~\eqref{eq:scaledtime}, and find
\begin{equation}
    \tlt(t) = \frac{1}{\td(t=0)}\int_0^t \dint t' \chi(t') (1 + \Delta\mu_0^{-1}\ln \chi(t'))^2,
\end{equation}
which, combined with Eq.~\eqref{eq:evaporationmodel}, yields
\begin{equation}\label{eq:tlttot}
    \tlt(t) = \tlt_\mathrm{s}\left[(1+ \Delta \mu^{-1}_0 \ln \chi(t))^3 - 1\right],
\end{equation}
expressed in terms of the reciprocal volume $\chi(t)$ defined by Eq.~\eqref{eq:evaporationmodel} for notational simplicity. Here, $\tlt_\mathrm{s} = {\Delta \mu_0}/{3 \alpha {\td}(0)}$ is a time scale that we interpret below. Unfortunately, inversion of Eq.~\eqref{eq:tlttot} does not produce a simple expression of the actual time $t$ in terms of the scaled time $\tlt$. For early times, a Taylor expansion yields $t/\td(0) = \tlt_\mathrm{s}/2 \sqrt{B} \left(\sqrt{\tlt/\tlt_\mathrm{s} + 1/4 B} - 1/2 \sqrt{B}\right)$, with a positively valued $B = (2 + \Delta \mu_0)/6$,
showing that the actual time $t$ increases sub-linearly with the scaled time $\tlt$. This is expected because solvent evaporation speeds up the nucleation process and the time $\tlt$ measures the time that passes from the point of view of the nucleating solute particles.

We next set out to find the time dependence of the non-stationarity index $\omega$ defined in Eq.~\eqref{eq:defI}, see also the list of symbols \ref{app:tablesymbols}. Here, we focus on $\omega$ instead of $\omega_*$, as the former is the relevant quantity with respect to the constraints in Eq.~\eqref{eq:constraints}. The two quantities $\omega$ and $\omega_*$ are related via the critical cluster size as $\omega_* = \omega(1-n_*^{-1})$, see Eq.~\eqref{eq:omegastar}. Although we have considered this quantity to be so weakly dependent on time that we can consider it a constant in our calculations, it can, by its definition Eq.~\eqref{eq:defI}, only be actually independent of time in special cases. Indeed, we find from the above analysis that
\begin{equation}\label{eq:Irealt}
    \omega(t) \equiv \td n_* \dpart{\ln \chi(t)}{t} = \omega(0)\left(1 + \Delta \mu_0^{-1} \ln\chi(t)\right)^{-5} = \omega(0)\left(\frac{\tlt(t)}{\tlt_\mathrm{s}} + 1\right)^{-5/3},
\end{equation}
where, $\omega(0) = \alpha {\td}(0) n_*(0)$. For high nucleation barriers $\beta\Delta W_*$ and sufficiently large critical cluster sizes $n_*$, we can express the time scale $\tlt_\mathrm{s}$ as the ratio of the height of the dimensionless nucleation barrier and the non-stationarity index $\omega$, both evaluated at $\tlt = 0$, as 
\begin{equation}\label{eq:tau_s}
\tlt_\mathrm{s} =  {2\beta \Delta W_*(0)}/{3 \omega(0)} \approx {2}/{3 \varepsilon^2 \omega(0)} \sim \mathcal{O}\left(\varepsilon^{3}\right),
\end{equation}
where we neglect a correction term of order $n_*^{-2/3}$, use the expression for the nucleation barrier Eq.~\eqref{eq:maxNucleationBarrier}, and make explicit that $\omega(0)$ implicitly depends on the initial barrier height via $n_*$ and $\td$ as $\omega(0)\sim \varepsilon^{-5}$, see Eqs.~\eqref{eq:criticalclustersize},~\eqref{eq:lifetime}~and~\eqref{eq:defI}. 

We are now in a position to interpret the dimensionless timescale $\tlt_\mathrm{s}$ as the timescale over which $\omega$ changes with time, which turns out to scale with the dimensionless nucleation barrier height as $\tlt_\mathrm{s} \sim \beta \Delta W_*^{-3/2}$. As expressed in Eq.~\eqref{eq:constraints}, our analytical results apply if the temporal drift in $\omega$ is small. Combining the constraint on $\omega$ of Eq.~\eqref{eq:constraints} with Eq.~\eqref{eq:Irealt}, we find 
\begin{equation}\label{eq:constraintomega}
   -\frac{1}{\omega}\ddif{\omega}{\tau} =\frac{5}{3} \left(\frac{1}{\tau_\mathrm{s} + \tlt}\right) \ll 1,
\end{equation}
indicating that the non-stationarity index decreases with time, since for solvent evaporation $\omega > 0$. Eq.~\eqref{eq:constraintomega} shows that our predictions should be valid if $\tlt_\mathrm{s} \gg 1$. That $\tau_\mathrm{s}$ must be large is actually implicit in Eq.~\eqref{eq:constraints}, as from Eq.~\eqref{eq:tau_s} we have $\tlt_\mathrm{s} \sim 1/\varepsilon^2 \omega$, which must be (much) larger than unity for our results to be internally consistent. Note that this also means that the constraints presented in Eq.~\eqref{eq:constraints} can be written as a single constraint: $\tlt_\mathrm{s} \gg 1$. 

Based on Eq.~\eqref{eq:constraintomega} we might conclude that our results should also apply for $\tlt \gg 1$ even if $\tlt_\mathrm{s}$ is not much larger than unity. Because we presume a fixed value for $\omega_* \equiv \omega_*(0)$ for our theoretical analysis, this in actual fact is not the case. Indeed, even if $\omega_*$ only drifts very weakly with time, the relative errors between the actual nucleation flux and our theoretical nucleation flux must accumulate with time. Eq.~\eqref{eq:constraintomega} actually suggests a natural way to correct for this: we could \textit{a posteriori} make the non-stationarity index $\omega_*$, the critical cluster size $n_*$ and $\varepsilon$ time-dependent, similar in spirit to how the quasi-steady-state nucleation flux is derived from the steady-state nucleation flux. For reasons of self-consistency these quantities should depend on the shifted time $\tlt - \tlt_\mathrm{i}(x)$, with $\tlt_\mathrm{i}(x)$ the incubation time, as was in fact suggested by Shneidman~\cite{Shneidman2010}. As a result of this approach we deduce from Eq.~\eqref{eq:constraintomega} that our results are expected to become accurate if $\tlt \gg 1$ also.

The question now arises how strongly solvent evaporation actually affects nucleation under experimentally relevant conditions. As discussed in Sec.~\ref{sec:CNT}, this depends on the conditions present at the moment that the first (supercritical) nuclei form and is hence related to our final assumption (iii): how the choice of the initial supersaturation affects our results. Considering that the induction time appears to vanish very close to the binodal, which is a consequence of the non-stationarity index $\omega_*$ becoming large very close to the binodal as discussed in the previous section, we only need to consider the effect of the initial conditions on the tQSS flux. To study how the choice for the initial conditions affect our results requires a reasonable estimate for the initial nucleation barrier height and a reasonable good understanding of the nucleation kinetics, so that we can estimate the attachment frequencies. These attachment frequencies can be determined by combining Eq.~\eqref{eq:ballisticmodel} with parameter estimates from either experimental or atomistic numerical studies~\cite{Kelton1991CrystalGlasses}. As far as we are aware, these are not known for the organic compounds of interest to us. Hence, at present, we cannot accurately estimate the instant in time we can set to $t=0$, and, consequently, neither the magnitude of the evaporation-related effects on nucleation under experimentally relevant conditions. 

In order to nevertheless make headway, we estimate a reasonable value for the smallest initial supersaturation required and subsequently discuss how changing this initial supersaturation changes our results qualitatively. Nucleation barriers for homogeneous nucleation observable on experimental time scales are generally estimated between 10 - 100 $\kBT$~\cite{Kashchiev2000}. For higher nucleation barriers, nucleation almost exclusively commences via heterogeneous nucleation~\cite{Kashchiev2000}. Hence, let us take 100 $\kBT$ as a reasonable estimate for the maximum barrier that results in nucleation on experimental timescales and we use a value for the dimensionless surface tension, see Eq.~\eqref{eq:maxNucleationBarrier}, $\sigma \approx 8.0 \pm 0.4$, corresponding with the nucleation of fullerene clusters in carbon disulfide~\cite{Aksenov2005KineticsSolutions}. This results in a maximum value for the initial supersaturation of $S(0) \approx 2.18 - 2.47$ with a maximum critical cluster size estimate of $n_* \approx 239 - 275$. 

As long as the constraint $\tlt_\mathrm{s} \gg 1$ presented in Eq.~\eqref{eq:tau_s} is valid for the conditions corresponding to the initial supersaturation, the expressions for the nucleation rate Eqs.~\eqref{eq:nucleationFluxResult}~and~\eqref{eq:correctedNucleationFluxResult} and the result for the induction time in Eq.~\eqref{eq:ResultInductionTime} are expected to be accurate. Upon changing the initial supersaturation for fixed values of the other kinetic and thermodynamic parameters, we find that the results shown in Fig.~\ref{fig:nucleationFlux} for the nucleation flux and in Fig.~\ref{fig:ResultInductionTime} for the induction time remain correct if we account for implicit changes in two model parameters. First, we note that the lifetime of a critical cluster scales linearly with the barrier height, $\td\sim \beta \Delta W_*$, indicating that if we were to double the initial barrier height, the initial nucleation time scale would also double. Since all results are presented in terms of the scaled time $\tlt$ instead of the actual time $t$, this change is actually implicit in the results presented. Second, the non-stationarity index $\omega_*$ depends on the initial nucleation barrier as $\omega_*\sim\beta \Delta W_*^{5/2}$, so for fixed volumetric evaporation rates $\alpha$ at different initial barrier heights, the non-stationarity index can change considerably. For example, if the initial barrier height changes from 100 $\kBT$ to 10 $\kBT$, $\omega_*$ decreases by a factor of about $3 \times 10^{2}$. As a result, the impact of solvent evaporation on nucleation decreases rapidly (superlinearly) with increasing initial degree of supersaturation.

This evidently presumes that homogeneous nucleation is the dominant nucleation mechanism for the experimental systems relevant to us. In a solution confined to a thin film geometry this might not always be the case, and homogeneous nucleation could be pre-empted by heterogeneous nucleation. This might occur if solvent evaporation is very rapid, resulting in accumulation of solute near the solution-air interface~\cite{Schaefer2017DynamicSolutions}. Alternatively, homogeneous nucleation becomes much less probable simply because the available space (volume) for homogeneous nucleation relative to the available space for heterogeneous nucleation (the substrate and solution-air interfaces) decreases with wet film thickness. As we discuss in Appendix~\ref{app:heterogeneousNucleation}, the extension of our work to heterogeneous nucleation is straightforward to carry out and all results in this manuscript apply with only minor modifications. We refer to Appendix~\ref{app:heterogeneousNucleation} for the details.

To further investigate the limitations of our theory, we return to numerics, reminding the reader that we compared our theory with numerical calculations in Fig.~\ref{fig:nucleationFlux}, without explaining how we set up the latter. This we do now, and subsequently re-examine the results presented in Fig.~\ref{fig:nucleationFlux}, returning in detail to the origins of the decreasing accuracy for the highest values of $\omega_*$ in Fig.~\ref{fig:nucleationFlux}. Finally, as we found in the previous section that the induction time vanishes at very small supersaturation, we expect that only the tQSS nucleation flux of Eq.~\eqref{eq:correctedNucleationFluxResult} is relevant for experimental situations. Hence, we next focus attention on the nucleation flux at times much later than the induction time and discuss the differences between the tQSS and the QSS fluxes.

For the numerical calculations we solve the discrete version of our nucleation equations, which are numerically more simple to solve. The discrete version of the nucleation Master equation Eq.~\eqref{eq:MasterEqContinuum} reads
\begin{equation}\label{eq:numericMaster}
  \dpart{\rho_n(t)}{t} = k^+_{n-1,n}(t) \rho_{n-1}(t) - \left(k^-_{n,n-1}(t) + k^+_{n,n+1}(t)\right) \rho_{n}(t) + k^-_{n+1,n}(t)\rho_{n+1}(t) + K_n(t).
\end{equation}
Here, the subscripts indicate the cluster size, with $k^+_{n,n+1}(t)$ the attachment frequency for a monomer to an $n$-mer, $k^-_{n,n-1}(t)$ the detachment frequency of a monomer from an $n$-mer and $K_n(t)$ the evaporation source term given by Eq.~\eqref{eq:EvaporationSourceTerm}. We associate the detachment frequencies to the attachment frequencies using the same detailed balance argument we used for the continuum nucleation equation. See the discussion of Eq.~\eqref{eq:Fluxinnspace} in Section \ref{sec:CNT}. To apply these within the discrete description, we write
\begin{equation}
    k^+_{n,n+1}(t) F_\mathrm{eq}(n,t) = k^-_{n+1,n}(t) F_\mathrm{eq}(n+1,t),
\end{equation}
with $F_\mathrm{eq}$ again the ``equilibrium" cluster densities \eqref{eq:ConstraintEquilibrium}. We solve the system of equations using a standard backward Euler method with time step size $\Delta \tlt = 0.05$, assuming $n_\mathrm{max} = 10000 \gg n_*$ as the maximum cluster size. This value for $n_\mathrm{max}$ is sufficiently large not to affect our results for the nucleation flux. 

In order to solve Eq.~\eqref{eq:numericMaster}, we need to set appropriate initial conditions. Here, we again use the arbitrarily chosen nucleation barrier of $\beta\Delta W_*(0) = 70$ and critical cluster size of $n_*(0) = 152$, and compare for different values for the non-stationarity index $\omega_*$. These values for $\beta\Delta W_*(0)$ and $n_*(0)$ ensure that the nucleation barrier is sufficiently high such that the high-barrier-assumption holds and that our analytical results should in principle be accurate. As long as the constraint presented in Eq.~\eqref{eq:constraintomega} is satisfied, other values for the initial conditions yield results that are qualitatively the same if we account for the change in the non-stationarity index $\omega_*\sim \beta \Delta W_*^{5/2}$ at fixed evaporation rate, as discussed above.

Before we can discuss the comparison of our theoretical and numerical results, we need to point out possible differences between the solutions to the continuous and the discrete nucleation equations~\cite{Shneidman1992, Wu1992continuum}. Specifically, they may result in slightly different induction times. The magnitude of these differences turn out to depend strongly on the aggregation model used. For the reaction-limited growth model we use, these differences are known to be very minor or even negligible, justifying a comparison between results from the discrete and continuum equations~\cite{Shneidman1992,Shneidman2010}. 

We can now re-examine the theoretical and numerical results shown in Fig.~\ref{fig:nucleationFlux}. As discussed in Section.~\ref{sec:nucleationflux}, the theoretical nucleation fluxes shown in the left panel of Fig.~\ref{fig:nucleationFlux} are in very good agreement with the numerical results for the values for $\omega_*$ shown, although the agreement decreases somewhat for the larger values of $\omega_*$. We can now explain this by considering the time scale $\tau_\mathrm{s}$. For the smallest values of $\omega_*$ shown ($\omega_* = 5\times 10^{-3}$), the time scale $\tau_\mathrm{s} =1.1 \times 10^{5}$ is very large, meaning that the assumptions underlying our analytical results are warranted. For the largest value of $\omega_*$ shown, the time scale $\tau_\mathrm{s} =1.1 \times 10^{1}$ is not particularly large, so we should expect that the theoretical expressions are less accurate. Still, we find that the deviations are about 20 \% or smaller for the tQSS flux $\tlt > 12$. 

\begin{figure}
    \begin{subfigure}[b]{0.49\textwidth}
        \includegraphics[width=\textwidth]{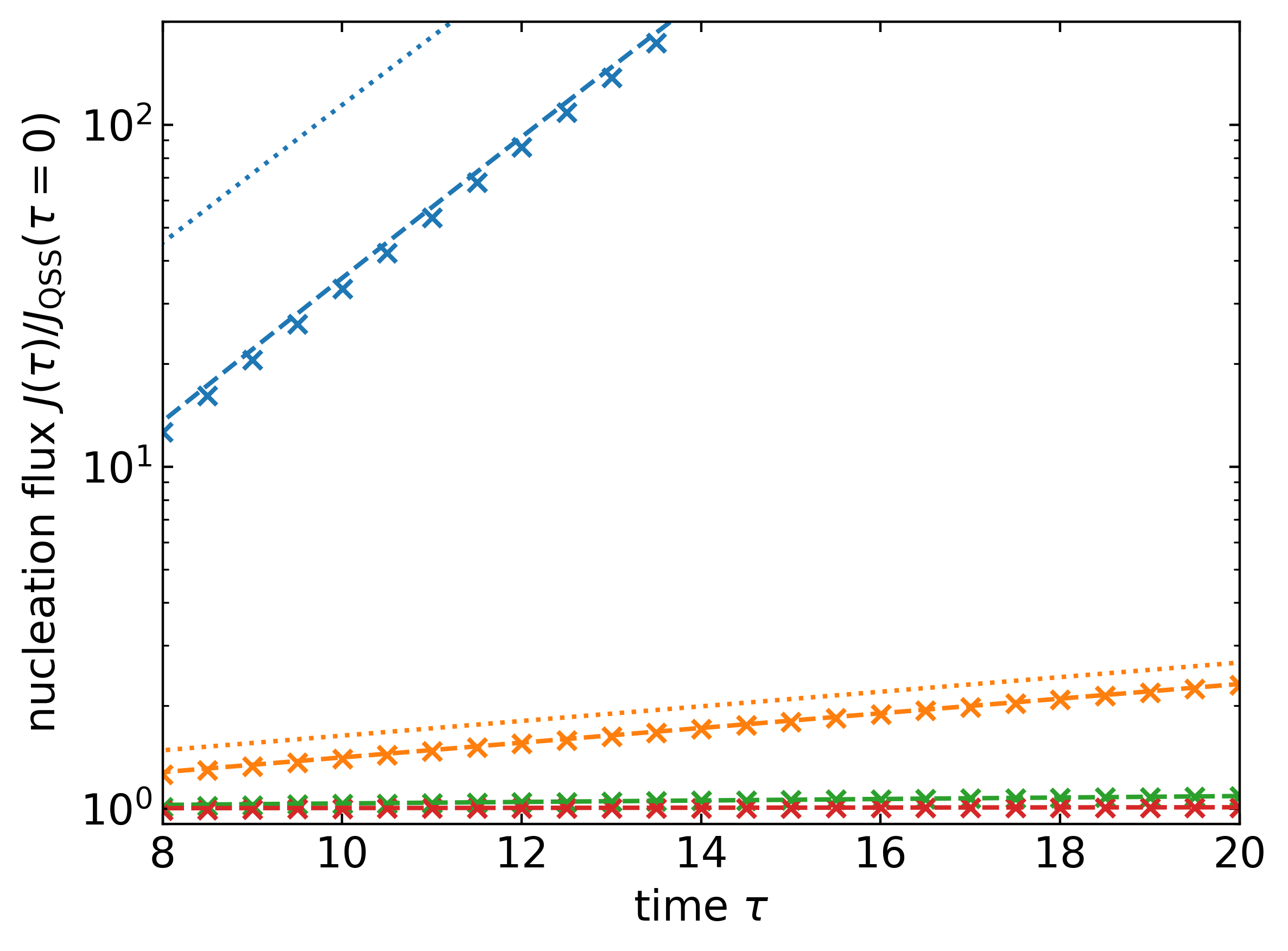}
    \end{subfigure}
    \caption{The nucleation flux $J(x,\tau)$ evaluated at the scaled cluster size $x = 1.5$, normalized by the $J_\mathrm{QSS}(\tlt=0)$ (see Eq.~\eqref{eq:QSS}) as a function of the scaled time $\tlt$. Crosses are the numerical solutions with time interval between the data points of $\Delta \tau = 1/4$, dotted lines are the QSS flux Eq.~\eqref{eq:QSS}, dashed lines the tQSS flux given by Eq.~\eqref{eq:correctedNucleationFluxResult}.  We set the critical cluster size $n_* = 152$ and initial nucleation barrier $\beta\Delta W_*(\tlt=0) = 70$. We show results for $\omega_*(0) = 5 \times 10^{-3}$, $\tau_\mathrm{s} =1.1 \times 10^{5}$ (red), $\omega_*(0) = 5 \times 10^{-2}$, $\tau_\mathrm{s} = 1.1 \times 10^{4}$ (green), $\omega_*(0) = 5\times 10^{-1}$, $\tau_\mathrm{s} =1.1 \times 10^{3}$ (orange) and $\omega_*(0) = 1.12$, $\tau_\mathrm{s} = 5.6\times 10^{1}$ (blue).}
    \label{fig:nucleationFlux2}
    \centering
\end{figure}

Again, we expect that only the tQSS flux and not the initial wind-up period can be observed under experimental conditions, where at time zero we cross the binodal during evaporation. The question might arise how much more accurate the tQSS flux is compared to the QSS flux. This we focus on in Fig.~\ref{fig:nucleationFlux2}, where we plot the numerical results from the same calculations as those shown in the left panel of Fig.~\ref{fig:nucleationFlux} (crosses), so $n_* = 152$ and $\beta \Delta W(0) = 70$ and compare them with the QSS flux Eq.~\eqref{eq:QSS} (dotted, color-matched), as well as the tQSS flux of Eq.~\eqref{eq:correctedNucleationFluxResult} (dashed, color matched). As discussed above, we let $\omega_*$, $n_*$ and $\varepsilon$ depend on the shifted time $\tlt - \tlt_\mathrm{i}(x)$ to account for the weak drift in these quantities. In Fig.~\ref{fig:nucleationFlux2} we do not include the initial time for $\tlt < 8$, but focus on the theoretical expressions that are accurate at later times. For these later times, we find that for low values of $\omega_*$ (red and green curves), the QSS and tQSS fluxes are, for all intends and purposes, identical. This we also expect based on the small $\omega_*$ limit of Eq.~\eqref{eq:correctedNucleationFluxResult}, see Eq.~\eqref{eq:transQSS}.  At $\tlt = 12$ we are (far) beyond the induction time, and find that at this instant in time for higher values of $\omega_*$, the QSS flux overpredicts the nucleation flux by about 70 \% for $\omega_* = 0.5$ (orange), and by about 90 \% for $\omega_* = 1.0$ (blue). The tQSS flux also overpredicts the nucleation flux, but does so to a significantly smaller extent, about 10 \% for $\omega_* = 0.5$, and about 20 \% for $\omega_* = 1.0$. 

Clearly, the tQSS nucleation flux (taking into account the time-drifting $\omega_*$, $n_*$ and $\varepsilon$) provides a significant improvement over the QSS approximation, even at relatively fast solvent evaporation where the assumptions required for its derivation are not necessarily justified. We attribute the eventual decrease in accuracy of the tQSS flux to the (relatively) small value of $\tlt_\mathrm{s}$, see Eq.~\eqref{eq:constraintomega}. The temporal drift in the non-stationarity index $\omega_*$, the critical cluster size $n_*$ and in $\varepsilon$ can in this case not be accounted for \textit{a posteriori}. For $\tlt_\mathrm{s}$ small, the temporal drift should actually be taken into account in the calculations presented in Sec.~\ref{sec:TNT},~\ref{sec:clustersizedistribution}~and~\ref{sec:nucleationflux}. This significantly complicates finding analytical solutions to the nucleation Master equation, and we are at present not aware of any method that allows one to achieve this. 

\section{Discussion and Conclusion}\label{sec:discussionandconclusion}
In this work we study the effect of solvent evaporation on the rate of homogeneous nucleation of a solute in a two-component solution to form a solid precipitate. Solvent evaporation causes the nucleation conditions to change continuously, implying that a description based on stationary nucleation theory cannot apply. In order to explicitly account for the effect of solvent evaporation on the rate at which supercritical clusters emerge, we extend classical nucleation theory for an instantaneous quench in the metastable region upon which nucleation commences under the conditions of on-going evaporation. Our findings demonstrate that additional timescales emerge that are not present in either stationary or quasi-steady-state approaches that have been put forward in the literature~\cite{Kashchiev2000}. We find that the ratio of two of those time scales dictates how strongly evaporation affects the nucleation process.

The three most important time scales we extract from our theory are (1) the timescale associated with solvent evaporation, (2) the time scale related to the lifetime of a critical cluster and (3) a generalized induction time, which is a measure for how quickly nucleation responds to an initial quench. The latter two we find to also be influenced by the rate of solvent evaporation. It turns out that the magnitude of the effect of solvent evaporation can be measured by a single quantity that we dubbed the \textit{non-stationarity index}, which is proportional to the ratio of the lifetime of a critical cluster and the evaporation timescale, and proportional to the number of molecules in a critical cluster. If the value of this quantity is very much smaller than unity, evaporation is slow and the relevant evaporation timescale is large. Hence, solvent evaporation only weakly affects nucleation and we retrieve a time-shifted, quasi-steady nucleation~\cite{Kashchiev2000}. Surprisingly, this time-shifted quasi-steady-state is never equivalent to the `standard' (not time-shifted) quasi-steady-state limit of nucleation, except in the trivial case of stationary nucleation~\cite{Kashchiev2000}. In the actual quasi-steady limit we must always account for a delayed response to ambient conditions, which, of course, is a direct consequence of the time lag of nucleation. The customary quasi-steady-state nucleation flux and time-shifted quasi-static flux we obtain are related via a factor that scales exponentially with the non-stationarity index. This is not only true for solvent evaporation, but can actually be generalized to arbitrary sources of non-stationarity~\cite{Shneidman2010} and should therefore have consequences in other fields where nucleation is non-stationary.

If the non-stationarity index is large, then solvent-evaporation strongly affects nucleation. The nucleation cascade cannot sufficiently rapidly accommodate changes in the ambient conditions, causing its rate to increase less rapidly than is to be expected presuming an instantaneous response. Faster evaporation causes this delay to increasingly strongly affect the nucleation rate. The delay in response to the ambient conditions we find to be related to a generalized induction time for which we have derived an analytical expression. Our prediction for the generalized induction time shows that it must \textit{decrease} with increasing evaporation rate. This agrees with our expectation that an increasing supersaturation due to solvent evaporation must result in continuously faster nucleation. The product of the evaporation rate and the delay time sets the magnitude of the effect of solvent evaporation on nucleation. Hence, the rate of nucleation is governed by two counteracting effects originating from the solvent evaporation: while a faster evaporation rate results in an increasingly larger influence of the delay time on the nucleation flux, this delay time itself effectively decreases with increasing evaporation rate. For sufficiently fast evaporation, which we argue to always occur sufficiently close to the binodal, our predictions indicate that the generalized induction time becomes negative, and in a sense becomes virtual. This we do not expect to actually happen because our theory should break down for a sufficiently large non-stationarity index. Under those circumstances, the key assumption underlying our theory that the timescale over which the nucleation barrier drifts with time is much larger than the nucleation time scale no longer applies. The reason is that the former is inversely proportional to the non-stationarity index.

Experimentally, producing an instantaneous, non-zero supersaturation upon which the solvent evaporation commences, seems difficult to achieve. Indeed, the inspiration of this work is the wet deposition of a spin-coated or slot-die cast thin film of a crystalline organic semiconductor used in, \textit{e.g.}, the production of organic transistors~\cite{Mei2013}, where the solution quickly but gradually changes from undersaturated to supersaturated~\cite{Mei2013}. For early times just after crossing the binodal, nuclei arguably do not emerge as nucleation is exceedingly slow very close to the binodal. Hence, the distinction between an initially undersaturated solution and an initially supersaturated solution becomes, in a sense, somewhat arbitrary. To connect with spin-coating or slot-dying experiments, we argue that in the theory the initial supersaturation should be set to the conditions corresponding to the time before the instant at which the first nuclei emerge. The associated effective supersaturation at this kind of ``time zero'' can even be calculated if a microscopic understanding of the underlying nucleation kinetics is available, informed for instance by experiments. Here, we steer away from attempting to precisely pinpoint this quantity, and suffice by arguing that any resulting large value of the nucleation barrier will do. Our results depend only weakly on the nucleation barrier at zero time, provided it is sufficiently large compared to the thermal energy.

Theoretically, the analysis of nucleation under time-dependent supersaturation has always been hindered by the presence of multiple time and reaction co-ordinate (the cluster size) scales that in general both also depend on time, and do so differently~\cite{Trinkaus1987NucleationSupersaturation}. This causes the kinetic nucleation equations to be exceedingly hard to solve analytically~\cite{Kashchiev2000,Trinkaus1987NucleationSupersaturation,Shneidman1987,Shneidman1988}. This is also true for the relevant time and reaction co-ordinate scales within our model. We solved this problem by making use of a co-ordinate system that is co-moving with respect to both the reaction co-ordinate and time, following the earlier work of Shneidman~\cite{Shneidman2010}. To obtain analytical predictions within the prescription of a co-moving co-ordinate frame, we presume that the rate of change of the nucleation barrier is constant in the co-moving time coordinate. We find that this is reasonable as long as the timescale over which the nucleation barrier drifts in time is much longer than the nucleation time scale. 

While this is strictly not true in reality, we can \textit{post hoc} account for such a drift in time, as long as this drift in time is sufficiently slow. This procedure is, of course, reminiscent of what is done to obtain the quasi-steady-state nucleation flux~\cite{Kashchiev2000,Shneidman2010}. The rate of change of the rate of change of the nucleation barrier that emerges sets a timescale that depends on the ratio of the nucleation barrier height and the non-stationarity index. If this timescale is sufficiently large (on the scale of the nucleation time) the drift of the nucleation barrier in terms of the co-moving time frame can be ignored. Our numerical calculations show that even if this timescale is not large, our theory turns out to be much more accurate than the usual quasi-steady-state theory. 

To conclude, we believe that our results increase the current understanding of the crystallization or solidification that occurs during the casting of solutions for thin film materials. The rate of solvent evaporation has been used experimentally to exert control over crystallization processes during the solution casting of functional thin films~\cite{Hu2020NucleationCells,Zeng2020ControllingCoatings}. How solvent evaporation affects the final crystal structure is not only determined by nucleation itself but also by the subsequent growth processes, which are material specific~\cite{Habraken2013Ion-associationPhosphate}. For example, for perovskite thin film solar cells, a higher evaporation rate results in more and smaller crystalline domains~\cite{Hu2020NucleationCells,Zeng2020ControllingCoatings}, whereas for experimental work on a fullerene-based solar cell, the rate of evaporation appears to have little-to-no effect on the mean size of the fiber-like crystal structure~\cite{vanFraneker2015PolymerFormation}. To be able to make tangible predictions on crystallized morphology based on the present contribution, we need to account for the relevant growth process as well. 
This can be achieved by, for example, combining our results with phase-field simulations~\cite{Michels2021PredictiveCoating} or incorporate it in a nucleation-and-growth model such as the Johnson-Mehl-Avrami-Kolmogorov theory~\cite{Cahn1995TheDomain}, which is often applied in combination with \textit{in-situ} diagnostics of the formation of functional polycrystalline thin films~\cite{Levitsky2021BridgingHeterojunction}. This would allow us to study not only the initial nucleation, but also the subsequent crystallite growth as well as the grain statistics of the polycrystalline final structure.

A possible extension of our work would be to account for the additional solutes and solvents that are commonly used to control the morphology and the performance of the devices~\cite{vanFraneker2015PolymerFormation,Franeker2015cosolvents}. Such additional components are known to affect the nucleation-and-growth process~\cite{vanFraneker2015PolymerFormation}, which is believed to originate from composition-dependent evaporation rates and surface tensions. These might also induce novel sources of non-stationarity and provide a possible route to gain better control over the nucleation process and the subsequent growth of the crystallites.

\section*{Acknowledgments}
R.d.B., J.J.M. and P.v.d.S. acknowledge funding by the Institute for Complex Molecular Systems at Eindhoven University of Technology. We are grateful to D. Reguera (University of Barcelona) for helpful discussions.

\section*{Data availability}
The data that support the findings of this study are available from the corresponding author upon reasonable request.

\section*{List of Symbols}\label{app:tablesymbols}
\renewcommand{\arraystretch}{0.2}
\begin{table}[H]
\caption{List of symbols}\label{tab:longtable}
\centering
 \begin{tabular}{||c | c | c | c |} 
 \hline
 Symbol & Explanation & Equation & Units \\ [0.3ex] 
 \hline\hline
 $\rho$ & dimensionless cluster densities & & [ - ] \\ 
 $F_\mathrm{eq}$ & dimensionless ``equilibrium'' cluster distribution & & [ - ] \\
 $\nu = \rho/F_\mathrm{eq}$ & reduced cluster densities & \eqref{eq:defnu} & [ - ] \\
 $v_0$ & molecular volume of solute molecules & & [ m$^3$ ] \\
 $J$ & nucleation flux & \eqref{eq:nucflux} & [ s$^{-1}$ ] \\
 $K$ & evaporation source nucleation master equation & & [ s$^{-1}$ ] \\
 $n$ & cluster size reaction coordinate (number of molecules in cluster) & \eqref{eq:EvaporationSourceTerm} & [ - ] \\
 $n_*$ & critical cluster size & \eqref{eq:criticalclustersize} & [ - ] \\
 $x$ & reduced cluster size reaction coordinate & & [ - ] \\
 $\tlt$ & reduced time co-ordinate & \eqref{eq:scaledtime} & [ - ]\\
 $k^+$ & attachment frequencies & & [s$^{-1}$] \\
  $\bar{k}^+$ & dimensionless attachment frequencies & & [ - ] \\
 $\beta$ & reciprocal thermal energy & & $[ J^{-1} ]$ \\
 $\Delta W$ & nucleation free energy & \eqref{eq:maxNucleationBarrier} & $[ J ]$ \\
 $\Delta \mu$ & chemical potential difference between phases & \eqref{eq:ChemPotDifF} & $[ J ]$ \\
 $\gamma_\infty$ & surface tension & & $[ J ]$ \\
 $A$ & area & & [ m$^{2}$ ] \\
 $\rho_\infty$ & solubility limit & & [ - ] \\
 $Z$ & Zeldovich Factor & \eqref{eq:zeldovichfactor} & [ - ] \\
 $\sigma$ & dimensionless surface tension & & [ - ] \\
 $S$ & supersaturation & \eqref{eq:ChemPotDifF} & [ - ] \\
 $V$ & volume in which the solution is contained & & [ m$^{3}$ ] \\
 $N$ & number of solute molecules in the solution & & [ - ] \\
 $\chi$ & dimensionless reciprocal volume & \eqref{eq:EvaporationSourceTerm} & [ - ] \\ 
 $\zeta$ & non-stationarity function & \eqref{eq:defI} & [ - ] \\
 $\omega$ & unmodified non-stationarity index & \eqref{eq:defI}, \eqref{eq:nonstatcoefficient} & [ - ] \\
 $\omega_*$ & non-stationarity index, defined as $\omega_* = \omega( 1-n_*^{-1} )$ &  & [ - ] \\
 $g$ & (deterministic) cluster growth rate & \eqref{eq:detGrowthRate}, \eqref{eq:detgrowthx} & [ - ] \\
 $\td$ & lifetime of a cluster & \eqref{eq:lifetime} & [ s ] \\
 $X$ & Stretched reduced cluster size & & [ - ] \\
 $\tlt_\mathrm{d}$ & subcritical cluster growth time & \eqref{eq:decaytime} & [ - ] \\
 $\tlt_\omega$ & timescale of the non-stationarity effects on the subcritical growth & \eqref{eq:tau_omega} & [ - ] \\ 
 $\psi$ & cluster size dependence of $\zeta$ & \eqref{eq:defI} & [ - ] \\ 
 $J_\mathrm{tQSS}$ & renormalized, time-shifted quasi-steady-state nucleation flux & \eqref{eq:correctedNucleationFluxResult} & [ - ] \\ 
 $J_\mathrm{QSS}$ & quasi-steady-state nucleation flux & \eqref{eq:QSS} & [ - ] \\ 
 $J_\mathrm{SS}$ & steady-state nucleation flux & \eqref{eq:QSS} & [ - ] \\ 
 $J_\mathrm{stat}$ &  nucleation flux under stationary conditions & \eqref{eq:stationaryFlux} & [ - ] \\ 
 $\tlt_\mathrm{i}$ &  incubation time & \eqref{eq:definductiontime} & [ - ] \\ 
 $\tlt_\mathrm{stat}$ &  part of the incubation time & \eqref{eq:constA} & [ - ] \\ 
 $\tlt_\mathrm{non-stat}$ &  non-stationary contribution to the lag time & \eqref{eq:constB} & [ - ] \\ 
 $\tlt_\mathrm{cor}$ &  time shift due to no-stationary effects & & [ - ] \\ 
 $\tlt_\mathrm{s}$ &  time scale over which $\omega$ drifts with time & \eqref{eq:tau_s} & [ - ] \\ 
 \hline
 \end{tabular}
\end{table}

\newpage
\appendix
\section{Matched Asymptotic Expansions}\label{app:technicalDetails}
The dimensionless nucleation master equation Eq.~\eqref{eq:MasterEqSolve} we solve subject to the boundary and initial conditions in Eq.~\eqref{eq:boundaryConditions}. We introduce the Laplace transform $\hat{\nu}(x,s)$ of $\nu(x,\tlt)$ with respect to the time variable, presuming that $\varepsilon$, $\zeta(x)$ and $\bar{k}^+(x)$ do not depend on time. This yields Eq.~\eqref{eq:MasterEqSolveLaPlace}. Let us treat $\varepsilon \ll 1$ as an asymptotically small parameter, implying a high nucleation barrier. The method we use is known as matched asymptotic expansions \cite{Nayfeh2011IntroductionTechniques,Shneidman1987,Shneidman1988,ShiSeinfeldOkuyama1990}. As discussed in the main text, we can distinguish three regions: Two outer regions for $x < 1$ and $x > 1$, and an inner region valid near $|x - 1| < \mathcal{O}(\varepsilon)$. We seek three piece-wise solutions to Eq.~\eqref{eq:MasterEqSolveLaPlace} that are valid in the three regions separately. 

We start with the two outer solutions, where we neglect the diffusive contribution in Eq.~\eqref{eq:MasterEqSolveLaPlace} that is proportional to $\varepsilon^2$. We introduce the order-by-order expansion in $\varepsilon$ as
\begin{equation}
    \hat{\nu}(x,s) = \hat{\nu}_0(x,s) + \varepsilon \hat{\nu}_1(x,s) + \varepsilon^2 \hat{\nu}_2(x,s) + \cdots.
\end{equation}
Inserting this into Eq.~\eqref{eq:MasterEqSolveLaPlace} and only retaining the lowest order terms yields
\begin{equation}
    \left(s + \zeta(x) \right)\hat{\nu}_0(x,s) = g(x)\dpart{\hat{\nu}_0(x,s)}{x}
\end{equation}
for both outer solutions. The solution of this differential equation is an exponential, reading
\begin{equation}
    \hat{\nu}_0(x_1,s) = \hat{\nu}_0(x_0,s)\exp\left[s \int_{x_0}^{x_1} \frac{\dint x'(s + \zeta(x'))}{g(x')}\right].
\end{equation}
Using the boundary conditions given in Eq.~\eqref{eq:boundaryConditions}, we find the leading order solutions for the two different outer regions to satisfy Eqs.~\eqref{eq:smaller}~and~\eqref{eq:larger}.

Next we focus on the inner solution $\hat{\nu}_\mathrm{inner}(x,s)$, valid in the region where $|x - 1| \ll 1$. In this region we cannot neglect the terms involving $\varepsilon$ as in this region, as we discuss in detail in the main text in Sec.~\ref{sec:clustersizedistribution}. Standard technique is now to introduce a so-called stretched variable, to magnify the changes in the solution in this inner region of width of order $\varepsilon$. In our case, the appropriate stretched variable is $X = (x-1)/\varepsilon$ \cite{ShiSeinfeldOkuyama1990,Shneidman1987,Shneidman1988}. In terms of the stretched variables, Eq.~\eqref{eq:MasterEqSolveLaPlace} reads
\begin{align}
    (s + \zeta(X))\hat{\nu}_\mathrm{inner}(X,s) & = \\  \nonumber  &\quad\left[\dpart{}{X}\bar{k}^+(X) + \bar{k}^+(X)\frac{3}{\varepsilon}\left(1 - (1 + \varepsilon X)^{-1/3}\right)\right] \dpart{\hat{\nu}_\mathrm{inner}(X,s)}{X} +\\\nonumber  &\quad \bar{k}^+(X) \dparth{\hat{\nu}_\mathrm{inner}(X,s)}{X}{2},
    \end{align}
where we have inserted explicitly the growth rate $g(X) = \bar{k}^+(X)\frac{3}{\varepsilon}\left(1 - (1 + \varepsilon X)^{-1/3}\right)$ expressed in stretched variables. Next, we introduce the order-by-order expansion for $\hat{\nu}_\mathrm{inner}(X,s)$, mirroring the expansion in $\varepsilon$ for the outer solutions. We apply the same expansion to the reduced attachment frequencies $\bar{k}^+(X)$ and the measure for the non-stationarity $\zeta(X)$. For both, only the zeroth-order terms survives, if we make use of the reasonable assumption that the attachment frequencies vary smoothly with the cluster size $n$. The resulting expression for the inner solution reads
\begin{align}\label{eq:HGF}
\dparth{\hat{\nu}_\mathrm{inner,(0)}(X,s)}{X}{2} + 2 X \dpart{\hat{\nu}_\mathrm{inner,(0)}(X,s)}{X} - 2 \left[s + \zeta(X = 0)\right] \hat{\nu}_\mathrm{inner,(0)}(X,s) = 0,
\end{align}
where we use
\begin{equation}
    \left(1 - (1 + \varepsilon X)^{-1/3}\right) \approx \frac{1}{3} \varepsilon X.
\end{equation}

The ordinary differential equation \eqref{eq:HGF} is a special form of the  differential equation for hypergeometric functions~\cite{2010NISTFunctions}, which for some values for $m = s + \zeta(X = 0)$ may yield analytical solutions. In our case, the only relevant solution reads
\begin{equation}
    \hat{\nu}_\mathrm{inner,(0)}(X,s) = A(m)\mathrm{i}^m \mathrm{erfc}(X) + B(m) \mathrm{i}^m\mathrm{erfc}(-X),
\end{equation}
for $m = s + \zeta(X = 0)$ is a natural number, and the repeated integral of the complementary error function is defined in Eq.~\eqref{eq:ierfc}. This restriction on $m$ to natural numbers is justified, as only the solutions for integer $m$ contribute when we return from the Laplace domain to the time domain. 

In order to obtain expressions for the functions $A(m)$ and $B(m)$, we apply a matching procedure, \textit{i.e.}, demand that at some suitably chosen intermediate point the inner and outer solution must be equal to each other. This we do for both outer solutions. The matching conditions are given in the main text in Eq.~\eqref{eq:matchlower} and Eq.~\eqref{eq:matchupper}. The matching condition Eq.~\eqref{eq:matchupper} straightforwardly yields $B(m) = 0$. We can obtain $A(m)$ by using the asymptotic limit for the repeated integral of the complementary error function 
\begin{equation}
    \lim\limits_{X\to-\infty} \mathrm{i}^{m}\mathrm{erfc}(X) \sim \frac{2(-X)^m}{\Gamma(m+1)},
\end{equation}
and the asymptotic limit of the outer solution for $x< 1$ as
\begin{equation}
    \lim\limits_{x\to1^-} \hat{\nu}_0(x < 1,s) \sim (1-x)^m \exp\{s \tlt_\mathrm{stat} + \omega_*\tlt_\mathrm{non-stat}\},
\end{equation}
with $\tlt_\mathrm{stat}$ given by Eq.~\eqref{eq:defCconstant} and $\tlt_\mathrm{non-stat}$ by Eq.~\eqref{eq:defDconstant}. Inserting these expressions in Eq.~\eqref{eq:matchupper} yields 
\begin{equation}
    A(m) = s^{-1} \varepsilon^{m}\Gamma(m+1)\exp\{s \tlt_\mathrm{stat} + \omega_*\tlt_\mathrm{non-stat}\},
\end{equation}
resulting in the inner solution
\begin{equation}\label{eq:app:innersol}
    \hat{\nu}_\mathrm{inner}(x,s) = \frac{1}{2 s}\varepsilon^{m} \Gamma(m+1)\exp\{s \tlt_\mathrm{stat} + \omega_*\tlt_\mathrm{non-stat}\} \mathrm{i}^m \mathrm{erfc}\left(\frac{x-1}{\varepsilon}\right),
\end{equation}
which is only valid near $x = 1$.

Next, we focus on the inverse Laplace transform of the nucleation flux derived in the main text in Eq.~\eqref{eq:laplacedomflux3}. For ease of notation, we write it as
\begin{equation}\label{eq:laplacedomfluxAPP}
    \widehat{J}(X,s) = G(X,s) \widehat{J}_\mathrm{stat}(X,s).
\end{equation}
with
\begin{equation}\label{eq:laplaceG}
G(X,s) = \left[\frac{s}{(s - \omega_*)}\exp\{-\omega_*\left(\tlt_\mathrm{stat} - \tlt_\mathrm{non-stat}\right)\}\right]. 
\end{equation}
To execute the inverse Laplace transform, we make use of the convolution theorem, which allows us two write the Laplace inverted Eq.~\eqref{eq:laplacedomfluxAPP} as
\begin{equation}\label{eq:ConvolutionInverse}
J(X,\tlt) = \int_0^{\tlt} g(X,\tlt') J_\mathrm{stat}(X,\tlt - \tlt') \dint \tlt',
\end{equation}
where $g(X,t)$ reads
\begin{equation}\label{eq:ConvolutionF1}
    g(X,\tlt) = \left(\delta(\tlt) + \omega_* \mathrm{e}^{\omega_* \tlt}\right)\exp\{-\omega_*\left(\tlt_\mathrm{stat} - \tlt_\mathrm{non-stat}\right)\}, 
\end{equation}
with $\delta(\tlt)$ the delta function, and $J_\mathrm{stat}(X,\tlt - \tlt')$ is the stationary nucleation flux defined in Eq.~\eqref{eq:stationaryFlux}.

We first rewrite Eq.~\eqref{eq:laplacedomfluxAPP} as
\begin{equation}
    J(X,\tlt)\exp\{\omega_*\left(\tlt_\mathrm{stat} - \tlt_\mathrm{non-stat}\right)\} = J_\mathrm{stat}(X,\tlt) +  \int_0^{\tlt} \omega_* \mathrm{e}^{\omega_* \tlt} J_\mathrm{stat}(X,\tlt') \dint \tlt'.
\end{equation}
Next, we use integration by parts to obtain 
\begin{align}\label{eq:ConvolutionInverse}
&J(X,\tlt)\exp\{\omega_*\left(\tlt_\mathrm{stat} - \tlt_\mathrm{non-stat}\right)\} = \\
&\quad J_\mathrm{stat}(X,\tlt) + \left[\mathrm{e}^{\omega_* \tlt'}J_\mathrm{stat}(X,\tlt - \tlt') \right]_{\tlt' = 0}^{\tlt' = \tlt} - \int_0^{\tlt} \mathrm{e}^{\omega_* \tlt'} \dpart{}{\tlt'}J_\mathrm{stat}(X,\tlt - \tlt') \dint \tlt'. \nonumber
\end{align}
This reduces to
\begin{align}\label{eq:ConvolutionInverse2}
J(X,\tlt)\exp\{\omega_*\left(\tlt_\mathrm{stat} - \tlt_\mathrm{non-stat}\right)\} = \left[\mathrm{e}^{\omega_* \tlt}J_\mathrm{stat}(X,0) \right] - \int_0^{\tlt} \mathrm{e}^{\omega_* \tlt'} \dpart{}{\tlt'}J_\mathrm{stat}(X,\tlt - \tlt') \dint \tlt'.
\end{align}
We set the first term equal to zero, as the initial stationary flux just after an instantaneous quench into a metastable state should be zero. In reality, we find that $J_\mathrm{stat}(X,0)$ contains a (minor) inconsistency as it is not actually equal to zero. It is sufficiently small that we are nevertheless justified in setting it to zero. See the discussion near Eq.~\eqref{eq:stationaryFlux}. The remaining integral reads
\begin{equation}\label{eq:ConvolutionInverse3}
J(X,\tlt) = \exp\{-\omega_*\left(\tlt_\mathrm{stat} - \tlt_\mathrm{non-stat}\right)\} \int_0^{\tlt} \mathrm{e}^{\omega_* \tlt'} \dpart{}{\tlt'}J_\mathrm{stat}(X,\tlt - \tlt') \dint \tlt'.
\end{equation}
We now use that $J_\mathrm{stat}(X,\tlt) = J_\mathrm{SS}f(X,\tlt)$, with $J_\mathrm{SS}$ the steady-state nucleation flux and $f(X,\tlt)$ a function that accounts for the initial induction period, see Eq.~\eqref{eq:stationaryFlux}. Next, we define $J_\mathrm{QSS}(0) = J_\mathrm{SS}$ to correspond the the (quasi-)steady-state flux using the initial conditions. This allows us to write
\begin{equation}\label{eq:ConvolutionInverse4}
J(X,\tlt) =  \int_0^{\tlt} {J_\mathrm{QSS}(0)}\mathrm{e}^{\omega_* \tlt'}\exp\{-\omega_*\left(\tlt_\mathrm{stat} - \tlt_\mathrm{non-stat}\right)\} \dpart{}{\tlt'}\frac{J_\mathrm{stat}(X,\tlt - \tlt')}{J_\mathrm{QSS}(0)} \dint \tlt'.
\end{equation}
Using Eq.~\eqref{eq:QSS} and presuming that the time drift of the pre-exponential in the quasi-steady-state nucleation flux is small relative to the contributions due to the exponential, this can be rewritten as
\begin{equation}\label{eq:ConvolutionInverseFinal}
J(X,\tlt) =  \int_0^{\tlt} J_\mathrm{QSS}\left(\tlt' - \tlt_\mathrm{stat.} + \tlt_\mathrm{non-stat}\right)\frac{J_\mathrm{stat}(X,\tlt - \tlt')}{J_\mathrm{QSS}(0)} \dint \tlt'.
\end{equation}
After re-introducing the non-stretched variable $x$ and using the definition $\tk = \tlt_\mathrm{non-stat} - \tlt_\mathrm{stat.}$, we find Eq.~\eqref{eq:RealDomain}.

\section{Heterogeneous nucleation}\label{app:heterogeneousNucleation}
\begin{figure}
    \begin{subfigure}[b]{0.49\textwidth}
        \includegraphics[width=\textwidth]{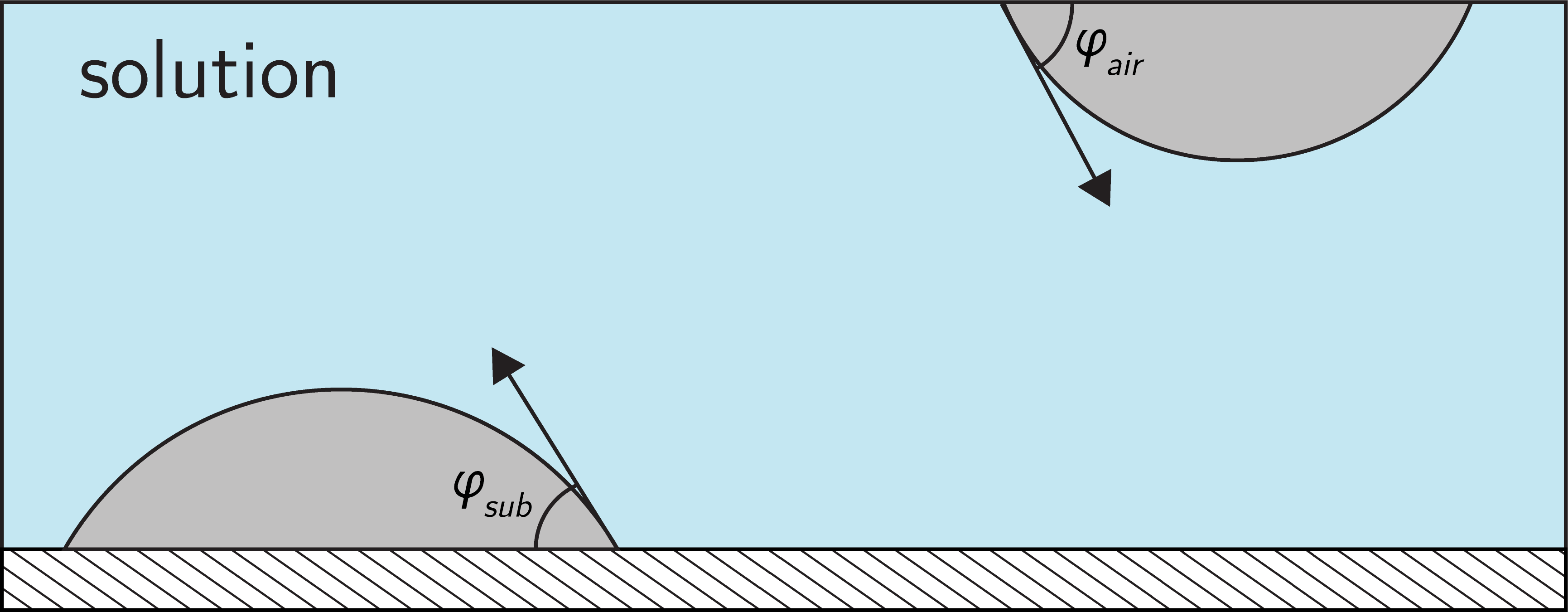}
    \end{subfigure}
    \caption{Schematic representation of cluster geometries during heterogeneous nucleation. The substrate is located at the bottom and dashed. The solution-air interface is located at the top. Nuclei attach to either the solution-air interface (top-right) or solution substrate interface (bottom-left). The clusters are cap-shaped with contact angles $\phi_\mathrm{sub}$ and $\phi_\mathrm{air}$ for the solution-substrate and solution-air interfaces, respectively. }
    \label{fig:heterogeneousNucleation}
    \centering
\end{figure}

The extension of our work to heterogeneous nucleation is straightforward, as we show next. Let us consider the generic case where the nuclei that form partially wet the solution-air interface or the substrate-solution interface, which we both presume to be flat. Here, for the sake of simplicity, we presume that the nuclei are cap-shaped, see Fig.~\ref{fig:heterogeneousNucleation} for a schematic representation. While such cluster shapes seem unphysical for heterogeneous crystal nucleation due to the underlying crystal symmetry and associated anisotropic interfacial tension, it turns out that the phenomenology of this model is still applicable to crystal nucleation~\cite{Turnbull1952NucleationCatalysis.,Kelton2010NucleationBiology,Djikaev2003ThermodynamicsNucleation}. Hence, we use it for illustrative purposes. 

Focusing first on the thermodynamics of heterogeneous nucleation, where we neglect any contribution from a line tension, the nucleation free energy is given by \cite{Kashchiev2000}
\begin{equation}\label{eq:HetNucBar}
\beta \Delta W_\mathrm{Het} = - n \beta \Delta \mu + f(\phi) \sigma n^{2/3},
\end{equation}
where $\sigma$ is the dimensionless surface tension between the solution and the nucleus (see also Eq.~\eqref{eq:NucleationFreeEnergy} in the main text), and  the scaling function $f(\phi)$ is a function that describes (i) the change in volume to surface ratio present in heterogeneous nucleation, and (ii) the change in surface energetics as the different type of interfaces have different surface tensions~\cite{Kashchiev2000,Kelton2010NucleationBiology}. In general, additional physics, such as that arising from the presence of a line tension, can be incorporated in the scaling function $f(\phi)$, in which case the scaling function might also depend on the cluster size $n$. Eq.~\eqref{eq:HetNucBar} shows that the difference between homogeneous and heterogeneous nucleation can, in general, be accounted for by a renormalized surface tension. This can lower~\cite{Turnbull1952NucleationCatalysis.,Kelton2010NucleationBiology} or increase~\cite{Djikaev2003ThermodynamicsNucleation} the nucleation barrier with respect to homogeneous nucleation. We refer to Refs.\cite{Turnbull1952NucleationCatalysis.,Kashchiev2000,Kelton2010NucleationBiology} for details on the scaling function $f(\phi)$. 

Note that we do not need to subtract the self-consistent correction from the free energy to set the free energy to form a monomer to zero as we did in the case of homogeneous nucleation, see Section.~\ref{sec:CNT} in the main text. For heterogeneous nucleation, we must distinguish between a monomer in free solution and a monomeric cluster, that is, a monomer that is adsorbed onto a surface, adding an adsorption free energy to the free energy barrier. The change in chemical potential $\Delta \mu$ we presume similar to that in homogeneous nucleation, that is, we presume the chemical potential to be given by Eq.~\eqref{eq:ChemPotDifF}, where the density $\rho(t)$ acts as the local solute density at substrate or solution-gas interface and $\rho_\infty$ the solubility limit. All thermodynamic quantities such as the critical nucleation barrier, the critical cluster size, and the Zeldovich factor presented in the main text for homogeneous nucleation apply for heterogeneous nucleation, if we account for the renormalized (effective) surface tension by the factor $f(\phi)$ and if we exclude the self-consistent monomer correction. These quantities are given by
\begin{equation}
    \beta\Delta W^*_\mathrm{Het} = \frac{1}{3} n_*^{2/3} f(\phi) \sigma = 3 \varepsilon^{-2},
\end{equation}
\begin{equation}
    n^*_\mathrm{Het} = \left(\frac{2 f(\phi)\sigma}{3 \ln S(t)}\right)^{3},
\end{equation}
and 
\begin{equation}
    Z_\mathrm{Het} = \sqrt{- \frac{1}{2 \pi} \dparth{\Delta W}{n}{2}}\Bigg|_{n=n_*} = \sqrt{\frac{ f(\phi) \sigma}{\pi}}\frac{1}{3}{n_*}^{-2/3}.
\end{equation}
The expression for the equilibrium distribution does change somewhat, and is now given by 
\begin{equation}
    F_\mathrm{eq}(n,t) = \rho_0 e^{-\beta \Delta W_\mathrm{Het}},
\end{equation}
where $\rho_0$ is the \textit{surface} density of nucleation sites, which for a flat interface is constant in time as evaporation does not affect this quantity.

The kinetics of heterogeneous nucleation are slightly different, since the continuity equation does not need to be supplemented by a source term. The reason is of course that the number of clusters per surface area does not change in time due to solvent evaporation. Hence, we find 
\begin{equation}
    \dpart{\rho(n,t)}{t} = - \dpart{J(n,t)}{n},
\end{equation}
where $J(n,t)$ is still given by the Zeldovich form Eq.~\eqref{eq:Fluxinnspace}.
Although solvent evaporation is not explicitly present in the continuity equation, it remains present as a supersaturation in the nucleation barrier and hence a model for solvent evaporation is still required. 

Combining the thermodynamic and kinetic descriptions of heterogeneous nucleation in the same way as we did in the main body of the article, we find
\begin{align}\label{eq:appendixMasterSolve}
    \dpart{\nu(x,\tlt)}{\tlt} & + \nu(x,\tlt)\zeta(x,\tlt) = \\  \nonumber  &\quad\left[\frac{\varepsilon^2}{2}\dpart{\bar{k}^+(x)}{x}- \varepsilon^2 \bar{k}^+(x) \dpart{\beta \Delta W_\mathrm{Het}(x,\tlt)}{x}  + x \dpart{\ln n_*(\tlt)}{\tlt}\right] \dpart{\nu(x,\tlt)}{x} +\\\nonumber  &\quad\frac{\varepsilon^2}{2} \bar{k}^+(x) \dparth{\nu(x,\tlt)}{x}{2},
\end{align}
expressed in the co-moving $\{x,\tlt\}$ co-ordinate system, see Sec.~\ref{sec:TNT}, and must be supplemented by the same boundary conditions as in the homogeneous nucleation case, Eq.~\eqref{eq:boundaryConditions}. Here, we define the function $\zeta(x,\tlt)$ that describes the non-stationarity
\begin{equation}
    \zeta(x,\tlt) = -\dpart{}{\tlt} \beta \Delta W_\mathrm{Het}(x,\tlt) = \omega \psi(x),
\end{equation}
see Eq.~\eqref{eq:defI} in the main text. For the non-stationarity coefficient $\omega$ we must account not only for the changing supersaturation, but also consider that the function $f(\phi)$ might, in principle, depend on time. For instance, for crystal nucleation the surface tensions of the different facets might depend on the (time-dependent) solute concentration. This causes the effective surface tension in Eq.~\eqref{eq:HetNucBar} to become time-dependent. If we for simplicity use the illustrative example of hemispherical nuclei shown in Fig.~\ref{fig:heterogeneousNucleation}, a time-dependent surface tension could can be related to a time-dependent surface-nuclei contact angle. This results in a non-stationarity coefficient that can be shown to be given by 
\begin{equation}
    \omega =  n_* \dpart{\ln \chi(\tlt)}{\tlt} -  \sigma n_*^{2/3}\dpart{f}{\phi}\dpart{\phi}{\tlt},
\end{equation}
Here, the first term is due to the non-stationarity originating from the time-dependent supersaturation and is also present in homogeneous nucleation, see \eqref{eq:defI}. The second term accounts for the change in the contact angles. Similar contributions can be included if the scaling function $f(\phi)$ turns out to depend on time. 

Presuming that $\omega$ remains approximately constant within the co-moving co-ordinate system described in the main text, the solution to Eq.~\eqref{eq:appendixMasterSolve} for the nucleation flux is equivalent to the expression found for homogeneous nucleation Eqs.~\eqref{eq:nucleationFluxResult}~and~Eq.~\eqref{eq:correctedNucleationFluxResult}. We must, however, replace the critical nucleus size, the critical nucleation free energy, the Zeldovich factor and the non-stationarity factor $\omega$ to  by their respective expressions presented in this Appendix. Moreover, the nucleation attachment mode might be different, so another description for the attachment frequencies might be required. The same holds for the induction time Eq.~\eqref{eq:ResultInductionTime}. 

Hence, we conclude that the general theory presented in the main text of this work applies also to heterogeneous nucleation, \textit{mutatis mutandis}.

\section*{References}
\bibliography{bibliography.bib}

\end{document}